\documentclass[12pt]{article}
\usepackage{graphicx,amssymb,amsmath,empheq}
\usepackage[dvipsnames]{xcolor}
\usepackage{color}
\usepackage{authblk}
\usepackage{braket}
\usepackage{amsthm}
\usepackage{units}
\usepackage{empheq}
\usepackage{booktabs}
\usepackage{subfig} % caution: not compatible with revtex4
\usepackage{bigints}
\usepackage[mathscr]{euscript}

\usepackage{hyperref}
\hypersetup{colorlinks = true, linkcolor=black, citecolor=black, urlcolor=blue}
\usepackage{url}
\usepackage{bbm}
\theoremstyle{plain}

\usepackage{tikz}
\usepackage{tikz-cd}
\usetikzlibrary{arrows}
\usetikzlibrary{intersections}
\usetikzlibrary{shapes.geometric}
\usetikzlibrary{decorations.pathmorphing, patterns,shapes}
\usetikzlibrary{decorations.markings}

\tikzset{
  mid arrow/.style={postaction={decorate,decoration={
        markings,
        mark=at position .575 with {\arrow[#1]{stealth}}
      }}},
  near arrow/.style={postaction={decorate,decoration={
        markings,
        mark=at position .275 with {\arrow[#1]{stealth}}
      }}},
   far arrow/.style={postaction={decorate,decoration={
        markings,
        mark=at position .800 with {\arrow[#1]{stealth}}
      }}},
}

\pgfdeclarepatternformonly{south west lines}{\pgfqpoint{-0pt}{-0pt}}{\pgfqpoint{3pt}{3pt}}{\pgfqpoint{3pt}{3pt}}{
        \pgfsetlinewidth{0.4pt}
        \pgfpathmoveto{\pgfqpoint{0pt}{0pt}}
        \pgfpathlineto{\pgfqpoint{3pt}{3pt}}
        \pgfpathmoveto{\pgfqpoint{2.8pt}{-.2pt}}
        \pgfpathlineto{\pgfqpoint{3.2pt}{.2pt}}
        \pgfpathmoveto{\pgfqpoint{-.2pt}{2.8pt}}
        \pgfpathlineto{\pgfqpoint{.2pt}{3.2pt}}
        \pgfusepath{stroke}}

\usepackage[nosort]{cite}

\renewcommand{\bar}{\overline}
\renewcommand{\tilde}{\widetilde}

\renewcommand{\leq}{\leqslant}
\renewcommand{\geq}{\geqslant}
\renewcommand{\Re}{\operatorname{Re}}
\renewcommand{\Im}{\operatorname{Im}}

\newcommand{\Tr}{\operatorname{Tr}}

\newcommand{\const}{\operatorname{const}}

\newcommand{\SL}{\operatorname{SL}}

\newcommand{\CC}{\mathbb{C}}
\newcommand{\RR}{\mathbb{R}}

\newcommand{\ZZ}{\mathbb{Z}}

\newcommand{\calE}{\mathscr{E}}%upright Euler script 

\newcommand{\calO}{\mathcal{O}}

\newcommand{\calT}{\mathcal{T}}

\newcommand*{\wideboxed}[1]{\setlength{\fboxsep}{1ex}%
	\fbox{\m@th$\displaystyle#1$}}

\newcommand{\fraksl}{\mathfrak{sl}}

\newcommand{\eqnref}[1]{Eq.~\eqref{#1}}
\newcommand{\figref}[1]{Fig.~\ref{#1}}
\newcommand{\appref}[1]{Appendix.~\ref{#1}}

\title{
Emergent Spatial Structure and Entanglement Localization in Floquet Conformal Field Theory
}

\author[1]{Ruihua Fan}
\author[1]{Yingfei Gu}
\author[1]{Ashvin Vishwanath}
\author[2]{Xueda Wen}

\affil[1]{\normalsize\it Department of Physics, Harvard University, Cambridge MA 02138, USA}
\affil[2]{\normalsize\it Department of Physics, Massachusetts Institute of Technology, Cambridge, MA 02139, USA}

\begin{document}

\maketitle

\begin{abstract}
We study the energy and entanglement dynamics of $(1+1)$D conformal field theories (CFTs) under a Floquet drive with the sine-square deformed (SSD) Hamiltonian. Previous work has shown this model supports both a non-heating and a heating phase. Here we analytically establish several robust and `super-universal' features of the heating phase which rely on conformal invariance but not on the details of the CFT involved.  First, we show the energy density is concentrated in two peaks in real space, a chiral and anti-chiral peak, which leads to an exponential growth in the total energy.  The peak locations are set by fixed points of the M\"obius tranformation. 
Second, all of the quantum entanglement is shared between these two peaks. 
In each driving period, a number of Bell pairs are generated, with one member pumped to the chiral peak, and the other member pumped to the anti-chiral peak.
These Bell pairs are localized and accumulate at these two peaks, and can serve as a source of quantum entanglement. 
Third, in both the heating and non-heating phases we find that the total energy is related to the half system  entanglement entropy by a simple relation $E(t)\propto c \exp \left( \frac{6}{c}S(t) \right)$ with $c$ being the central charge. 
%This formula holds for both the heating and non-heating phases. 
%The structure of the phase diagram can be understood by studying a periodically driven harmonic oscillator. 
In addition, we show that the non-heating phase, in which the energy and entanglement oscillate in time, is unstable to small fluctuations of the driving frequency in contrast to the heating phase. Finally, we point out an analogy to the periodically driven harmonic oscillator which allows us to understand global features of the phases, and introduce a quasiparticle picture to explain the spatial structure, which can be generalized to setups beyond the SSD construction.   
%The heating phase, where the energy grows exponentially, shows robust features. %against many perturbations. 
%In particular, in the heating phase, the system will develop a chiral and an anti-chiral energy peaks that grow with time exponentially.  These phenomena are universal for any $(1+1)$D CFTs and have a geometric interpretation that can be generalized to other setups beyond this SSD construction. We also find that the total energy is related to the entanglement entropy by a simple relation $E(t)\propto c \exp \left( \frac{6}{c}S(t) \right)$ with $c$ being the central charge. This formula holds for both the heating and non-heating phases. 

\end{abstract}

\tableofcontents

\section{Introduction}
\label{sec:intro}

Floquet driving sets up a new stage in the search for novel systems that 
may not have an equilibrium analog, such as Floquet topological phases\cite{Jiang:2011xv,demler2010prb,rudner2013anomalous,sondhi2016,else2016prb,ashvin2016prx,roy2016prb,Po:2016qlt,roy2017prb,roy2017prl,po2017radical,yao2017floquetspt,po2017timeglide,ashvin2019prb} and time crystals\cite{khemani2016phase, else2016timecrystal, sondhi2016phaseI, sonhdi2016phaseII, Else:2017ghz, normal2017timecrystal, lukin2017exp, zhang2017observation, normal2018classical}. 
It is also one of the simplest protocols to study non-equilibrium phenomena, such as localization-thermalization transitions, prethermalization,
dynamical Casimir effect, etc\cite{rigol2014prx, abanin2014mbl, abanin2014theory, abanin2015prl, abanin2015rigorous, abanin2015effectiveh,Law1994,Dodonov1996,martin2019floquet}. However, exactly solving Floquet many-body systems is, in general, a formidable task. Usually, we have to resort to numerical methods limited to small system size. This makes an analytical understanding of Floquet dynamics extremely valuable. Conformal field theories provide an ideal platform for such a purpose\cite{Berdanier:2017kmd,wen2018floquet}. In particular, for $(1+1)$D CFTs, the conformal symmetry is enlarged to the full Virasoro symmetry, which makes the calculation even more tractable \cite{belavin1984infinite,francesco2012conformal}. In this paper, we  focus on $(1+1)$D CFTs. Generalization to other dimensions should be possible and left to future work.

However, a CFT as a gapless many-body system is expected to be vulnerable to a generic driving. If we start from the ground state of the original Hamiltonian, then Floquet driving might lead it to an infinite temperature state easily. This thermalization process is an interesting problem but not the focus of this paper. Our goal is to explore what type of phenomena and structures can be engineered with a Floquet many-body system that may not be realized in a simple way with a static Hamiltonian. To avoid thermalization, we need to choose special protocols. In this paper, we are going to use the Virasoro symmetry generators as our driving Hamiltonian so that we can  take maximal advantage of the conformal symmetry to constrain the system. As one of the most canonical choices, we will use the $\fraksl(2,\RR)$ subalgebra, the exact meaning of which will be discussed later. Although this choice may look a bit special, it is powerful enough to reveal some universal features of the problem that apply more generally. We will also discuss one generalization of this simplest protocol.

We will follow the setup used in \cite{wen2018floquet}, where the authors consider an open chain and implement the driving with the sine-square deformed Hamiltonian\cite{Nishino2011prb,2011freefermionssd,katsura2012sine,ishibashi2015infinite,ishibashi2016dipolar, Okunishi:2016zat, Wen:2016inm, Tamura:2017vbx, Tada:2017wul, Wen:2018vux,tada2019time}. It was shown that if we start from the ground state and turn on the  Floquet drive, we can identify a non-heating phase in the high-frequency driving regime and a heating phase in the low-frequency driving regime by looking at the entanglement entropy growth. The fact that we have these two phases has an algebraic reason which can be understood by using a quantum mechanical model, as we will discuss later.

The main part of this paper will present a more detailed study on what happens in the Floquet dynamics, paying special attention to the spatial structure that emerges, that has not previously been discussed.
%For example we show that although in the heating regime the system continuously draws energy from the drive and heats up exponentially, it does not evolve into a featureless infinite temperature state. In fact, there are robust spatial features associated with both the energy density and the entanglement structure that emerge.} We will discuss these universal features and their robustness. We will also draw intuition from this special setup and make a few comments on what to expect for the case of more general \AV{Floquet drives, initial conditions etc. ???}.

Let us summarize the main phenomena.  In the heating phase, although the total energy and entanglement keep growing, the system does {\em not} evolve into a featureless state. We find that in this phase, the system heats up in a very non-uniform way. The energy pumped in concentrates at two points, one of which has purely chiral excitations and the other one only has anti-chiral excitations. The entanglement entropy also comes from the entanglement between the excitations at these two points.
%All the energy pumped in  accumulates at two points, one of which has purely chiral energy excitations and the other one only has anti-chiral excitations. All the entanglement entropy also comes from the entanglement between the excitations at these two points. %Second, since energy and entanglement are both carried by these chiral and anti-chiral excitations, it is not hard to deduce the energy and entropy growth are related to each other. 
Furthermore, the energy and entanglement entropy are related by a simple formula. 
All these features above are universal and only depend on the central charge of the CFT. In the non-heating phase, if we do a stroboscopic measurement, we can find that energy excitation will move back and forth in the system with the total energy and entanglement entropy oscillating in time.

Since a real experiment will inevitably have noise, we are also interested in the question that how stable those phenomena are to noise. For example, we could start from an excited state or have local perturbation during evolution. Furthermore, the driving frequency could have a small fluctuation. By combining analytical and numerical analysis, we will argue that the non-heating phase is delicate but all the reported features in the heating phase are quite robust to these perturbations. For the non-heating phase, an arbitrarily tiny noise in the driving frequency will eventually heat the system. The dimensionless heating rate is proportional to $\alpha^2$, where $\alpha$ characterizes the magnitude of randomness.

The paper is organized as follows. In Sec.\ref{sec:review}, we will briefly review the set-up in \cite{wen2018floquet}, summarize the method of studying the evolution of operators and see how to interpret the Hamiltonian by $\fraksl(2,\RR)$ algebra. In particular, in Sec.\ref{sec:phases}, we discuss the phase diagram from a different angle using the mapping to a driven harmonic oscillator. In Sec.\ref{sec:features}, we will present our main result of this paper on various features of the heating phase. We will focus on how the energy is absorbed, how the entanglement is generated and their relation. We will also draw intuition from this special setup and make a few comments on what to expect for the case of more general initial conditions, boundary conditions and Floquet drives. In Sec.\ref{sec: Randomness_nonheating}, we will analyze the stability of these phenomena against driving with random periods. In Sec.\ref{sec:generalization}, we introduce one generalization of the simplest case and study how the spatial structure of the energy and entanglement gets modified. Finally in Sec.\ref{sec:conclusions}, we give some conclusions and outlook. 

\section{Setup for a Floquet CFT}
\label{sec:review}

In this section, in the interest of completeness, we review the set-up and some results of prior work  in \cite{wen2018floquet} that are relevant to our discussion. 

\subsection{Floquet driving} 

We start with a $(1+1)$D CFT with an open boundary condition. Let us denote its total length by $L$, and its central charge by $c$. 
%The system is a 1+1D CFT with the central charge $c$. We put it on an open 1D chain of length $L$ with the conformally invariant boundary condition. The Floquet driving is designed as
We will consider the following time-dependent Hamiltonian
\begin{equation}
	H(t) = \left\{
	\begin{array}{ll}
		H_1 & 0<t<T_1 \\
		H_0 & T_1<t<T_1+T_0
	\end{array}
	\right. \,,
\end{equation}
where
$H_0$ is the ordinary Hamiltonian %for a canonically quantized CFT. 
that can be written as an integral of energy density $T_{00}(x)$ along the real space as follows,
\begin{equation}
	\label{eqn:H0}
	H_0 = \int_0^L dx~ T_{00}(x).
\end{equation}
 $H_1$ is the so-called sine-square deformed (SSD) Hamiltonian\cite{Nishino2011prb,2011freefermionssd,katsura2012sine,ishibashi2015infinite,ishibashi2016dipolar, Okunishi:2016zat, Wen:2016inm, Tamura:2017vbx, Tada:2017wul, Wen:2018vux,tada2019time,Ryu2019SSD}
\begin{equation}
	\label{eqn:H1}
	H_1 = 2\int_0^L dx \sin^2
	\left(\frac{\pi x}{L}\right) T_{00}(x).
\end{equation}
For simplicity, the initial state $|\psi_0\rangle$ is chosen to be {\em the ground state} of $H_0$, i.e. $\ket{\psi_0} = \ket{GS}$.%Although it looks like a special choice, many features we find still apply for other generic initial states.
%which pumps or takes energy from our system.

It is also useful to introduce the Floquet operator $F=F_0F_1 = e^{-iH_0T_0} e^{-iH_1T_1}$ to characterize the unitary evolution for a single cycle. In the stroboscopic measurement, the Floquet dynamics is determined by the state $\ket{\psi(nT)} = F^n\ket{GS}$. For example, the two point function of local operators $O_1(x_1)$ and $O_2(x_2)$ after $n$ cycles is given by
$
     \braket{\psi(nT)|\calO_1(x_1)\calO_2(x_2)|\psi(nT)}.
$ 
In the ``Heisenberg'' picture, the calculation amounts to determining the operator evolution $\calO(x,nT)= F^{-n}\calO(x) F^n$. For general Floquet drives, this is a difficult problem. However, for the SSD Hamiltonian defined in \eqref{eqn:H1}, the operator evolution has a simple expression in terms of the M\"obius transformation.

\subsection{Operator evolution and Mobius transformation}
\label{sec: mobius}

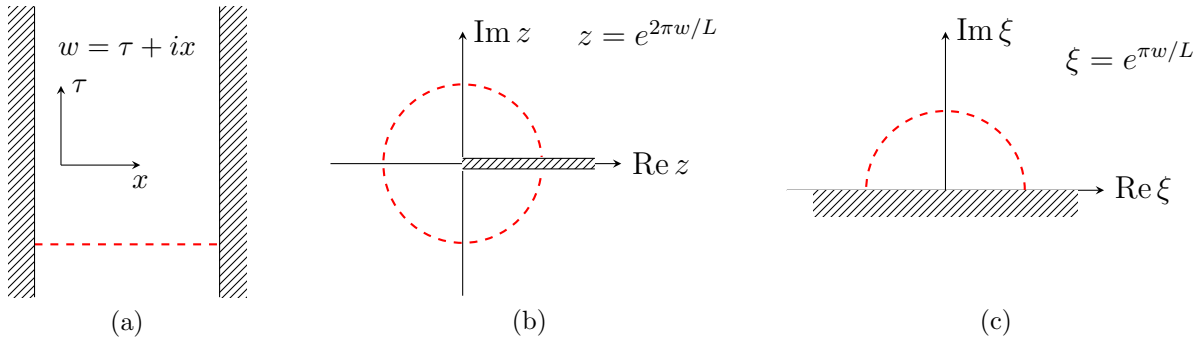
\begin{figure}[t]
    \centering
    \subfloat[]{
   \begin{tikzpicture}[scale=1.0, baseline={(current bounding box.center)}]
    \draw[->,>=stealth](10pt,40pt)--(40pt,40pt) node[below] {\small $x$};
    \draw[->,>=stealth](10pt,40pt)--(10pt,70pt) node[right] {\small $\tau$};
    \draw[dashed, thick, red] (0pt,10pt) -- (70pt,10pt);
    \node at (35pt,85pt) {\small $w=\tau+ix$};
   \draw (0,-10pt)-- (0pt,100pt);
    \fill[pattern = south west lines] (0,-10pt) rectangle (-10pt,100pt); 
    \draw (70pt,-10pt) -- (70pt,100pt);
    \fill[pattern= south west lines] (70pt,-10pt) rectangle (80pt,100pt); 
   \end{tikzpicture}
   }
   \hspace{20pt}
   \subfloat[]{
   \begin{tikzpicture}[scale=1.0, baseline={(current bounding box.center)}]
   \draw[->,>=stealth] (-50pt,0pt) -- (60pt,0pt) node[right] { $\Re z$};
      \draw[->,>=stealth] (0pt,-50pt) -- (0pt,50pt) node[right] {$\Im z$};
      \draw[thick, dashed, red] (0pt,0pt) circle (30pt);
      \fill[white] (0pt,-2.5pt) rectangle (50pt,2.5pt);
      \draw (0pt,2.0pt) -- (50pt,2.0pt);
      \draw (0pt,-2.0pt) -- (50pt,-2.0pt);
      \fill[pattern= south west lines] (0pt,-2.0pt) rectangle (50pt,2.0pt);
        \node at (70pt,50pt) {$z=e^{2\pi w/L}$};        
   \end{tikzpicture}
   }
   \hspace{10pt}
   \subfloat[]{
   \begin{tikzpicture}[scale=1.0, baseline={(current bounding box.center)}]
   \draw[->,>=stealth] (-60pt,0pt) -- (60pt,0pt) node[right] {$\Re \xi$};
      \draw[->,>=stealth] (0pt,-40pt) -- (0pt,60pt) node[right] {$\Im \xi$};
      \draw[thick, dashed, red] (0pt,0pt) circle (30pt);
      \fill[white] (-60pt,-0.1pt) rectangle (50pt,-40pt);
      \fill[pattern= south west lines] (-50pt,-0.1pt) rectangle (50pt,-10pt);
       \node at (70pt,50pt) {$\xi=e^{\pi w/L}$};    
   \end{tikzpicture}
   }
	\caption{\label{fig:three_geometries} Schematic plot of the three geometries. (a) Strip geometry where the $(1+1)$D CFT with open boundary is defined on. (b) $z$-plane where we compute the operator evolution. (c) $\xi$-plane where we compute the operator expectation values.}
\end{figure}

In this section, we will derive the explicit expressions for the operator evolution. It is convenient to work in Euclidean coordinates, and the Lorentzian correlator can be obtained by analytic continuation. We will use three coordinates in this paper, denoted by
\begin{equation}
    w=\tau+ ix\,, \quad z=e^{2\pi w/L}\,, \quad \xi = e^{\pi w/L} \,.
\end{equation}
They correspond to the stripe geometry, complex plane and upper half-plane respectively. See Fig.~\ref{fig:three_geometries} for an illustration.

\begin{figure}[t]
    \centering
    \subfloat[]{
   \begin{tikzpicture}[scale=1.0, baseline={(current bounding box.center)}]
   \draw[->,>=stealth] (-55pt,0pt) -- (55pt,0pt) node[right] {$\Re z$};
      \draw[->,>=stealth] (0pt,-55pt) -- (0pt,55pt) node[right] {$\Im z$};
      \draw[thick, dashed, red, near arrow] (35pt,0pt) arc (0:180:35pt);
      \draw[thick, dashed, red] (-35pt,0pt) arc (180:360:35pt);
      \fill[fill=white] (0pt,-1.0pt) rectangle (50pt,1.0pt);
      \draw (0pt,1.0pt) -- (50pt,1.0pt);
      \draw (0pt,-1.0pt) -- (50pt,-1.0pt);
      \fill[pattern= south west lines] (0pt,-1.0pt) rectangle (50pt,1.0pt);
   \end{tikzpicture}
   }
   \hspace{10pt}
   \subfloat[]{
   \begin{tikzpicture}[scale=1.0, baseline={(current bounding box.center)}]
   \draw[->,>=stealth] (-55pt,0pt) -- (55pt,0pt) node[right] {$\Re z$};
      \draw[->,>=stealth] (0pt,-55pt) -- (0pt,55pt) node[right] {$\Im z$};
      \draw[thick, dashed, red, near arrow] (35pt,0pt) arc (0:180:35pt);
      \draw[thick, dashed, red] (-35pt,0pt) arc (180:360:35pt);
      \draw[thick, dashed, red, far arrow] (-15pt,0pt) arc (180:00:15pt);
      \draw[thick, dashed, red] (-15pt,0pt) arc (180:360:15pt);
      \filldraw[fill=black] (12pt,20pt) circle (1pt) node[right] {$O$};
      \fill[fill=white] (0pt,-1.0pt) rectangle (50pt,1.0pt);
      \draw [red, mid arrow, dashed, thick] (15pt,2.5pt)-- (35pt,2.5pt);
      \draw [red, mid arrow, dashed, thick] (35pt,-2.5pt)-- (15pt,-2.5pt);
      \draw (0pt,1.0pt) -- (50pt,1.0pt);
      \draw (0pt,-1.0pt) -- (50pt,-1.0pt);
      \fill[pattern= south west lines] (0pt,-1.0pt) rectangle (50pt,1.0pt);
   \end{tikzpicture}
   }
    \hspace{10pt}
   \subfloat[]{
   \begin{tikzpicture}[scale=1.0, baseline={(current bounding box.center)}]
   \draw[->,>=stealth] (-55pt,0pt) -- (55pt,0pt) node[right] {$\Re z$};
      \draw[->,>=stealth] (0pt,-55pt) -- (0pt,55pt) node[right] {$\Im z$};
      \draw[thick, dashed, red, near arrow] (27pt,20pt) arc (0:180:15pt);
      \draw[thick, dashed, red] (-3pt,20pt) arc (180:360:15pt);
      \filldraw[fill=black] (12pt,20pt) circle (1pt) node[right] {$O$};
      \fill[fill=white] (0pt,-1.0pt) rectangle (50pt,1.0pt);
      \draw (0pt,1.0pt) -- (50pt,1.0pt);
      \draw (0pt,-1.0pt) -- (50pt,-1.0pt);
      \fill[pattern= south west lines] (0pt,-1.0pt) rectangle (50pt,1.0pt);
   \end{tikzpicture}
   }
    \caption{(a) Integral contour for $H_0$ and $H_1$ on the $z$ coordinate. (b) Dashed circles are the integral contour for the commutator $[H_0,\calO]$, $[H_1,\calO]$; red horizontal lines are attached to circles to make it close. (c) The deformed contour.}
    \label{fig:cauchy_integral}
\end{figure}
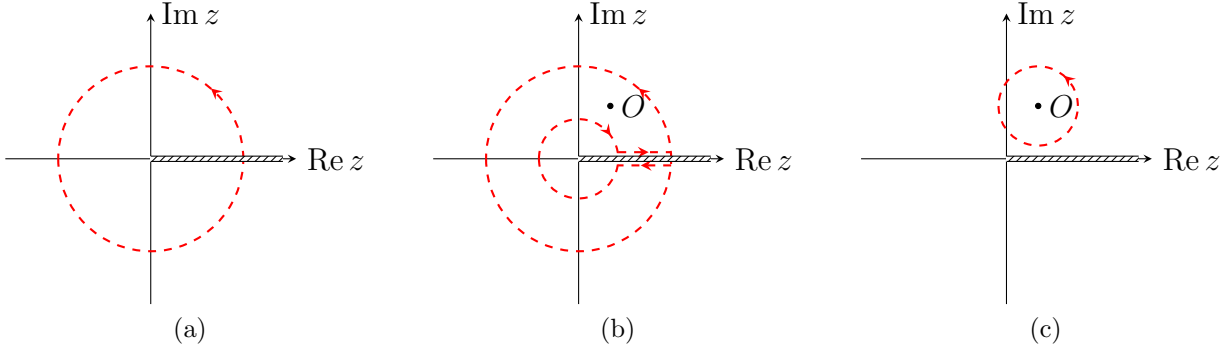

In the imaginary time, the Floquet operator is given by $F = e^{-\tau_0 H_0} e^{-\tau_1 H_1}$. 
Let us first check how the operator evolves after one cycle, namely
\begin{equation}
    \label{eqn:one_cycle_evolution}
	e^{H_1\tau_1}e^{H_0 \tau_0} \calO(w,\bar w) e^{-H_0\tau_0} e^{-H_1 \tau_1}.
\end{equation}
Here we assume $\calO(w,\bar w)$ to be a primary operator with the conformal dimension $(h,\bar h)$. 
On the strip $w$, the algebraic relations between $H_{0,1}$ and $\calO$ are complicated. It will be easier to work in $z = e^{2\pi w/L}$ coordinate instead, where $H_{0,1}$ are expressible as contour integrals of the stress tensor. More explicitly, we have
\begin{equation}
    \label{eqn:H on z plane}
    \begin{aligned}
    H_0 = & \frac{2\pi}{L} \int_{C} \frac{dz}{2\pi i} zT(z) - (z\rightarrow\bar z) - \frac{c\pi}{6L}\\
	H_1 =& \frac{2\pi}{L} \int_{C} \frac{dz}{2\pi i} \left(-\frac{1}{2} + z - \frac{z^2}{2}\right)T(z) - (z\rightarrow\bar z) - 
	\frac{c\pi}{6L}\,.
    \end{aligned}
\end{equation}
The term $-c\pi^2/6L$ comes from the Schwarzian derivative and will not affect the operator evolution. The contour $C$ is shown in \figref{fig:cauchy_integral}~(a). 
The subtlety is that $C$ is not closed due to the branch cut arising from the open boundary condition. 
%Strictly speaking,  the contour $C$ is not closed due to the branch cut along the positive real axis, where the physical boundaries are located. 
The branch cut can be treated as follows. First, we use the Baker-Campbell-Hausdorff formula to expand \eqnref{eqn:one_cycle_evolution} and write it as commutators. Each term can be depicted as a double contour integral shown in \figref{fig:cauchy_integral}~(b). The conformal boundary condition requires $T(z)=\bar T(\bar z)$ right above and below the branch cut, respectively. Thus, we can attach two horizontal lines along with the branch cut (as the red horizontal lines in \figref{fig:cauchy_integral}(b)) for free since the contributions exactly cancel. After the above manipulations, the new contour can be deformed to enclose operator $O$ as shown in \figref{fig:cauchy_integral}(c). 
Therefore, on the $z$-plane, the Floquet operator acts on the operators as if there is {\em no} branch cut. 

As a consequence, the operator evolution that is driven by the stress tensor will be determined by a two-step conformal transformation 
\begin{equation}
    \label{eqn:one_cycle}
	e^{H_1\tau_1}e^{H_0 \tau_0} \calO(w,\bar w) e^{-H_0\tau_0} e^{-H_1 \tau_1} = \left(\frac{\partial z}{\partial w}\right)^h \left(\frac{\partial \bar z}{\partial \bar w}  \right)^{\bar h}
	\left(\frac{\partial z_1}{\partial z}\right)^h \left(\frac{\partial \bar z_1}{\partial \bar z}  \right)^{\bar h}\calO(z_1,\bar z_1),
\end{equation}
where $(\partial z/\partial w)$ corresponds to the transformation from the strip $(w)$ to the complex plane $(z)$ and $(\partial z_1/\partial z)$ is the transformation generated by the Floquet dynamics $F$.

To determine the map $z_1(z)$, we notice that without the branch cut, $H_0$ and $H_1$ in \eqnref{eqn:H on z plane} can be written as Virasoro generators $L_{0,\pm 1}$ and their anti-holomorphic patterns as follows,
\begin{equation}
    \tilde{H_0} =\frac{2\pi}{L} \left( L_0 + \bar{L}_0 \right) \,, \quad \tilde{H_1} = \frac{2\pi}{L} \left( L_0 -\frac{L_{-1}+L_{1}}{2} + \bar{L}_0 - \frac{\bar{L}_{-1}+\bar{L}_{1}}{2} \right)
\end{equation}
we use tilde to emphasize that the identification only works for operator evolution. 
The generators $L_{0,\pm 1}$ form an $\fraksl(2,\RR)$ algebra. Therefore, the corresponding Floquet operator $F$ generates a M\"obius transformation on $z$, namely
\begin{equation}
    	\label{eqn:mobius}
	z_1 = f(z) = \frac{a z + b}{c z + d}, \quad \begin{pmatrix}
	a & b \\
	c & d
	\end{pmatrix} \in \SL(2,\RR) \,.
\end{equation}
The coefficients $a,b,c,d$ are determined by the dimensionless driving periods $\tau_0/L$ and $\tau_1/L$ as follows,
\begin{equation} \label{eqn: mobius z1 abcd}
\begin{aligned}
    a &= \left(1+\frac{\pi\tau_1}{L}\right)e^{\frac{\pi \tau_0}{L}}\,, \qquad
    b = - \frac{\pi\tau_1}{L} e^{-\frac{\pi\tau_0}{L}}\,, \\
    c &= \frac{\pi\tau_1}{L} e^{\frac{\pi\tau_0}{L}}\,, \quad \qquad \qquad 
    d = \left(1-\frac{\pi\tau_1}{L}\right) e^{-\frac{\pi\tau_0}{L}}.
\end{aligned}
\end{equation}
More explicitly, the evolution induced by $H_0$ acts as a dilation on the $z$-plane, namely $z$ goes to $\tilde z = e^{2\pi \tau_0/L}z$, which explains the $e^{\pi\tau_0/L}$ factors. The evolution by $H_1$ is also a dilation but in a different coordinate $\chi$.\footnote{ Since the $H_1$ acts on $z$ coordinate in a complicated way, we can instead look for a new coordinate $\chi$, on which $H_1$ acts as a simple dilation. Namely we assume a coordinate change $\chi(z)$ and accordingly $T(z) = \chi'^2 T(\chi)$,
\begin{equation}
	H_1 = \frac{2\pi}{L} \oint \frac{d\chi}{2\pi i} \frac{-(1-z)^2}{2} \chi' T(\chi).
\end{equation}
Requiring $H_1$ generates a dilation amounts to the following condition,
\begin{equation}
	-\frac{(1-z)^2}{2} \chi' = \chi \Rightarrow
	\frac{1}{2}\log \chi = \frac{1}{z-1} + \const.
\end{equation} 
Under the evolution of $H_1$, $\chi$ goes to $\chi e^{2\pi \tau_1/L}$ and correspondingly the $\tilde z$ transforms as,
\begin{equation}
	\frac{1}{z_1-1} = \frac{1}{\tilde z-1} + \frac{\pi \tau_1}{L}.
\end{equation}
Inserting $\tilde{z}= e^{2\pi \tau_0/L}z$, we get \eqnref{eqn: mobius z1 abcd}. 
}

For the Floquet problem, we would like to study the operator evolution for $n$ repeated cycles of M\"obius transformations, namely $z_n = f(f\ldots f(z))$ and we will denote it as,
\begin{equation}
    \label{eqn: zn formula}
    z_n=f^n(z)=\frac{A z+B}{Cz +D}\,.
\end{equation}
The successive application of M\"obius transformation is better described using the fixed points $f(\gamma)=\gamma$ and the  ``rotations'' $\eta$ relative to the fixed points
\begin{equation}
    \label{eqn:gamma1 gamma2}
    \gamma_1 = \frac{a-d-\sqrt{(a-d)^2+4bc}}{2c}\,,\quad
    \gamma_2 =  \frac{a-d+\sqrt{(a-d)^2+4bc}}{2c}\,,\quad
    \eta = \frac{c\gamma_2 + d}{c\gamma_1 + d}\,.
\end{equation}
With these new variables, \eqnref{eqn: zn formula} can be rearranged into the following form.
\begin{equation}
    \frac{z_n-\gamma_1}{z_n-\gamma_2} = \eta^n \, \frac{z-\gamma_1}{z-\gamma_2}\,.
    \label{eqn: fixed point}
\end{equation}
For our physical application, $c=\frac{\pi\tau_1}{L}e^{\pi \tau_0/L}$ is non-zero and there are in general three possible scenarios depending on the position of the fixed point:
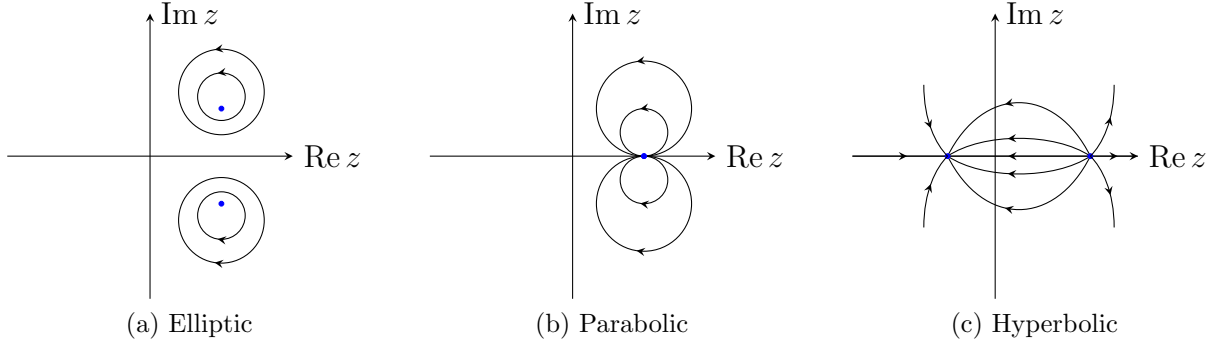
\begin{figure}
	\centering
	\subfloat[Elliptic]{
		\begin{tikzpicture}[scale=0.9]
		\draw [->,>=stealth] (-60pt,0pt) -- (60pt,0pt) node[right]{$\Re z$};
		\draw [->,>=stealth] (0pt,-60pt) -- (0pt,60pt) node[right]{$\Im z$};
		\filldraw[blue] (30pt,20pt) circle (1pt);
		\filldraw[blue] (30pt,-20pt) circle (1pt);
		\draw[near arrow] (30pt,25pt) circle (10pt);
		\draw[near arrow] (30pt,27pt) circle (18pt);
		\draw[near arrow, yscale=-1] (30pt,25pt) circle (10pt);
		\draw[near arrow, yscale=-1] (30pt,27pt) circle (18pt);
		\end{tikzpicture}
	}
	\hspace{10pt}
	\subfloat[Parabolic]{
		\begin{tikzpicture}[scale=0.9]
		\draw [->,>=stealth] (-60pt,0pt) -- (60pt,0pt) node[right]{$\Re z$};
		\draw [->,>=stealth] (0pt,-60pt) -- (0pt,60pt) node[right]{$\Im z$};
		%\filldraw[blue] (30pt,20pt) circle (1pt);
		%\filldraw[blue] (30pt,-20pt) circle (1pt);
		\draw[near arrow] (30pt,20pt) circle (20pt);
		\draw[near arrow] (30pt,10pt) circle (10pt);
		\draw[near arrow, yscale=-1] (30pt,20pt) circle (20pt);
		\draw[near arrow, yscale=-1] (30pt,10pt) circle (10pt);
		\filldraw[blue] (30pt,0pt) circle (1pt);
		\filldraw[blue] (30pt,0pt) circle (1pt);
		\end{tikzpicture}
	}
	\hspace{10pt}
	\subfloat[Hyperbolic]{
		\begin{tikzpicture}[scale=0.9]
		\draw [->,>=stealth] (-60pt,0pt) -- (60pt,0pt) node[right]{$\Re z$};
		\draw [->,>=stealth] (0pt,-60pt) -- (0pt,60pt) node[right]{$\Im z$};
		\draw[mid arrow] (40pt,0pt) -- (-20pt,0pt);
		\draw[mid arrow] (40pt,0pt) ..controls (25pt,10pt) and (-5pt,10pt).. (-20pt,0pt);
		\draw[mid arrow, yscale=-1] (40pt,0pt) ..controls (25pt,10pt) and (-5pt,10pt).. (-20pt,0pt);
		\draw[mid arrow, yscale=3] (40pt,0pt) ..controls (25pt,10pt) and (-5pt,10pt).. (-20pt,0pt);
		\draw[mid arrow, yscale=-3] (40pt,0pt) ..controls (25pt,10pt) and (-5pt,10pt).. (-20pt,0pt);
		\draw[mid arrow] (-60pt,0pt) -- (-20pt,0pt);
		\draw[mid arrow] (40pt,0pt) -- (60pt,0pt);
		\filldraw[blue] (-20pt,0pt) circle (1pt);
		\filldraw[blue] (40pt,0pt) circle (1pt);
		\draw[mid arrow] (-30pt,30pt) .. controls (-30pt,15pt) and (-25pt,5pt) .. (-20pt,0pt);
		\draw[mid arrow, yscale=-1] (-30pt,30pt) .. controls (-30pt,15pt) and (-25pt,5pt) .. (-20pt,0pt);
		\draw[mid arrow] (40pt,0pt) .. controls (45pt,5pt) and (50pt,15pt) .. (50pt,30pt);
		\draw[mid arrow, yscale=-1] (40pt,0pt) .. controls (45pt,5pt) and (50pt,15pt) .. (50pt,30pt);
		\end{tikzpicture}
	}
	\caption{Three classes of M\"obius transformation: (a) Elliptic, where two fixed points are two conjugate roots, the orbits are circulating the fixed points;  (b) Parabolic, where the two fixed points coincide; (c) Hyperbolic, where two fixed points are two real roots. }
	\label{fig: three types}
\end{figure}
\begin{enumerate}
    \item Elliptic class: the quadratic equation $f(\gamma)=\gamma$ has two distinct roots that are conjugate to each other $\gamma_1=\gamma_2^*$, the rotation parameter is determined by the following formula
    \begin{equation}
        \label{eqn:eta definition}
        \eta=\dfrac{c\gamma_2+d}{c\gamma_1+d}\,.
    \end{equation}
    In this case, $\eta$ is a pure phase, namely $|\eta|=1$. See \figref{fig: three types}~(a) for an illustration. 
    The corresponding Mobius transformation $z_n=f^n(z)$ can be represented as an $\SL(2,\RR)$ matrix as follows,
    \begin{equation}
        \begin{pmatrix}
         A & B \\
         C & D
        \end{pmatrix}=
        \begin{pmatrix}
              \gamma_1-\eta^n \gamma_2 & -(1-\eta^n) \gamma_1 \gamma_2 \\
              1-\eta^n & -(\gamma_2-\eta^n \gamma_1)\,.
        \end{pmatrix}
    \label{eqn:2 roots sl2}
    \end{equation}
         
    \item Hyperbolic class: the two distinct roots are purely real and the parameter $\eta$ defined above is also a real number. 
    See \figref{fig: three types}~(c) for an illustration. The parameter $0<\eta<1$ represents the rescaling near the fixed points. The Mobius transformation matrix is in the same form as \eqref{eqn:2 roots sl2}. 
    
    \item Parabolic class: two roots are merged together $\gamma_1=\gamma_2=\gamma$. Therefore \eqref{eqn: fixed point} does not apply. For this case, we introduce a new parameter $\beta= \frac{a-d}{2c}$ such that
    \begin{equation}
        \frac{1}{z_n-\gamma} = \frac{1}{z-\gamma} + n \beta\,.
    \end{equation}
    The corresponding transformation matrix is 
    \begin{equation}
		\begin{pmatrix}
		A & B \\
		C & D
		\end{pmatrix}=
		\begin{pmatrix}
		1+n\beta \gamma & -n\beta \gamma^2 \\
		n\beta & 1-n\beta \gamma
		\end{pmatrix}\,
    \label{eqn: 1 root sl2}     
	\end{equation}
	which can not be diagonalized. The parabolic class may be thought as the marginal case of either elliptic or hyperbolic class, see \figref{fig: three types}~(b) for an illustration.
\end{enumerate}

We remark here that the M\"obius transformation also applies to quasi-primaries such as the stress tensor. In that case, although we will obtain a Schwarzian derivative term when transforming between different geometries, the operator evolution driven by M\"obious transformation on the complex plane is still determined by \eqnref{eqn:one_cycle}. 
More explicitly, the stress tensor on the strip after $n$-cycle driving becomes
\begin{equation}
    \label{eqn:Tw evolution}
	F^{-n} T(w) F^n = \left( \frac{\partial z}{\partial w} \right)^2 \left( \frac{\partial z_n}{\partial z} \right)^2 T(z_n) - \left(\frac{2\pi}{L}\right)^2 \frac{c}{24}.
\end{equation}

Finally, for the operator evolution in real (Lorentzian) time, we perform the analytic continuation $\tau_0 \rightarrow iT_0$, $\tau_1 \rightarrow i T_1$. In real time, a space-time position $(x,t)$ on the strip maps to $z=e^{i2\pi(x+t)/L}$ on the $z$-plane, 
which is always on the unit circle. 
%\YG{[YG: I don't have a better explanation for this fact. Here are some preliminary thoughts: in general, we may complexify $x,t$ and allow them to live on a two dimensional complex manifold (i.e. four real dimensions). The particular physical problem we consider only concerns a one dimensional sub-manifold (two real dimensions). So there are some degrees of freedom to choose the parametrization. Choosing this particular parametrization $z=\exp...$ is convenient in the sense that the mapping preserves the unit circle.]}
Therefore, the operator evolution is geometrically related to the automorphism of a unit circle under the conformal mapping. 
Although the M\"obius transformations after the analytic continuation generally belong to $\SL(2,\CC)$, the basic structure remains the same. (Naively we may expect an additional class known as loxodromic class shows up where $\eta$ is a general complex number, not necessarily a phase or purely real. However, the physical parameters that appear in the Floquet setting do not fall into such class.)

\subsection{Parametric oscillator (swing) analogy}
\label{sec:phases}

The last section explained the relation between the operator evolution and the Mobius transformation, which is further classified into three classes: elliptic, hyperbolic and parabolic. The corresponding Floquet dynamics are also classified into the non-heating, heating, and the critical classes respectively, and the phase diagram was first obtained in ~\cite{wen2018floquet}. For reader's convenience, we reproduce the phase diagram in \appref{app:phase diagram cft}.

\begin{figure}[t]
    \centering
    \subfloat[Swing]{
    \begin{tikzpicture}[scale=1.2]
    \draw (-50pt,0pt) -- (50pt,0pt);
    \fill[pattern= south west lines] (-50pt,-5.0pt) rectangle (50pt,0.0pt);
    \filldraw (0pt, 100pt) circle (1pt);
    \draw (0,100pt) -- (0pt, 10pt);
    \draw[thick] (-5pt,10pt) -- (5pt,10pt);
    \draw[densely dotted] (0pt,100pt) -- (-30.7818 pt,15.4277 pt);
    \draw[thick] (-35.4803pt,17.1378pt) -- (-26.0833pt,13.7176pt );
    \filldraw[fill=blue] (0pt,15pt) circle (3pt);
    \filldraw[fill=blue] (-27.3616pt,24.8246pt) circle (3pt);
       \draw[densely dotted] (0pt,100pt) -- (30.7818 pt,15.4277 pt);
       \draw[thick] (35.4803pt,17.1378pt) -- (26.0833pt,13.7176pt );
        \filldraw[fill=blue] (27.3616pt,24.8246pt) circle (3pt);
    \end{tikzpicture}
    }
    \hspace{20pt}
    \subfloat[Phase diagram of a Mathieu oscillator]{
    \includegraphics[width=7cm,height=5cm]{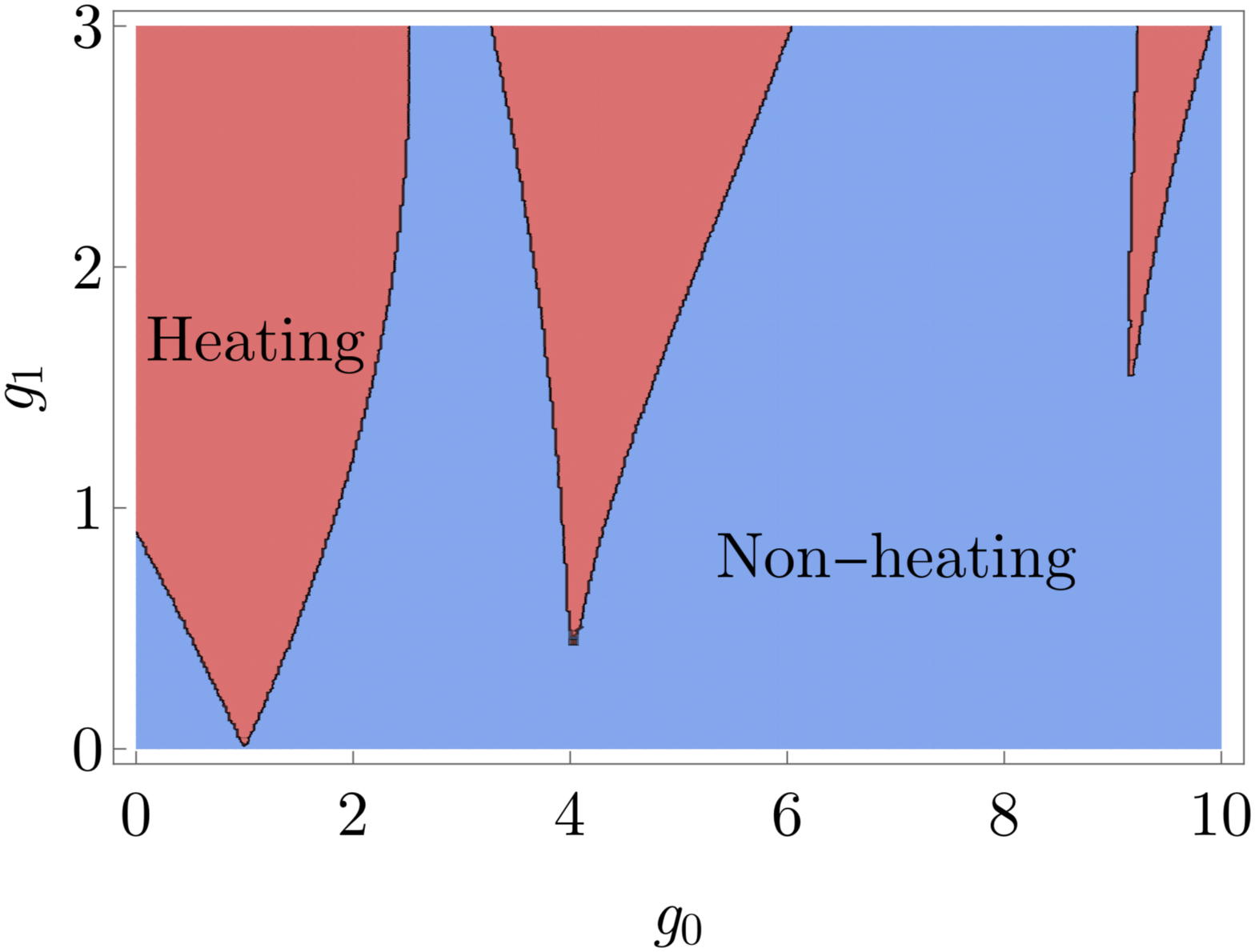}
    }
    \caption{(a)A simple example of a Mathieu oscillator is a child pumping a swing by periodically standing and squatting to increase the amplitude of the oscillation. The pump motion of a skilled child is typically at twice the frequency of the swing's oscillations which belongs to the heating phase. (b) In the red regime, energy keeps growing exponentially. In the blue regime, energy only oscillates. The black curve corresponds to the phase boundary, where energy grows quadratically.}
    \label{fig:swing}
\end{figure}

It is instructive and amusing to gain  intuition into this  classification in a more elementary setting with the same $\SL(2,\RR)$ structure. The example we would like to use is the parametric oscillator with the following Hamiltonian\cite{Perelomov1969,Gritsev:2017zdm},
\begin{equation}
	H(t) = f(t) \frac{p^2}{2} +  g(t) \frac{x^2}{2},
\end{equation} 
where $f(t) = f(t+T)$ and $g(t)=g(t+T)$ are periodic functions. 
One familiar example is the Mathieu oscillator, which corresponds to $f(t)=1, g(t) = g_0-2g_1\cos (2 t)$. 
Classically, they are useful in explaining the motion of a playground swing,
 see Fig.~\ref{fig:swing}(a) for a classical picture. Furthermore, the recognition of the $\SL(2,\RR)$ structure in the problem also has interesting consequence in the ultra-cold quantum gases, e.g. see Ref.~\cite{EfimovExpansion,superEfimov}.

For the quadratic Hamiltonian, the Heisenberg operators $(x(t), p(t))$ evolve under a $\SL(2,\RR)$ transformation that preserves the commutation relation $[x,p]=i$,\footnote{Another familiar example is the Bogoliubov transformation for bosons}
\begin{equation}
    \begin{pmatrix}
		x(t) \\ p(t)
	\end{pmatrix}=
	\begin{pmatrix}
		c_{11} & c_{12} \\
		c_{21} & c_{22}
	\end{pmatrix}
	\begin{pmatrix}
		x \\ p
	\end{pmatrix},\quad
	\begin{pmatrix}
		c_{11} & c_{12} \\
		c_{21} & c_{22}
	\end{pmatrix} \in \SL(2,\RR)\,.
	\label{eqn: sl2 matrix}
\end{equation}
Therefore the stroboscopic evolution of $(x,p)$ is represented by a $\SL(2,\RR)$ transformation $F_{(x,p)}$, whose classification determines the 
stroboscopic trajectory of $(x(nT),p(nT))$.
More explicitly, to compare with the M\"obius transformation used in \eqnref{eqn:mobius} we treat $(x,p)$ as a point on the complex projective plane $\mathbb{CP}^1$ which can be more conveniently parametrized by $z=x/p$. Then the $\SL(2,\RR)$ action on the point $(x,p)$ shown in \eqnref{eqn: sl2 matrix} is equivalent to the M\"obius transformation \eqnref{eqn:mobius} and also have three classes. The fixed points $\gamma_{1,2}$ and the rotation angle $\eta$ of the M\"obius transformation can be translated to the eigenvectors $v_{1,2}$ and the ratio of eigenvalues $\lambda_{1,2}$ of the $\SL(2,\RR)$ matrix, respectively. Their correspondence is given explicitly below.
\begin{center}
	\begin{tabular}{ |c|c|c|c|c| } 
		\hline
		& \multicolumn{2}{|c|}{M\"obius transformation} & \multicolumn{2}{|c|}{$\SL(2,\RR)$ matrix $F_{(x,p)}$} \\
		\hline
		Classification & Fixed points & $\eta$ & Eigenvectors & Eigenvalues \\
		\hline
		Elliptic & $\gamma_1,\gamma_2 \in \CC$ & $\eta=e^{i\theta}$ & $v_1,v_2 \in \CC$ & $\lambda_1=\lambda_2=e^{i\theta/2} $ \\
		\hline
		Hyperbolic & $\gamma_1,\gamma_2 \in \RR$ & $0<\eta<1$ & $v_1,v_2\in \RR$ & $\lambda_1 = 1/\lambda_2 < 1$ \\
		\hline
		Parabolic & $\gamma_1=\gamma_2 \in \RR$ & 1 & $v_1 = v_2 \in \RR$ & $\lambda_1 = \lambda_2 = 1$\\
		\hline
	\end{tabular}
\end{center}
 
This explains the different dynamics. In the elliptic class, $(x,p)$ as a real vector only keeps rotating on the $x-p$ plane. The energy measured by $p^2/2+x^2/2$ just oscillates $E(nT) \sim \cos(n\theta + \varphi)$ with a period controlled by the angle $\theta$ of the eigenvalue. In the hyperbolic class, $F_{(x,p)}$ has two real right eigenvectors: $v_1$ with an eigenvalue $\lambda>1$ and $v_2$ with an eigenvalue $1/\lambda<1$. Therefore unless the initial condition $(x_0,p_0)$ is along $v_2$, $(x(nT),p(nT))$ will flow to infinity along $v_1$ exponentially fast, which causes the energy to grow exponentially in the long time limit
\begin{equation}
    \label{eqn:oscillator hyperbolic}
    E(nT) \propto (p_0 v_{2,1} - x_0 v_{2,2})^2 \lambda^{2n}\,,\quad n\gg 1 \,.
\end{equation}
In the parabolic class, $F_{(x,p)}$ only has one right eigenvector $v$ with eigenvalue $1$ thus becomes singular. To determine the dynamics, we can look at the Jordan normal form of $F_{(x,p)}$.\footnote{
    Since $F_{(x,p)}$ is singular, its Jordan normal form has a nonzero off-diagonal element
    \begin{align*}
	F_{(x,p)}=
	P\begin{pmatrix}
		1 & 1 \\
		0 & 1
	\end{pmatrix}
	P^{-1}.
    \end{align*}
    The off-diagonal element will increase linearly with the driving cycles, i.e.  $F_{(x,p)}^n=P\begin{pmatrix}
		1 & n \\
		0 & 1
	\end{pmatrix}
	P^{-1}$.
} Unless the initial condition $(x_0,p_0)$ is along $v$, $(x(nT),p(nT))$ will flow to infinity linearly, which causes the energy to grow quadratically
\begin{equation}
    \label{eqn:oscillator parabolic}
    E(nT)\propto
		\left(p_0 v_1 - x_0 v_2 \right)^2 n^2.
\end{equation}
As a concrete example, the Mathieu oscillator introduced at the beginning of this section can support all of the three different dynamics, and its phase diagram is presented in \figref{fig:swing}. 
 
%\AV{We may want to point out that the matrix is singular. We could add a sentence in the footnote regarding generalized eigenvectors associated with that Jordan normal form} . 
%To determine the dynamics, we can look at the Jordan normal form of the stroboscopic $\SL(2,\RR)$ matrix, which tells us $(x(nT),p(nT))$ will flow to infinity but in a linear fashion. Therefore the energy grows quadratically with time, which is a critical case between the non-heating and heating dynamics. \AV{Can we list the total energy growth expressions and compare later with the CFT?}

The Floquet CFT studied in the current paper is richer than its oscillator analog. In particular, the $(1+1)$D CFT has locality in space, which will lead to features in the energy density and entanglement that are the focus of the following sections. 

\section{Energy and Entanglement}
\label{sec:features}

Energy and entanglement are the most straightforward and fundamental diagnostics of states evolving under Floquet driving. Fortunately, both can be studied in $(1+1)$D CFT analytically using the operator evolution method we have discussed. In this section, we will present the stroboscopic measurement of the energy and entanglement under Floquet driving. We will also provide a semi-classical picture of the phenomenon and point out an interesting relation between energy and entanglement.

\subsection{Energy density and total energy}
\label{sec:energy dynamics}

The energy of a state under the Floquet evolution can be measured by the expectation value of the stress tensor $T_{00}=T+\bar{T}$, whose time evolution can be obtained by \eqnref{eqn:Tw evolution}. 
The $n$ dependence of the energy arises from the first term. 
%The second term, coming from the Schwarzian, is a constant thus can be dropped. 
To compute $\braket{G|T(z_n)|G}$, we need to perform another conformal transformation to the upper half-plane via $\xi=\sqrt{z}$. This mapping generates a Schwarzian derivative
\begin{equation}
    \frac{c}{12}\operatorname{Sch}\left(\xi,z\right) =\frac{c}{12}\left(\frac{\xi'''(z)}{\xi'(z)} - \frac{3}{2}\left(\frac{\xi''(z)}{\xi'(z)} \right)^2 \right)= \frac{c}{32z^2}
\end{equation}
and leaves a second term $\braket{G|T(\xi)|G}$. On the one hand, Ward identity and scale invariance constrains $\braket{G|T(\xi)|G} \propto 1/\xi^2$. On the other hand, it is invariant under the horizontal translation. Therefore this term has to vanish and all the contribution comes from the Schwarzian term, namely,
\begin{equation}
    \label{eqn:Tchiral}
	\braket{GS|F^{-n} T(w) F^n|GS} =  \left( \frac{\partial z}{\partial w} \right)^2 \left( \frac{\partial z_n}{\partial z} \right)^2
	\frac{c}{32z_n^2} \,.
\end{equation}
where $w=\tau+ix$ is the complex coordinate for the stress tensor $T$ on the strip, $z_n$ is the coordinate on the $z$-plane after $n$-cycle driving and $c$ is the central charge. The initial value has been subtracted and will be ignored in the rest of discussion in this section.
%\AV{Maybe useful to  note that the result of the Schwatzian calculation as a separate equation, for the  mapping from the plane to the half plane via $\xi = \sqrt{z}$.}
After analytic continuation $\tau_0 \rightarrow i T_0$, $\tau_1 \rightarrow i T_1$, the expectation value of $T$ has the following form
\begin{equation}
    \label{eqn:Echiral}
    \braket{T}(x,t=nT)  = 
    \left(\frac{2\pi}{L} \right)^2
    \frac{c}{32}
    \frac{(AD-BC)^2z^2}{(Az+B)^2(Cz+D)^2}\,,
\end{equation}
with $A,B,C,D$ depending on $T_0/L,T_1/L$ and $n$ through the prescription described in section~\ref{sec: mobius}. Replacing $z$ with $\bar z$ gives us the expectation value of $\bar T$. 

\begin{figure}[t]
    \centering
	\subfloat[Non-heating phase]{\includegraphics[width=2.1in]{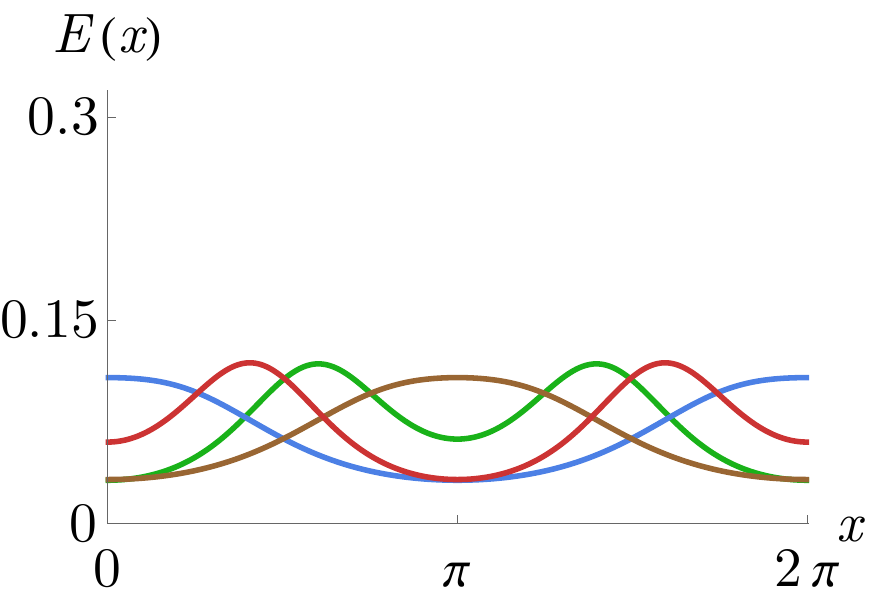}}
	\hspace{5pt}
	\subfloat[Critical]{\includegraphics[width=2.1in]{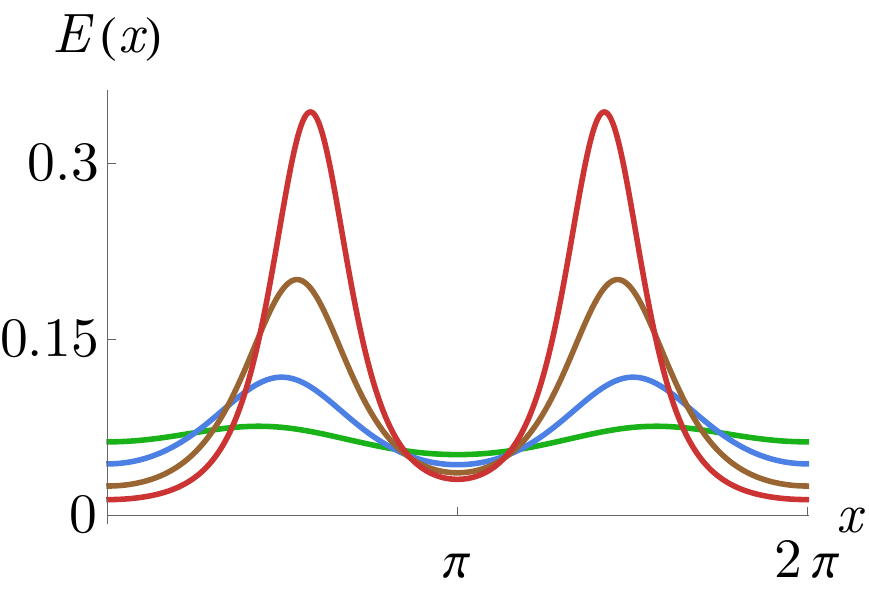}}
	\hspace{5pt}
	\subfloat[Heating phase]{\includegraphics[width=2.1in]{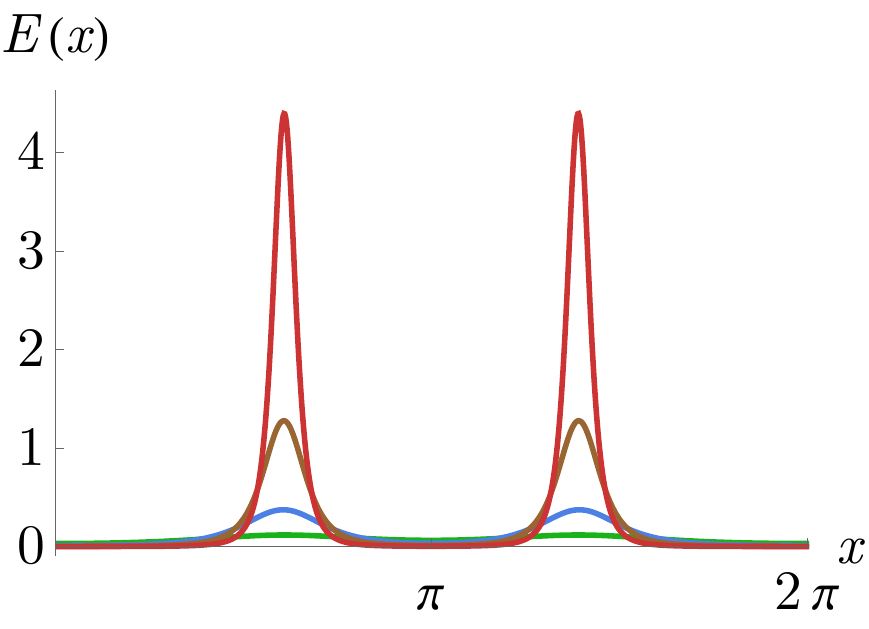}}
    \caption{The evolution of energy density profile in different phases. In the plot, we choose the system size $L=2\pi$ and the central charge $c=1$. Different colors means different times. The green, blue, brown and red curve corresponds to $t=T, 2T, 3T, 4T$ respectively. (a)  $T_0=0.5L,T_1=0.1L$ which is in the non-heating phase. The energy density only oscillates (b) $T_0=0.9L$, and $T_1$ is tuned to make the system right at the phase boundary. The position of the peaks in the plot are not given by $\gamma$. This is because we are not at the late time regime. One can check that as we increase $n$, the peaks will move towards $\log \gamma$. (c) $T_0=0.9L, T_1=0.1 L$ which is in the heating phase. We can clearly see the formation and growth of two peaks.}
    \label{fig:energydensity}
\end{figure}

These formulae allow us to look at the evolution of energy density directly, which is found to have different behaviors in different phases, as shown in \figref{fig:energydensity}. In the non-heating phase, the energy density just fluctuates without a definite period. In the heating phase, the energy density quickly develops two peaks, which grows with time vary fast. The positions of the energy peaks are determined by the unstable fixed points of the M\"obius transformation, i.e. $e^{2\pi i x_{\text{peak}}/L}=\gamma_2$ or $\gamma_2^*$. At the phase boundary, there are also two energy peaks but growing much slower.  

These phenomena can be understood from the perspective of the fixed points of M\"obius transformation. The $n$ dependence enters \eqnref{eqn:Tchiral} through two parts: (a) the Schwarzian term $c/32z_n^2$, which has a constant magnitude due to the fact that $|z_n|=1$ in the real-time; (b) the rescaling factor $(\partial z_n/\partial z)^2$, whose different behaviors in three phases explain the feature shown in \figref{fig:energydensity}. 
\begin{figure}[t]
    \centering
    \subfloat[Non-heating phase]{
    	\begin{tikzpicture}
    	\draw[->,>=stealth] (-60pt,0pt) -- (60pt,0pt) node[right]{$\Re z$};
    	\draw[->,>=stealth] (0pt,-60pt) -- (0pt,60pt) node[right]{$\Im z$};
    	\draw (0pt,0pt) circle (40pt);
    	\filldraw (20pt,-20pt) circle (1pt) node[left]{\scriptsize $\gamma_1$};
    	\filldraw (35pt,-35pt) circle (1pt) node[right]{\scriptsize $\gamma_2$};
    	\end{tikzpicture}
    }
    \hspace{40pt}
    \subfloat[Heating phase]{
    \begin{tikzpicture}
     \draw[->,>=stealth] (-60pt,0pt) -- (60pt,0pt) node[right]{$\Re z$};
     \draw[->,>=stealth] (0pt,-60pt) -- (0pt,60pt) node[right]{$\Im z$};
     \draw (0pt,0pt) circle (40pt);
     \filldraw (20pt, 34.641pt) circle (1pt) node[right]{\scriptsize $\gamma_2$};
      \filldraw (34.641pt,-20pt) circle (1pt) node[right]{\scriptsize $\gamma_1$};
      \draw[mid arrow] (40pt,0pt)  arc (0:-30:40pt) ;
      \draw[mid arrow] (0pt,-40pt)  arc (-90:0:40pt) ;
      \draw[far arrow] (40pt,0pt)  arc (0:90:40pt) ;
      \draw[mid arrow] (0pt,40pt)  arc (90:0:40pt) ;
    \end{tikzpicture}
    }
    \caption{Fixed point distribution in real time. (a) In non-heating phase, one fixed point is inside the unit circle while the other one is outside. (b) In the heating phase, both fixed points are on the unit circle. One is attractive and the other is repulsive.}
    \label{fig:fixed point distribution}
\end{figure}
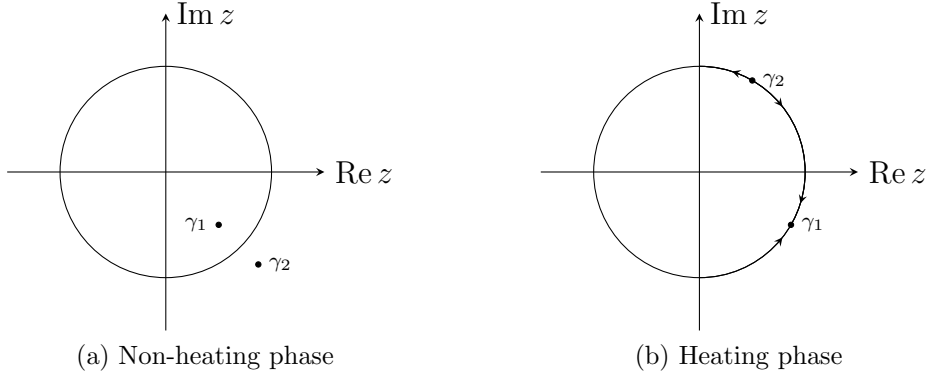

\begin{enumerate}
    \item Non-heating phase: The two fixed points sit on different sides of the unit circle. In our convention, $\gamma_1$ is inside the unit circle and $\gamma_2$ is outside the unit circle, as depicted in \figref{fig:fixed point distribution}~(a). Since $z_n$ is constrained on the unit circle, it cannot flow to either of them but just keeps rotating around them. That is the reason that energy density fluctuates in this phase. Since the rotation angle $\eta$, defined by \eqnref{eqn:eta definition}, is not a rational phase, these fluctuations do not have a definite period.
    \item Heating phase: Both two fixed points are now on the unit circle. In our convention, $\gamma_1$ is a stable fixed point and $\gamma_2$ is an unstable fixed point. For the chiral stress tensor, when $z=\gamma_2$, although $z_n$ doesn't change the rescaling factor $(\partial z_n/\partial z)^2 = \eta^{-2n}$ will grow exponentially with $n$. For the anti-chiral stress tensor, the same thing happens at $\bar z=\gamma_2$. Therefore we observe two energy peaks at two symmetric positions. On the other hand, for a generic position $z,\bar z \neq \gamma_2$, $z_n$ will flow to the stable fixed point and the rescaling factor $(\partial z_n/\partial z)^2$ will decrease exponentially with $n$. Therefore the stress tensor shrinks, making the two peaks sharper and sharper. In a lattice system, the energy peaks are also observed and consistent with the CFT prediction in the short time. In the late time, they will saturate and oscillate due to having only a finite number of degrees of freedom, as detailed in \appref{app:spatial structure lattice}.
    \item Critical line: The two fixed points merge to $\gamma$ on the unit circle, which is a marginal case. One can show that the maximal value of the rescaling factor keeps growing but in a power-law fashion, which explains the slowly growing peaks. The position for the maximum gradually moves to the position corresponding to $\gamma$.
\end{enumerate}
Note the phenomena here do not rely on the initial state, as long as it is not a common eigenstate of $H_0$ and $H_1$. For example we may consider a generic initial state $|\phi\rangle$ with the expectation value of stress tensor $\langle \phi | T(z) | \phi \rangle = \calE_{\phi}(z)$, then the \eqnref{eqn:Tchiral} generalizes to
\begin{equation}
	\braket{\phi|F^{-n} T(w) F^n|\phi} =  \left( \frac{\partial z}{\partial w} \right)^2 \left( \frac{\partial z_n}{\partial z} \right)^2 
	\calE_\phi(z)\,,
	%\left(\calE_{\phi}(z) + \frac{c}{32 z^2} \right)\,,
\end{equation}
and the discussions above still hold. In this scenario, the boundary condition is also irrelevant since the operator evolution discussed in Sec.~\ref{sec: mobius} is independent of the choice of boundary conditions.\footnote{Indeed, in Sec.~\ref{sec: mobius} we have reduced the operator evolution with open boundary condition to the one with periodic boundary condition using a contour deformation trick. The reason we start with open boundary condition is that the ground state of $H_0$ is also an eigenstate of $H_1$ for periodic boundary condition but not for open boundary condition.
}

The idea of relating the fixed points to the heating/non-heating phenomena also applies to more general setups. For example, we can use $H_0$ and $H_2 = 2\int_0^L dx \sin^2 \frac{2\pi x}{L} T_{tt}(x)$ to generate the Floquet dynamics. $H_2$ is related to $L_{\pm2}$ and thus the operator evolution is still a conformal transformation but with four fixed points. When none of the fixed points are on the unit circle, the system is in the non-heating phase without energy peaks. In certain parameter regime, there are two unstable fixed points locating on the unit circle, which implies heating dynamics and correspondingly four growing energy peaks. Furthermore, we can define a Hamiltonian by a generic deformation $H = \int_0^L dx f(x) T_{tt}(x)$. As long as $f(x)$ is a smooth real function and has a Fourier decomposition, $H$ can be represented as a linear combination of Virasoro generators and the operator evolution can be written as a conformal transformation. \footnote{Given $f(x)$, we can use its Fourier decomposition to rewrite it in terms of $z=e^{2\pi i x/L}$ as $f(x) = \tilde{f}(z)$. Then one can use the same technique as the footnote 1 to show that $H$ generates a dilation in the coordinate $\chi = e^{\int \frac{dz}{z\tilde{f}(z)}}$.} If $f(x)=\sin^2 \frac{k\pi x}{L},k\ge 1$, the conformal mapping is essentially the same as what we discussed here, which supports a non-heating and heating phase. However, for a generic $f(x)$, determination of fixed points and the corresponding dynamics is a hard problem, which we leave for a future study.

Besides the energy density, we can also look at the total energy
\begin{equation}
	E(t) = \int_0^L \frac{dx}{2\pi} \left( \braket{T} + \braket{\bar T} \right).
\end{equation}
For stroboscopic measurement, we can plug in the \eqnref{eqn:Tchiral} and have
\begin{equation}
    E(t=nT) = 
    \frac{2\pi}{L} \frac{c}{16}
	\frac{AD+BC}{AD-BC}.
\end{equation}
In either non-heating or heating phase, the M\"obius transformation has two fixed points and we need to use \eqnref{eqn:2 roots sl2} to get,
\begin{equation}
	E(t=nT) = \frac{2\pi}{L} \frac{c}{16}
	\frac{AD+BC}{AD-BC} 
	= \frac{2\pi}{L} \frac{c}{16}
	\frac{-2\gamma_1\gamma_2 +(\gamma_1+\gamma_2)^2\eta^n -2\gamma_1\gamma_2 \eta^{2n}}{\eta^n(\gamma_1-\gamma_2)^2}.
\end{equation}
For the non-heating phase, since $\eta$ is a pure phase the total energy will oscillate with time. Generally, $\eta$ is a non-rational phase factor, thus we do not expect any periodicity. Since the energy is oscillating, we cannot talk about the long-time behavior itself but the average, 
\begin{equation}
    \label{eqn:E non-heating}
    \bar{E}_{\text{non-heating}}
    :=
    \lim_{n\rightarrow\infty}
    \frac{1}{n}\sum_{k=1}^n E(t=kT)
    =
    \frac{2\pi}{L} \frac{c}{16}
	\left(\frac{\gamma_1+\gamma_2}{\gamma_1-\gamma_2}\right)^2\,,
\end{equation}
which is a finite number. For the heating phase, $0<\eta<1$ and $\eta^n$ becomes exponentially small at the late time regime. Therefore, to the leading order we can drop the $\eta^n$ and $\eta^{2n}$ terms in the numerator and find the energy grows exponentially,
\begin{equation}
    \label{eqn:E heating}
    E_{\text{heating}}(t=nT) \approx \frac{2\pi}{L} \frac{c}{16} \frac{-2\gamma_1\gamma_2}{(\gamma_1 - \gamma_2)^2}\eta^{-n},\, \quad \text{for} \quad n\gg 1.
\end{equation}

When the system is at the boundary between the non-heating and heating phase, the two fixed point merges together and we need to use \eqnref{eqn: 1 root sl2}. Noting that $\beta\gamma$ is pure imaginary, the total energy can be written as
\begin{equation}
    \label{eqn:E critical}
    E_{\text{phase boundary}}(t=nT) = \frac{2\pi}{L}\frac{c}{16} (1+2|\beta\gamma|^2n^2).
\end{equation}
The total energy grows quadratically in cycle number $n$.

This long-time asymptotics of the total energy provides a direct diagnostic of the different phases. The oscillation, exponential and quadratic growth behavior matches the simple picture obtained in the driven harmonic oscillators, as shown in Sec.~\ref{sec:phases}. In particular, noticing that $\eta = \lambda^{-2}$, the heating rate in \eqnref{eqn:oscillator hyperbolic} and \eqnref{eqn:E heating} are exactly the same. This is because the growth behavior only depends on the underlying algebra and not its detailed realization.

\subsection{Entanglement pattern in the heating phase}
\label{sec:entanglement}

Besides the total energy, the entanglement entropy of the left/right half system also has different behaviors in different phases, which was shown in \cite{wen2018floquet}. Here, we will focus on the spatial structure of entanglement that has not previously been discussed. in particular we examine the heating phase and discuss the relation between the energy peaks observed above and the entanglelment. 
%The main conclusions are listed below and the details are left in  \appref{app:entanglement}.

%Let us make the entanglement cut at the position $x \in (0,L)$, and look at the time evolution of the entanglement entropy of the subsystem $A=[0,x]$, whose reduced density matrix is denoted by $\rho_{A}$. We will follow the Calabrese-Cardy  prescription \cite{Calabrese:2004eu} to compute the von Neumann entropy $S_{A}$. That is to say, we insert the twist operator $\calT_m$ at $x$ to compute the $m$-th R\'enyi entropy
%\begin{equation}
%	S_A^{(m)} = \frac{1}{1-m} \log \Tr \rho_A^m,
%\end{equation}
%whose $m \rightarrow 1$ limit gives the von Neumann entropy. 

\begin{figure}[t]
    \centering
    \includegraphics[width=0.4\textwidth]{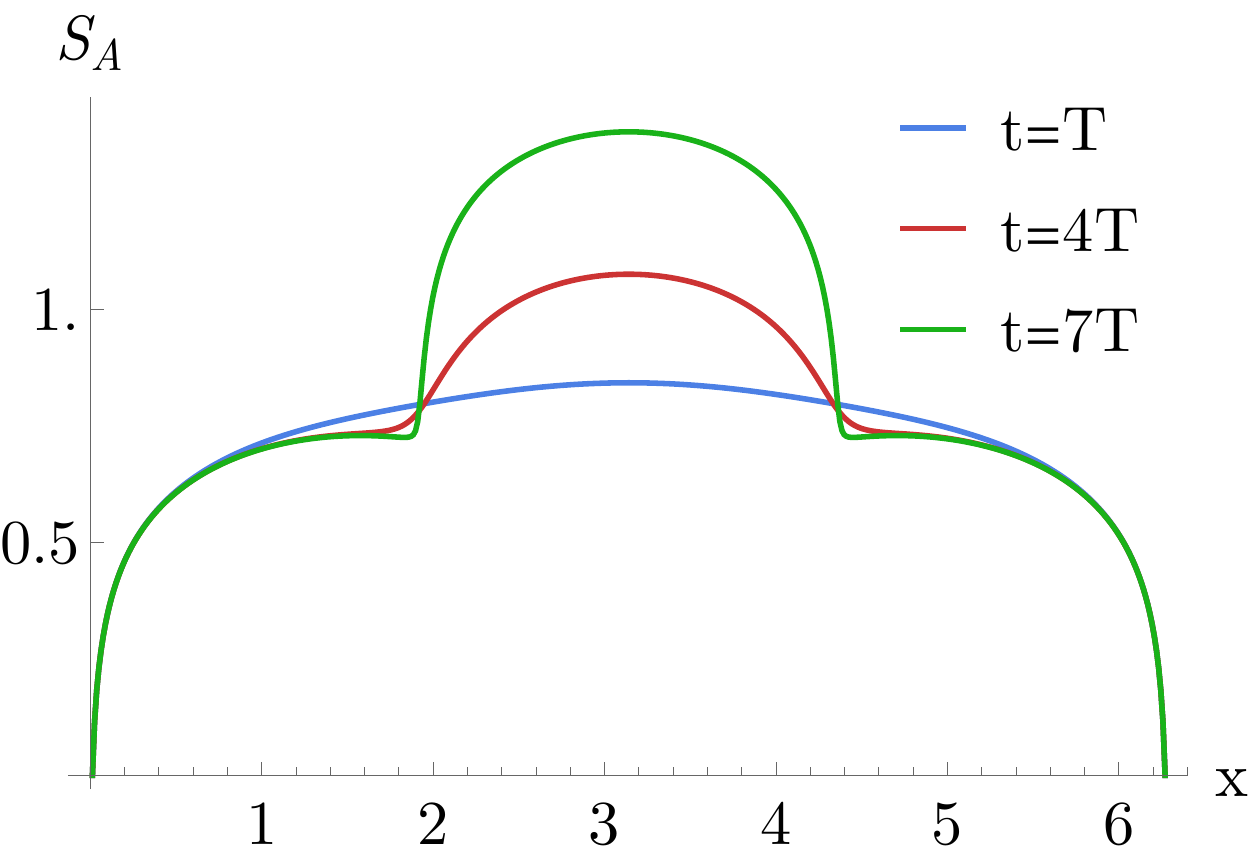}
    \caption{Entanglement entropy of the subsystem $A=[0,x]$ as a function of entanglement cut $x$. Different colors represent different times. As time increases, the curve develops two kinks. Only curves between the two kinks grows with time. The positions of the kinks are the same as the positions of the energy peaks shown in \figref{fig:energydensity}. In this plot, we choose the system size $L=2\pi$ and $T_0=0.9L, T_1=0.1L$.
    %\XW{[XW:It seems there is a cooling of the background. This phenomenon also exists in the cavity fields with oscillating boundary. This may be very useful if we start from a thermal initial state. The finite temperature $\beta$ will be decreased in the region away from the
    %peaks. We may study this in detail in the future work.
    %]}
    }
    \label{fig:entanglement pattern}
\end{figure}

Given a pure state $\ket{\Psi}$, the reduced density matrix of a subsystem $A$ is defined by the partial trace $\rho_A = \Tr_{\bar{A}} \ket{\Psi}\bra{\Psi}$ and its von Neumann entanglement entropy is given as $S_A = -\Tr \rho \log \rho$. In our setting, we consider a time dependent state $\ket{\psi(nT)} = F^n \ket{GS}$ that evolves under the Floquet driving and study the corresponding entanglement entropy $S_A(nT)$ as a function of driving cycle $n$. %and so is the entanglement.

For a subsystem $A=[0,x]$ starting from the left end and end at position $x \in (0,L)$, we plot the results in \figref{fig:entanglement pattern} and keep the details of the calculation in \appref{app:entanglement}. The entanglement entropy has a background value from the initial state. As time increases, the curve quickly develops two kinks, the positions of which exactly coincide with the energy peaks. Only the curve between the two kinks grows with time while the curve outside does not. This implies only when the subsystem includes one of the energy peaks, does the entanglement grow with time. If the subsystem includes either none or both peaks, the entanglement remains at its background value and does not grow at all.

%\AV{the background value of entanglement that does not grow in time is from the entanglement of the ground state of H0? Maybe we should mention the static background?}

\begin{figure}[t]
    \centering
    \subfloat[$x_1<x_C$]{
   \begin{tikzpicture}[scale=1.0, baseline={(current bounding box.center)}]
    \draw[thick] (-60pt,0pt) -- (60pt,0pt);
   \draw[thick] (-60pt,-5pt)-- (-60pt,5pt);
    \fill[pattern = south west lines] (-70pt,-5pt) rectangle (-60pt,5pt); 
    \draw[thick] (60pt,-5pt) -- (60pt,5pt);
    \fill[pattern= south west lines] (60pt,-5pt) rectangle (70pt,5pt); 
    %\draw (-60pt,0pt) .. controls (-30pt,0pt) and (-30pt,40pt) ..(0pt,40pt)..controls (30pt,40pt) and (30pt,0pt) .. (60pt,0pt);
    \draw [thick](-31pt,0pt)..controls (-30pt,2pt) and (-29pt,5pt)..(-28pt,40pt)..controls (-27pt,5pt) and (-26pt,2pt)..(-25pt,0pt);
    \draw [thick](31pt,0pt)..controls (30pt,2pt) and (29pt,5pt)..(28pt,40pt)..controls (27pt,5pt) and (26pt,2pt)..(25pt,0pt);
    
    \node at (-28pt,-7pt){${x_C}$};
     \node at (28pt,-7pt){${x_A}$};
     \draw[thick] (-50pt,-3pt)-- (-50pt,3pt);
     \draw[thick] (10pt,-3pt)-- (10pt,3pt);
      \node at (-50pt,7pt){${x_1}$};
       \node at (10pt,7pt){${x_2}$};
   \end{tikzpicture}
   }
   \hspace{10pt}
    \subfloat[$x_C<x_1<x_A$]{
   \begin{tikzpicture}[scale=1.0, baseline={(current bounding box.center)}]
    \draw[thick] (-60pt,0pt) -- (60pt,0pt);
   \draw[thick] (-60pt,-5pt)-- (-60pt,5pt);
    \fill[pattern = south west lines] (-70pt,-5pt) rectangle (-60pt,5pt); 
    \draw[thick] (60pt,-5pt) -- (60pt,5pt);
    \fill[pattern= south west lines] (60pt,-5pt) rectangle (70pt,5pt); 
    %\draw (-60pt,0pt) .. controls (-30pt,0pt) and (-30pt,40pt) ..(0pt,40pt)..controls (30pt,40pt) and (30pt,0pt) .. (60pt,0pt);
    \draw [thick](-31pt,0pt)..controls (-30pt,2pt) and (-29pt,5pt)..(-28pt,40pt)..controls (-27pt,5pt) and (-26pt,2pt)..(-25pt,0pt);
    \draw [thick](31pt,0pt)..controls (30pt,2pt) and (29pt,5pt)..(28pt,40pt)..controls (27pt,5pt) and (26pt,2pt)..(25pt,0pt);
        \node at (-28pt,-7pt){${x_C}$};
     \node at (28pt,-7pt){${x_A}$};
          \draw[thick] (-10pt,-3pt)-- (-10pt,3pt);
     \draw[thick] (10pt,-3pt)-- (10pt,3pt);
      \node at (-10pt,7pt){${x_1}$};
       \node at (10pt,7pt){${x_2}$};
   \end{tikzpicture}
   }
   \hspace{10pt}
       \subfloat[$x_1>x_A$]{
   \begin{tikzpicture}[scale=1.0, baseline={(current bounding box.center)}]
    \draw[thick] (-60pt,0pt) -- (60pt,0pt);
   \draw[thick] (-60pt,-5pt)-- (-60pt,5pt);
    \fill[pattern = south west lines] (-70pt,-5pt) rectangle (-60pt,5pt); 
    \draw[thick] (60pt,-5pt) -- (60pt,5pt);
    \fill[pattern= south west lines] (60pt,-5pt) rectangle (70pt,5pt); 
    %\draw (-60pt,0pt) .. controls (-30pt,0pt) and (-30pt,40pt) ..(0pt,40pt)..controls (30pt,40pt) and (30pt,0pt) .. (60pt,0pt);
    \draw [thick](-31pt,0pt)..controls (-30pt,2pt) and (-29pt,5pt)..(-28pt,40pt)..controls (-27pt,5pt) and (-26pt,2pt)..(-25pt,0pt);
    \draw [thick](31pt,0pt)..controls (30pt,2pt) and (29pt,5pt)..(28pt,40pt)..controls (27pt,5pt) and (26pt,2pt)..(25pt,0pt);
        \node at (-28pt,-7pt){${x_C}$};
     \node at (28pt,-7pt){${x_A}$};
          \draw[thick] (40pt,-2pt)-- (40pt,2pt);
    \draw[thick] (10pt,-3pt)-- (10pt,3pt);
     \node at (40pt,7pt){${x_1}$};
     \node at (10pt,7pt){${x_2}$};
   \end{tikzpicture}
   }
    \caption{Entanglement cuts for different cases. In (a) and (c), the subsystem includes only one energy peak. In (b), the subsystem doesn't include any energy peak.}
    \label{fig: Entanglement cut}
\end{figure}
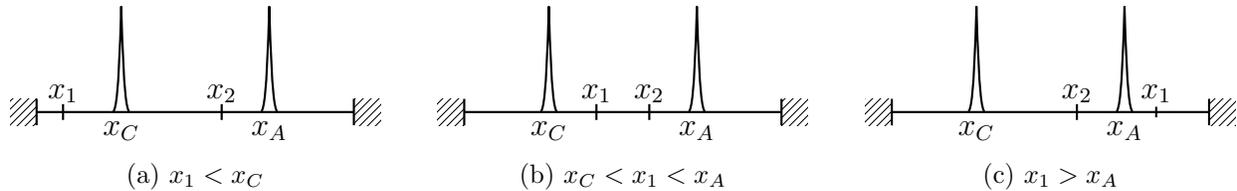

This statement can be further verified by studying the entanglement of the subsystem $A'$ with ending points $x_1,x_2 \in (0,L)$. %The R\'enyi and von Neumann entropy are then related to an equal-time correlation function of two twist operators inserted at $x_1$ and $x_2$, respectively. 
We fix $x_2$ to sit between the two energy peaks and study how the long-time behavior of entanglement growth depends on the choice of $x_1$. Without loss of generality, we assume the chiral energy peak is on the left and the anti-chiral peak is on the right in the following discussion. In general, there are three different choices of $x_1$:
\begin{enumerate}
    \item $0<x_1<x_C$. In this case, the subsystem $A'$ only includes the chiral peak, as depicted in \figref{fig: Entanglement cut}~(a). The entanglement entropy is,
    \begin{equation}
        \label{eqn:SA heating}
	    S_{A'}(x_1,x_2,t) = 
	    -\frac{c}{6}n\log \eta + \text{(non-universal)}
    \end{equation}
    where the first term grows linearly with time (i.e. the driving cycle $n$), which is consistent with the result in \cite{wen2018floquet}. As long as $x_1<x_C$, the slope only depends on the central charge and the characteristic constant $\eta$ but not on the positions of entanglement cuts. This behavior is universal and does not depend on the operator content. The non-universal terms are sub-leading in the $n\gg 1$ limit. 
%    information is encoded into the sub-leading terms denoted by "$\cdots$" in \eqnref{eqn:SA heating}, which is not important to the entanglement growth here. 
    \item $x_C<x_1,x_2<x_A$. In this case, the subsystem is between the chiral and anti-chiral peak, as depicted in \figref{fig: Entanglement cut}~(b). To the leading order, one can show that it saturates to an $\calO(1)$ value, which depends on the operator content and position of insertion. The exact value is not relevant but the most important is that the entanglement entropy does not have interesting time dependence in the long time limit.  
    \item $x_A<x_1$. In this case, the subsystem $A'$ only includes the anti-chiral peak, as depicted in \figref{fig: Entanglement cut}~(c). The entanglement entropy grows linearly as in \eqnref{eqn:SA heating}.
\end{enumerate}
We also provide lattice calculation to further check  these statements. The results can be found in \appref{app:spatial structure lattice}.

Using the results above, we can also infer the bipartite mutual information between the two energy peaks. Let us choose two disjoint regions $X$ and $Y$, with $X$ and $Y$ only including the left and right peak respectively. We call their complement as $Z$, which is composed of three disjoint regions $Z_1$, $Z_2$ and $Z_3$, located on the left of the chiral peak, between the two peaks and on the right of the anti-chiral peak, respectively. Since we are studying a pure state, the mutual information between $X$ and $Y$ is
\begin{equation}
\begin{tikzpicture}[scale=0.8, baseline={([yshift=-6pt]current bounding box.center)}]
    \draw[thick] (-60pt,0pt) -- (60pt,0pt);
   \draw[thick] (-60pt,-5pt)-- (-60pt,5pt);
    \draw[thick] (60pt,-5pt) -- (60pt,5pt);
    \draw [thick](-31pt,0pt)..controls (-30pt,2pt) and (-29pt,5pt)..(-28pt,40pt)..controls (-27pt,5pt) and (-26pt,2pt)..(-25pt,0pt);
    \draw [thick](31pt,0pt)..controls (30pt,2pt) and (29pt,5pt)..(28pt,40pt)..controls (27pt,5pt) and (26pt,2pt)..(25pt,0pt);
    \draw[thick] (-45pt,-3pt)-- (-45pt,3pt);
    \draw[thick] (-12pt,-3pt)-- (-12pt,3pt);
    \draw[thick] (12pt,-3pt)-- (12pt,3pt);
    \draw[thick] (45pt,-3pt)-- (45pt,3pt);
    \node at (-52.5pt,-10pt) {\scriptsize $Z_1$};
    \node at (-29pt,-10pt) {\scriptsize $X$};
    \node at (0pt,-10pt) {\scriptsize $Z_2$};
    \node at (29pt,-10pt) {\scriptsize $Y$};
    \node at (52.5pt,-10pt) {\scriptsize $Z_3$};
   \end{tikzpicture} \hspace{40pt}
    I(X;Y)=S_X + S_Y - S_{XY} = S_X + S_Y - S_{Z}.
\end{equation}
Following the prescriptions in \appref{app:entanglement}, the calculation of $S_{XY}$ requires a four-point function of twist operators in the boundary CFT which is in general unknown. Instead, we can use the subadditivity of entanglement to bound $S_{Z}$
\begin{equation}
    0\leq S_{Z} \leq S_{Z_1} + S_{Z_2} + S_{Z_3} \sim \calO(1)
\end{equation}
where in the last step we use the fact that none of $S_{Z_j}, j=1,2,3$ grows with time and saturates to an $\calO(1)$ value. Therefore $S_{Z}$ itself can only be $\calO(1)$ value and does not make an important contribution to the time dependence of the mutual information. On the other hand, since each of $X$ and $Y$ includes one peak, $S_{X}$ and $S_{Y}$ grows linearly with time  as shown in \eqnref{eqn:SA heating}. Thus the mutual information also linearly grows with time,
\begin{equation}
    I(X;Y) = -\frac{c}{3}n \log \eta + \text{(non-universal)},
\end{equation}
where the non-universal terms are sub-leading in the $n\gg 1$ limit. 

All of the results above provide strong evidence that the state prepared by this Floquet driving only contains bipartite entanglement. We can think of the entanglement pattern as being described by many EPR pairs accumulating at the two peaks, i.e. one member of the pair is at one peak and the other member of the pair is at the other peak. In each Floquet cycle, there are $\frac{c}{3} \log_2 \frac1\eta$ pairs created.

This suggests a quasi-particle picture which is developed in the next section and will help us understand the phenomena outlined by the  calculations.

\subsection{The quasi-particle picture}
\label{Sec: Quasi-particle}

In this section, we provide a quasi-particle picture to understand the formation of the peaks and the entanglement pattern similar to the discussions in Calabrese and Cardy \cite{Calabrese:2004eu, Calabrese:2005in}. It is not surprising that such a quasi-particle picture exists since our analysis above should apply to any $(1+1)$D CFT, including the one realized by $(1+1)$D massless free fermion. What is interesting is that the predictions from the quasi-particle picture agree quantitatively with the CFT calculations.

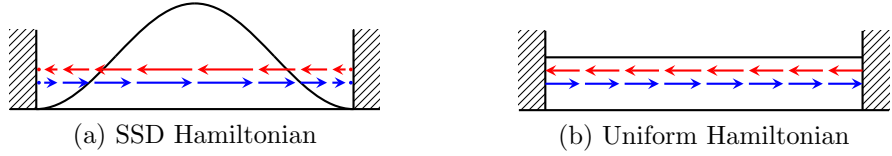
\begin{figure}
    \centering
    \subfloat[SSD Hamiltonian]{
   \begin{tikzpicture}[scale=1.0, baseline={(current bounding box.center)}]
    \draw[thick] (-70pt,0pt) -- (70pt,0pt);
   \draw[thick] (-60pt,0pt)-- (-60pt,30pt);
    \fill[pattern = south west lines] (-70pt,0pt) rectangle (-60pt,30pt); 
    \draw[thick] (60pt,0pt) -- (60pt,30pt);
    \fill[pattern= south west lines] (60pt,0pt) rectangle (70pt,30pt); 
    %\draw (-60pt,0pt) .. controls (-30pt,0pt) and (-30pt,40pt) ..(0pt,40pt)..controls (30pt,40pt) and (30pt,0pt) .. (60pt,0pt);
     \draw[thick] (-30pt,20pt) sin (0,40pt);
     \draw[thick] (-30pt,20pt) sin (-60pt,0pt);
     \draw[xscale=-1,thick] (-30pt,20pt) sin (0,40pt);
     \draw[xscale=-1,thick] (-30pt,20pt) sin (-60pt,0pt);
     \draw[->,>=stealth, thick, blue] (-57pt,10pt) -- (-52pt,10pt); \draw[->,>=stealth, thick, blue] (-50pt,10pt) -- (-40pt,10pt);
     \draw[->,>=stealth, thick, blue] (-38pt,10pt) -- (-24pt,10pt);
     \draw[->,>=stealth, thick, blue] (-22pt,10pt) -- (-1pt,10pt);
     \draw[->,>=stealth, thick, blue] (52pt,10pt) -- (57pt,10pt); \draw[->,>=stealth, thick, blue] (40pt,10pt) -- (50pt,10pt);
     \draw[->,>=stealth, thick, blue] (24pt,10pt) -- (38pt,10pt);
     \draw[->,>=stealth, thick, blue] (1pt,10pt) -- (22pt,10pt);
     \filldraw[blue] (-59pt,10pt) circle (0.5 pt);
     \filldraw[blue] (59pt,10pt) circle (0.5 pt);
     
     \draw[->,>=stealth, thick, red] (57pt,15pt) -- (52pt,15pt); \draw[->,>=stealth, thick, red] (50pt,15pt) -- (40pt,15pt);
     \draw[->,>=stealth, thick, red] (38pt,15pt) -- (24pt,15pt);
     \draw[->,>=stealth, thick, red] (22pt,15pt) -- (1pt,15pt);
     \draw[->,>=stealth, thick, red] (-52pt,15pt) -- (-57pt,15pt); \draw[->,>=stealth, thick, red] (-40pt,15pt) -- (-50pt,15pt);
     \draw[->,>=stealth, thick, red] (-24pt,15pt) -- (-38pt,15pt);
     \draw[->,>=stealth, thick, red] (-1pt,15pt) -- (-22pt,15pt);
     \filldraw[red] (-59pt,15pt) circle (0.5 pt);
     \filldraw[red] (59pt,15pt) circle (0.5 pt);
   \end{tikzpicture}
   }
   \hspace{40pt}
    \subfloat[Uniform Hamiltonian]{
   \begin{tikzpicture}[scale=1.0, baseline={(current bounding box.center)}]
   \draw[white] (0pt,40pt) circle (1pt);
    \draw[thick] (-70pt,0pt) -- (70pt,0pt);
   \draw[thick] (-60pt,0pt)-- (-60pt,30pt);
    \fill[pattern = south west lines] (-70pt,0pt) rectangle (-60pt,30pt); 
    \draw[thick] (60pt,0pt) -- (60pt,30pt);
    \fill[pattern= south west lines] (60pt,0pt) rectangle (70pt,30pt); 
    \draw[thick] (-60pt,20pt) -- (60pt,20pt);
    \draw[->,>=stealth, thick, blue] (-60pt,10pt) -- (-47pt,10pt); \draw[->,>=stealth, thick, blue] (-45pt,10pt) -- (-32pt,10pt);
     \draw[->,>=stealth, thick, blue] (-30pt,10pt) -- (-17pt,10pt);
     \draw[->,>=stealth, thick, blue] (-15pt,10pt) -- (-1pt,10pt);
     \draw[->,>=stealth, thick, blue] (1pt,10pt) -- (15pt,10pt); \draw[->,>=stealth, thick, blue] (17pt,10pt) -- (30pt,10pt);
     \draw[->,>=stealth, thick, blue] (32pt,10pt) -- (45pt,10pt);
     \draw[->,>=stealth, thick, blue] (47pt,10pt) -- (60pt,10pt);
     \draw[->,>=stealth, thick, red] (60pt,15pt) -- (47pt,15pt); \draw[->,>=stealth, thick, red] (45pt,15pt) -- (32pt,15pt);
     \draw[->,>=stealth, thick, red] (30pt,15pt) -- (17pt,15pt);
     \draw[->,>=stealth, thick, red] (15pt,15pt) -- (1pt,15pt);
     \draw[->,>=stealth, thick, red] (-1pt,15pt) -- (-15pt,15pt); \draw[->,>=stealth, thick, red] (-17pt,15pt) -- (-30pt,15pt);
     \draw[->,>=stealth, thick, red] (-32pt,15pt) -- (-45pt,15pt);
     \draw[->,>=stealth, thick, red] (-47pt,15pt) -- (-60pt,15pt);
   \end{tikzpicture}
   }
   \hspace{20pt}
    \caption{Quasi-particle picture: in both (a) and (b), black lines are the cartoon profile for the Hamiltonians and the red/blue arrows are the corresponding velocities for the left/right moving quasi-particles.}
    \label{fig: quasi-particle}
\end{figure}

From the quasi-particle picture, in each Floquet driving cycle, when we suddenly change the Hamiltonian, we expect that there will be quasi-particle excitations emitting from different points. The pairs of particles moving to the left and right from a given point are highly entangled. For example, at the beginning of each cycle, we change $H_0$ to $H_1$, which 
%introduces some 
creates EPR pairs in the system. Then they move together with all other quasi-particles that have been created in previous cycles with velocity $v(x)=2\sin^2(\pi x/L)$. The velocity is determined by the sin-square envelope we defined in \eqnref{eqn:H1}. In the second part of each driving cycle, the quasi-particles will be governed by $H_0$ and the velocity will now change to $v(x)=1$. Therefore, we can determine the distance that a quasi-particle travels in one cycle by the following formula:

\begin{equation}
    \text{Distance} = \int_0^{T_1} v(x) dt + T_0   \hspace{40pt}
    \begin{tikzpicture}[scale=1.0, baseline={([yshift=-4pt]current bounding box.center)}]
    \draw[thick] (-70pt,0pt) -- (70pt,0pt);
   \draw[thick] (-60pt,0pt)-- (-60pt,30pt);
    \fill[pattern = south west lines] (-70pt,0pt) rectangle (-60pt,30pt); 
    \draw[thick] (60pt,0pt) -- (60pt,30pt);
    \fill[pattern= south west lines] (60pt,0pt) rectangle (70pt,30pt); 
    \draw[thick,mid arrow] (-20pt,15pt)-- (60pt,15pt);
    \draw[thick,mid arrow] (60pt,20pt)-- (20pt,20pt);
    \filldraw (-20pt,15pt) circle (1pt) node[below] {$x_i$};
    \filldraw (20pt,20pt) circle (1pt) node[above] {$x_f$};
   \end{tikzpicture} \,
   \label{eqn: quasi-particle moving}
\end{equation}
This distance depends on $T_0$, $T_1$, and the initial 
position $x_i$ of the quasi-particle, thus uniquely determining the final position $x_f$
of the quasi-particle.
For the plot in \eqref{eqn: quasi-particle moving}, 
we assume the quasi-particle bounces back once from the boundary\footnote{The conformal boundary condition ensures the magnitude of the velocity remains the same after the bouncing.}.
In general, the number of bounces is $n$ if $(n-1)L<T_0<nL$, where $n$ is a positive integer. One can find that $T_1$ does not come in, because under $H_1$ the quasi-particles will never reach the boundary due to the vanishing of velocity at the boundary\footnote{More exactly, this statement is only true for a vanishing function at least faster than linear, because $t\sim \int_0^\epsilon \frac{dx}{x^a} $ diverges when $a\geq 1$.
Here for the SSD, we have $v(x)=2\sin^2(\pi x/L)2\sim x^2$ for $x\to 0$.}.

With this quasi-particle picture, we can determine the positions of the 
two energy peaks as observed in Fig.~\ref{fig:energydensity}.
Recall that in the long time limit $n\gg 1$, after each driving cycle, the positions of the two energy peaks stay the same.
There are only two possibilities as follows
(without loss of generality let us focus on the chiral energy peak and track its position $x_C$), 
\begin{enumerate}
    \item If the number of bounces $n$ is odd, then the chiral energy peak
will become an anti-chiral energy peak due to the bounces at the boundary.
To keep the positions of the chiral/anti-chiral energy peaks the same,
we have to do the switching: $x_C\leftrightarrow x_A$.
That is, we have $x_i=x_C$ and $x_f=x_A$ in \eqref{eqn: quasi-particle moving}.
    \item If the number of bounces $n$ is even, then the chiral energy peak
is still chiral after each driving cycle. Then one has $x_i=x_f=x_C$.
\end{enumerate}
The tracking of the anti-chiral energy peak can be analyzed in the same
way. Noting that the two energy peaks are symmetric 
about $x=L/2$, i.e., $x_A=L-x_C$, we can determine 
the positions of the two energy peaks by the ``quantization'' condition 
\begin{equation}
    \int_0^{T_1}v(x)dt+T_0=nL,\quad n\in \ZZ\,.
\end{equation}
Evaluating the above equation explicitly, one can find that 
$x_C$ and $x_A$ are determined by the following equation:
\begin{equation}
    \frac{2\pi T_1}{L}=\cot \frac{\pi(x^{\ast}+T_0-nL)}{L} - \cot\frac{\pi\, x^{\ast}}{L}\,,
        \label{peak_position}
\end{equation}
with $x_C=x^{\ast}$ and $x_A=L-x^{\ast}$.
One can check that the solution $x^*$ in \eqnref{peak_position}
matches the repulsive fixed point $\gamma_2 = e^{2\pi i x^*/L}$ of the M\"obius transformation in \eqnref{eqn:gamma1 gamma2} \emph{exactly}.

This quasi-particle picture turns out to be useful.    
On the one hand, it gives us a semi-classical explanation for the formation of the peaks.  In the heating phase, the system keeps absorbing energy by creating many EPR pairs. Due to the fixed point solution in the semi-classical motion, those EPR pairs will accumulate at the fixed point, with chiral part staying at $x_C=x^*$ and anti-chiral part staying at $x_A=L-x^*$. If we keep track of what happens within one cycle, we will find that the particles at the two peaks will switch their position after one cycle if the number of bounces $n$ is odd, but will stay the same if $n$ is even. In the non-heating phase, the system does not absorb much energy and there is no fixed point in the equation of motion. Therefore we only observe energy oscillation.

On the other hand, it also provides us insight into the growth of entanglement entropy. The EPR pairs generated, not only carry energy but also share entanglement. Therefore, as the system absorb energy, the entanglement also grows.
%, which was already studied in \cite{wen2018floquet}. AV[i removed the reference because as stated it seems like all results in the section were previously known. We could qualify that the half line entanglement was previously calculated but it may be too long. Anyway we have mentioned this previously]
Based on this semi-classical picture, it is not hard to conjecture that all the entanglement is shared by the two peaks. Since the energy and entanglement are carried by the same objects, it is also natural to expect some relationship between them. In the following section, we will derive such a relation.

\subsection{Energy-entanglement relation}
\label{sec:relation}

As we said before, since we have this interpretation that the energy and entanglement are all carried by those EPR pairs at the two peaks, it will be natural to ask whether there is any relation between them. The result for entanglement entropy always contains a divergent non-universal piece due to the absence of a UV cutoff in a field theory, which is absent on the energy side. Therefore, in the following, we only compare their universal time dependence and dispense with the non-universal part.

First, let us look at the results for the heating phase. By comparing \eqnref{eqn:E heating} with \eqnref{eqn:SA heating}, we can find the following equation,
\begin{equation}
    \label{eqn:entropy_energy}
	E_{\text{heating}}(t) \propto c\exp\left( \frac{6}{c}S(t) \right),
\end{equation}
which relates the total energy growth to the entanglement growth for the chiral or anti-chiral peaks. We use proportion instead of equality because we only keep the universal information and drop all the other non-universal details.

Then, let us see whether this relation also holds in the non-heating phase and the critical case. Because we already know that the entanglement comes from the two peaks, it is sufficient and technically easier if we just use the result for the left half and right half entanglement, as has been computed in \cite{wen2018floquet}. Since they used a different notation, we reproduce their results in \appref{app:entanglement} using our notations so that readers can compare with the total energy more conveniently. For the non-heating phase, the entanglement entropy keeps oscillating as $\#\eta^n$ around a non-zero average value, which matches this result. For the critical case, the entanglement entropy grows logarithmically in time $S\approx \frac{c}{3}\log n$, which also matches this result. 

Several remarks follow below. This equation only contains the central charge and thus is true for any CFT. Generically, we would not expect such a universal relation. It only holds here because the states prepared are related by a conformal transformation that makes the universal relation possible\cite{chen2016holographic}. It is also, to some extent, reminiscent of the Cardy formula in an equilibrium CFT, which says energy is proportional to the square of the entropy. However, what we found here is that the energy is the exponential of the entropy, thus much larger than the entropy, which suggests the state we prepare is far from the equilibrium state. 

\section{Effect of randomness in non-heating phase}
\label{sec: Randomness_nonheating}

In a real experiment, local perturbations and imperfections of the pulse sequences are inevitable. In the section, we will discuss the effect of having some small randomness on the driving period $T_0$ and $T_1$, i.e. in each cycle $T_0$ and $T_1$ are independently drawn from a certain distribution, the final results are obtained after doing a ``disorder" average. Here are some detailed explanations of the protocol, 
\begin{enumerate}
    \item The driving time for each cycle is given by
\begin{equation}
	\label{eqn:Random T0T1 a}
	T_0= \bar{T_0}+\delta T_0, \quad T_1= \bar{T_1}+\delta T_1,
\end{equation}
where $\delta T_0$ and $\delta T_1$ denote the deviation from the (constant) mean values $\bar{T_0}$ and $\bar{T_1}$. For simplicity, we consider the case that $\delta T_0$ and $\delta T_1$ are uniformly distributed in the following domain: 
\begin{equation}
	\label{eqn:Random T0T1 b}
	\delta T_0,\, \delta T_1\in \left[-\frac{\alpha L}{2}, \frac{\alpha L}{2}\right],
\end{equation}
where $\alpha$ characterizes the magnitude of randomness. $L$ is the total length of the system, which is the fundamental time scale of the system. 
    \item 
    Given a sequence of randomized driving time, the operator $\calO$ on the $z$-plane under $n$-cycle imaginary time evolution is given by the familiar formula
\begin{equation}
    \left( \frac{\partial z_n}{\partial z} \right)^h
    \left( \frac{\partial \bar{z}_n}{\partial \bar{z}} \right)^{\bar{h} }
    \calO(z_n,\bar{z}_n).
\end{equation}
The derivative terms are calculated using the chain rule,
\begin{equation}\label{Zn_ChainRule}
    \frac{\partial z_n}{\partial z}=\frac{\partial z_n}{\partial z_{n-1}}\cdot \frac{\partial z_{n-1}}{\partial z_{n-2}}\cdots 
    \frac{\partial z_2}{\partial z_1}\cdot 
    \frac{\partial z_1}{\partial z}\,,
\end{equation}
where (c.f. Eqs.\eqref{eqn:mobius} and \eqref{eqn: mobius z1 abcd})
\begin{equation}
    	\label{eqn:mobius_z_i}
	z_i = f(z_{i-1}) = \frac{a z_i + b}{c z_i + d}, \quad \begin{pmatrix}
	a & b \\
	c & d
	\end{pmatrix} \in \SL(2,\RR) \,.
\end{equation}
The difference comparing to the previous sections is that the parameters $a,b,c,d$ here are determined by the randomized driving cycles, in particular, the terms in the chain rule formula are independent. 
%dimensionless driving periods $\tau_0/L$ and $\tau_1/L$. They have the same expression as those in \eqnref{eqn: mobius z1 abcd}, but with $\tau_0$ and $\tau_1$ (after analytical continuation) replaced with the random ones in \eqnref{eqn:Random T0T1 a}.

We comment that in principle for the real-time evolution, we need to analytically continue %the $\tau_0$ and $\tau_1$ 
for each cycle in order to keep track of the trajectories of $z_n$ and $\bar{z}_n$ to determine the branch cut crossing. The disorder average is done after the analytic continuation. Here we only consider the stress tensor and the entanglement entropy of the left (right) half system. Both quantities are free of the branch cut issue and we can safely perform the analytic continuation.
\end{enumerate}

\subsection{Energy and entanglement growth}

%The first thing we need to check is the phase diagram with weak randomness. We first keep our main focus on the fate of the non-heating phase.
In this section, we present numerical results for the energy and entanglement growth affected by the randomness. We will focus on the non-heating phase and briefly comment on the heating phase. 

\begin{figure}[t]
	\centering
	\subfloat[Growth of total energy]{\includegraphics[width=3in]{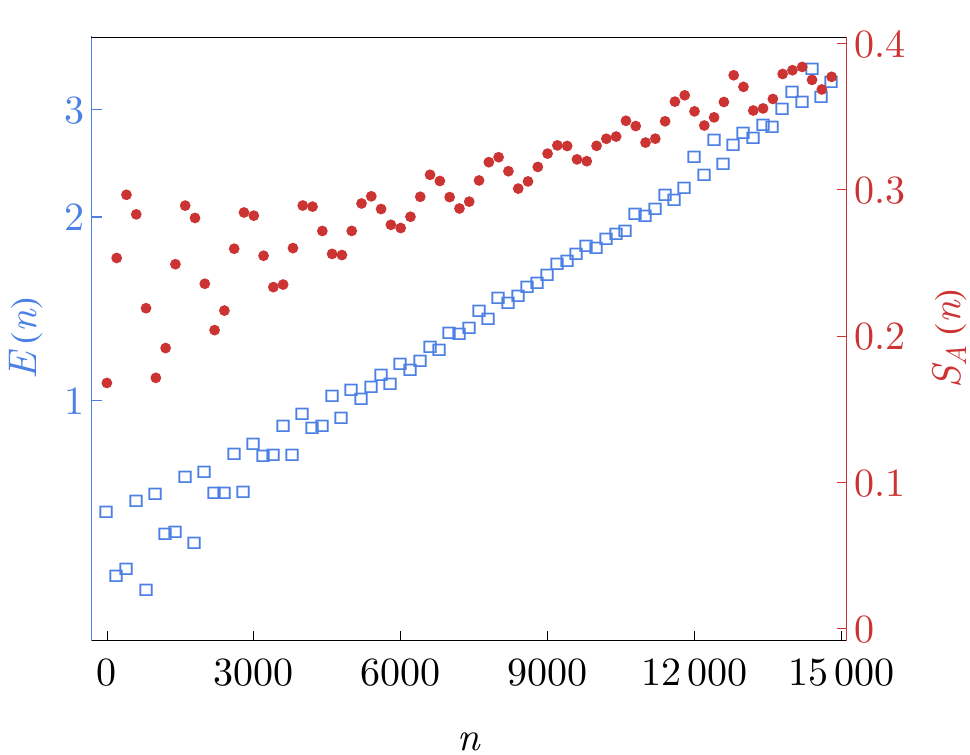}}
	\hspace{5pt}
	\subfloat[Growth of entanglement entropy]{\includegraphics[width=3in]{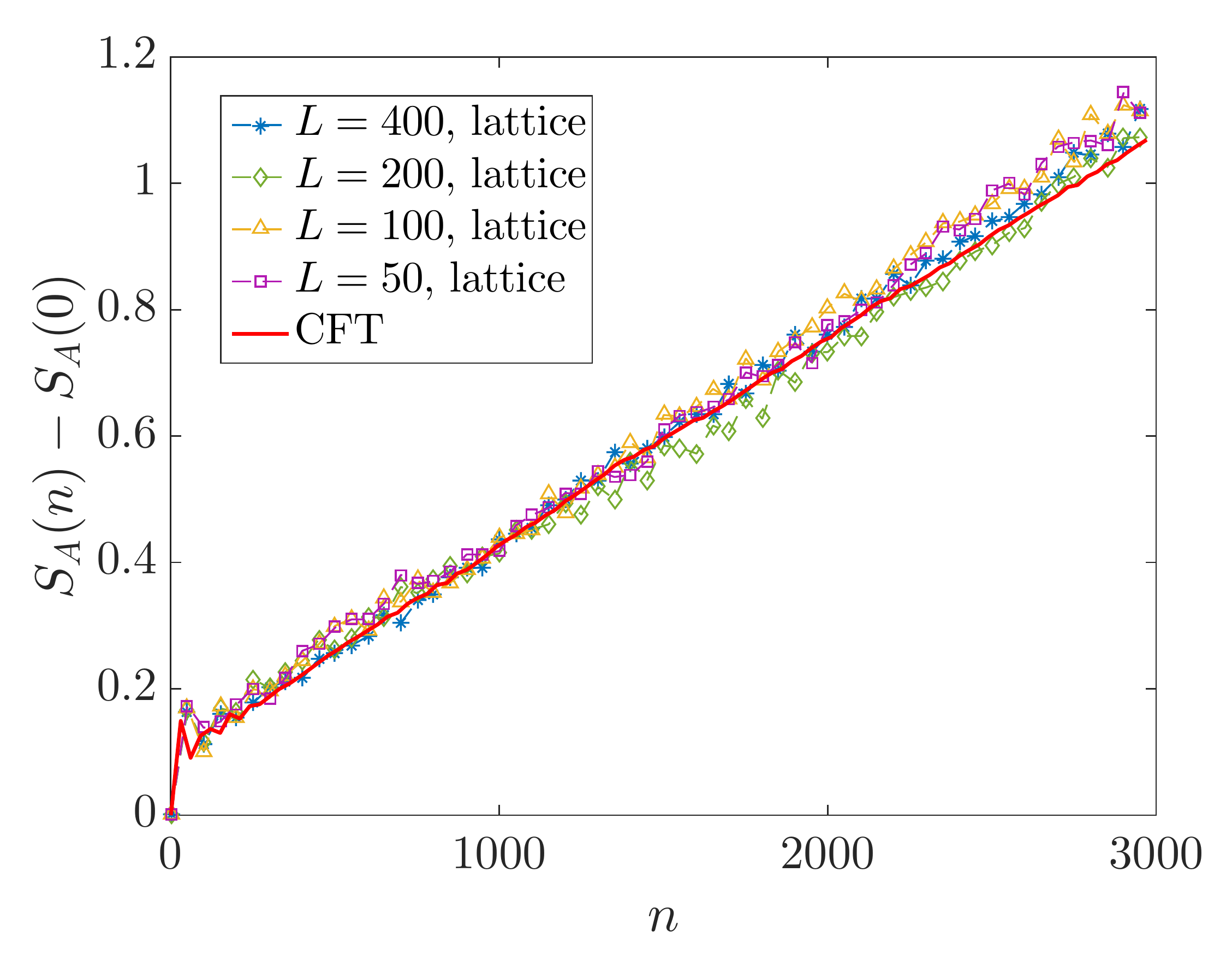}}
	\caption{The total energy and entanglement entropy of the subsystem $A=[0, L/2]$ with random driving. (a) We choose $L=2\pi$ and $\bar{T_0}=\bar{T_1}=\bar{T}$, $\bar{T}/L=0.1, \alpha=0.01$. $n$ is the number of driving cycles. Each data point is calculated by doing random average 1000 times. The energy (blue hollow squares) grows with time exponentially. The entanglement entropy (red dots) for the left half system grows with time linearly. (b) Entanglement entropy evolution for subsystem $A=[0, L/2]$ with different $L$. The lattice calculation is done with complex free fermions} and is averaged over $N_{\text{sample}}=200$, with $L=50$, $100$, $200$, and $400$, respectively. The randomness is chosen as $\bar{T_0}=\bar{T_1}=\bar{T}$, $\bar{T}/L=0.03, \alpha=0.06$.
	\label{fig:ee growth}
\end{figure}

The first result is presented in \figref{fig:ee growth}, where we choose parameters $\bar{T_0}$, $\bar{T_1}$ and $\alpha \ll 1$ such that each individual sample $(T_0,T_1)$ belongs to the non-heating phase, namely corresponds to an elliptic M\"obius transformation. With the randomized protocol, we find that the total energy of the system grows with time (even for $\alpha \ll 1$). Asymptotically, it grows exponentially with $n$ as numerically verified in \figref{fig:ee growth}~(a). At the same time, the entanglement entropy grows linearly with $n$, as shown in \figref{fig:ee growth}~(b). That is to say, the non-heating phase will disappear immediately for arbitrarily weak randomness and we are only left with the heating phase. The total energy and entanglement entropy grow in the same way as what we find for the heating phase without any randomness. As a side note, if we implement the random driving set-up for a Mathieu oscillator, small randomness also leads the energy to grow exponentially (see \appref{app:random mathieu} for details), which suggests that the phenomenon here might be a generic feature of the $\SL(2,\RR)$ algebra and not special to our CFT setting.

%To determine what happens to the non-heating phase under a weak randomness, we choose $\bar{T_0}$ and $\bar{T_1}$ to be deep inside the non-heating phase regime and fix $\alpha L/\bar{T}$ to be small. Thus, each $(T_0,T_1)$ in the distribution gives an elliptic M\"obius transformation and will lead to a non-heating dynamics without randomness. After turning on the tiny randomness, we find that the total energy of the system begins to grow with time. Even for a small $\alpha$, the total energy still grows eventually. At a very long time scale, it grows exponentially as numerically verified in \figref{fig:ee growth}~(a). At the same time, the entanglement entropy grows linearly with time, as shown in \figref{fig:ee growth}~(b). Therefore, the non-heating phase will disappear immediately for arbitrarily weak randomness and we are only left with the heating phase. The total energy and entanglement entropy grow in the same way as what we find for the heating phase without any randomness. If we implement the random driving set-up for a Mathieu oscillator, tiny randomness also leads energy to grow exponentially, as demonstrated in \appref{app:random mathieu}. Therefore, this phenomenon might be a generic feature of the algebra not specialized to our CFT setting.

\begin{figure}[t]
	\centering
	\subfloat[Growth rate $\kappa$ v.s. driving period]{\includegraphics[width=3.3in]{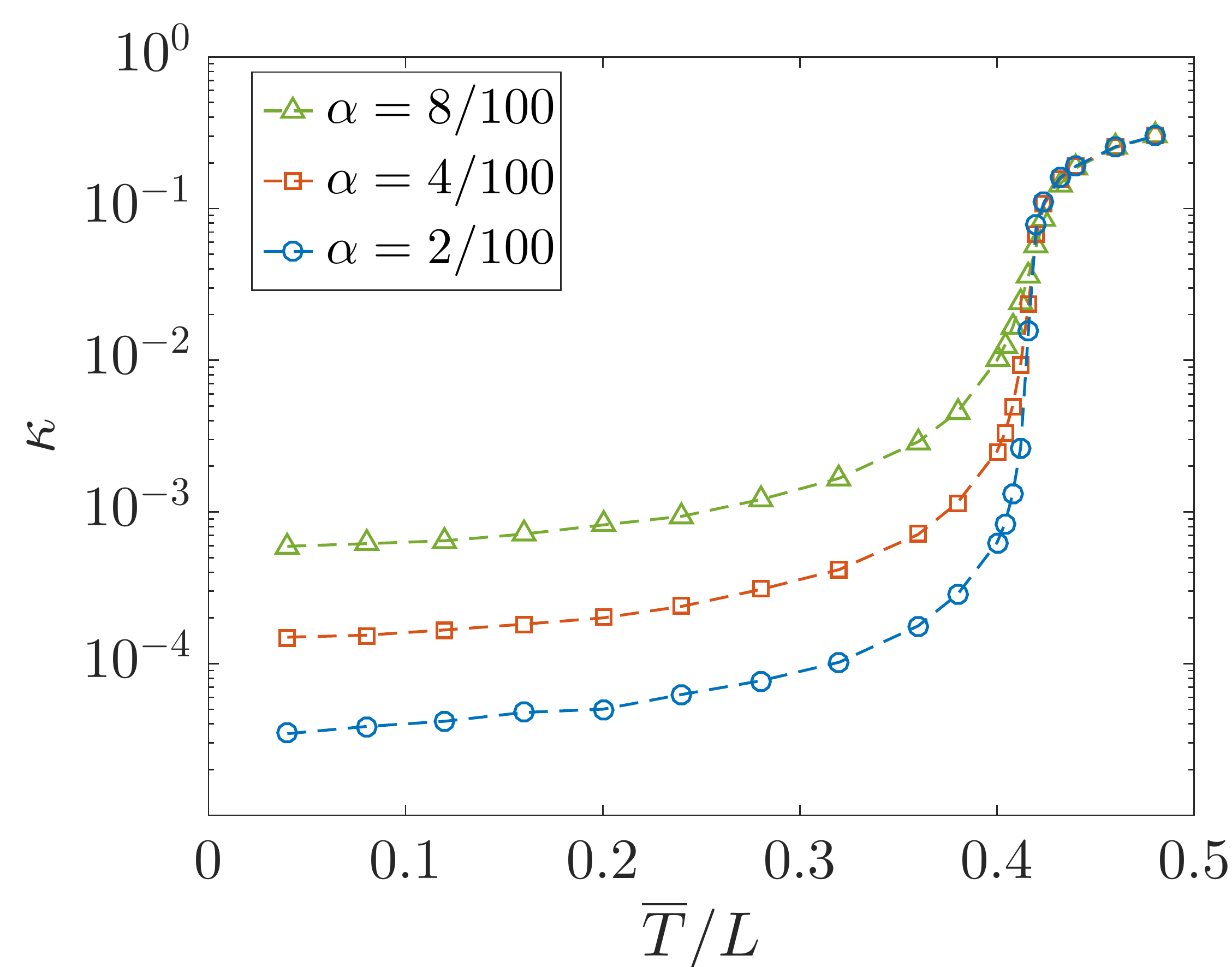}}
	\subfloat[Growth rate $\kappa$ v.s. $\alpha$]{\includegraphics[width=3.3in]{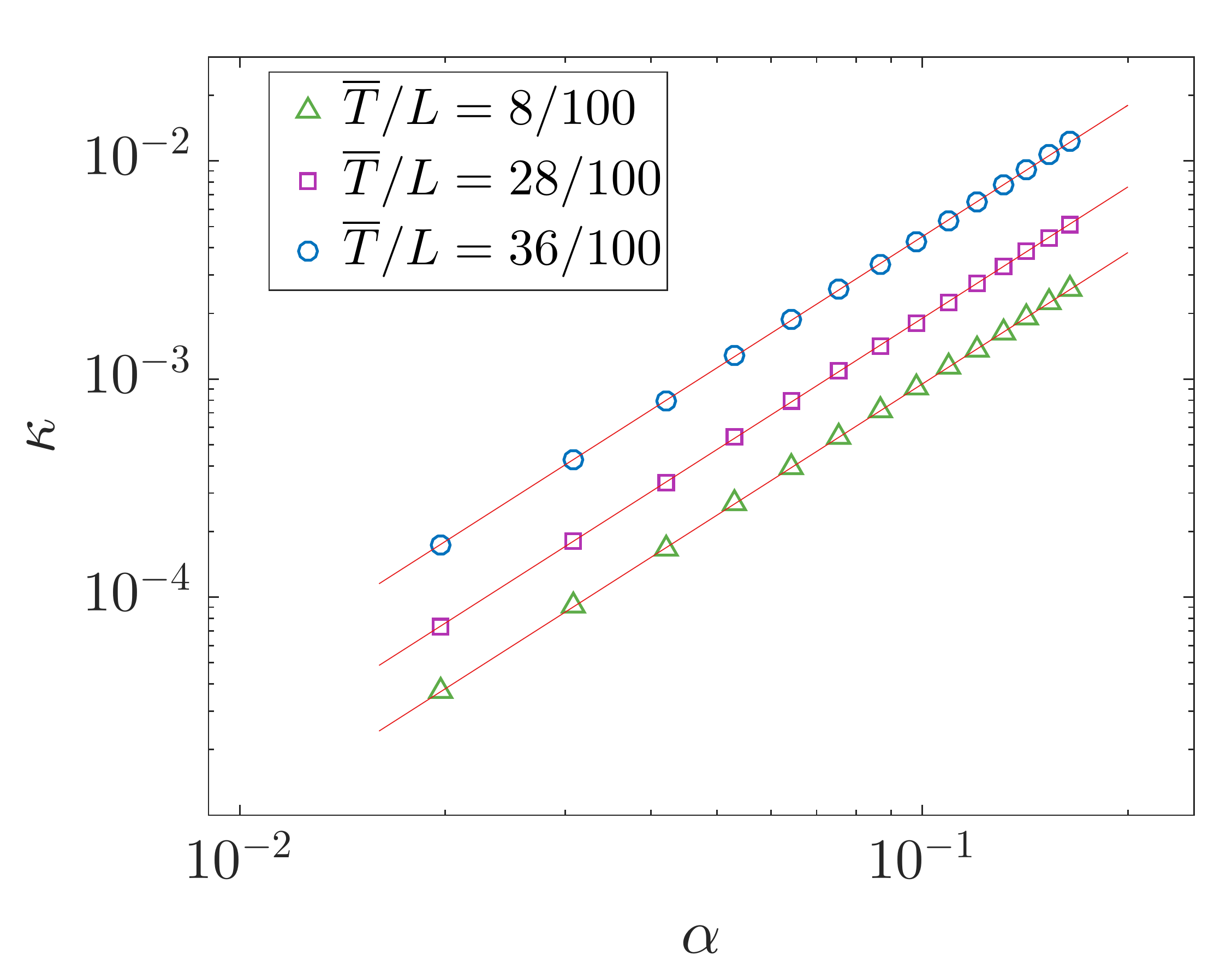}}
	\caption{(a) The slope $\kappa$ of entanglement entropy growth as a function of $\bar{T}/L$, for different magnitudes of randomness $\alpha$. Note that the phase transition from non-heating to heating phases with no randomness ($\alpha=0$) happens at $T^{\ast}/L\simeq 0.416$. (b) The slope $\kappa$ of entanglement entropy growth as a function of $\alpha$, for different $\bar{T}$. The red solid lines are fittings with $\kappa\propto \alpha^2$. Each $\kappa$ is extracted from the CFT calculation of $S_A(n)$ by averaging over $N_{\text{sample}}=1000$.
	}
	\label{fig:Heating rate}
\end{figure}

Then, it is desirable to ask how the heating rate is related to the magnitude of randomness $\alpha$. We will study this problem based on the entanglement entropy, as follows. With randomness, there are four dimensionless parameters in the calculation of entanglement, $l_A/L$, $\bar{T_0}/L$, $\bar{T_1}/L$, and $\alpha$. Recalling that $S(n)$ also depends on the total length $L$ of the system through its initial value, we consider the quantity $S_A(n)-S_A(0)$ only depends on the dimensionless ratios we introduced above.\footnote{
    Based on \eqnref{eqn: Trace rhoN} and \eqnref{eqn: Twist Single}, one can find the difference of $m$-th Renyi entropy as follows:
\begin{equation}\label{Entropy_Diff}
\small
\begin{split}
    S_A^{(m)}(n)-S_A^{(m)}(0)=&\frac{1}{1-m}\log\left[
    \left(\frac{\partial z_n}{\partial z}\right)^{h_m}\cdot  \left(\frac{\partial \bar{z}_n}{\partial \bar{z}}\right)^{h_m}\cdot 
    \frac{\mathcal{T}_m(z_n,\bar{z}_n)}{\mathcal{T}_m(z,\bar{z})}
    \right]\\
       =&\frac{1}{1-m}\log\left[
    \left(\frac{\partial z_n}{\partial z}\right)^{h_m}\cdot  \left(\frac{\partial \bar{z}_n}{\partial \bar{z}}\right)^{h_m}\cdot 
\left(\frac{\sqrt{z}\cdot \sqrt{\bar{z}}}{\sqrt{z_n}\cdot\sqrt{\bar{z}_n}}\right)^{h_m}\cdot
\left(
\frac{\sqrt{z}-\sqrt{\bar{z}}}{\sqrt{z_n}-\sqrt{\bar{z}_n}}
\right)^{2 h_m}
    \right],
    \end{split}
\end{equation}
where $\partial z_n/ \partial z$ and $\partial \bar{z}_n/\partial \bar{z}$ are calculated through the chain rule in Eq.~\eqref{Zn_ChainRule}. $z=e^{2\pi i\frac{l}{L}}$ only depends on the ratio $l/L$. $z_n$ ($\bar{z}_n$) are determined by Eq.~\eqref{Zn_ChainRule} and Eq.~\eqref{eqn:mobius_z_i}, which only depend on the dimensionless parameters $l/L$, 
$T_0/L$, and $T_1/L$ in \eqnref{eqn:Random T0T1 a}. It is noted that $T_0$ and $T_1$ in \eqnref{eqn:Random T0T1 a} are random
numbers. After doing average over Eq.~\eqref{Entropy_Diff}, the result will only
depend on $l/L$, $\bar{T_0}/L$, $\bar{T_1}/L$, and $\alpha$,
} 
Let us introduce the heating rate $\kappa$ and write the entropy growth as
\begin{equation}
	S_A(n)-S_A(0)=\kappa\cdot n+\text{const}.
\end{equation}
For simplicity, we consider the choice of $T_0$ and $T_1$ in \eqnref{eqn:Random T0T1 a} with $\bar{T_0}=\bar{T_1}=\bar{T}$. Then for $A=[0,L/2]$, $\kappa$ will only depend on two dimensionless parameters, i.e., $\bar{T}/L$ and $\alpha$.

As shown in \figref{fig:Heating rate}(a), we study $\kappa$ as a function of $\bar{T}/L$ with different $\alpha$. There are several interesting features: (i) Fixing $\alpha$, as $\bar{T}$ increases from $\bar{T}/L=0$, $\kappa$ will increase accordingly. In particular, $\kappa$ grows the fastest near the phase transition $T^{\ast}/L\simeq 0.416$ (note that the phase transition is defined for the case with no randomness). This indicates that the system is \textit{more} sensitive to the randomness near the phase transition. 
(ii) Fixing $\bar{T}<T^{\ast}$, one can find that $\kappa$ will increase
with the randomness $\alpha$. That is, with larger fluctuations in the 
driving periods, the system will be heated up more easily.
(iii) 
For $\bar{T}>T^{\ast}$, $\kappa$ collapse to the same curve, indicating that
the heating phase (defined before adding noise) is robust under the
effect of fluctuations in the driving periods.
%$\kappa$ is sensitive to $\alpha$ for $\bar{T}<T^{\ast}$, but is not sensitive to $\alpha$ for $\bar{T}>T^{\ast}$. This indicates the heating phase (before adding randomness) is stable to the randomness. \YG{[YG: I am not sure I understand the indications (ii) made here??]}

With the analysis above, now we are interested in how $\kappa$ 
depends on the magnitude of randomness $\alpha$ for a fixed $\bar{T}$ with $\bar{T}<T^{\ast}$, which corresponds to
the non-heating phase (before adding randomness).
As shown in \figref{fig:Heating rate} (b), it is found that $\kappa$ depends on $\alpha$ in the following way:
\begin{equation}
	\label{eqn:heating_rate}
	\kappa\propto \alpha^2,
\end{equation}
where the coefficient $\kappa/\alpha^2$ depends on $\bar{T}$, as can be seen in \figref{fig:Heating rate} (b). With this observation, one can alternatively plot $\kappa/\alpha^2$ as a function of $\bar{T}$. 
%based on the data in \figref{fig:Heating rate} (a), as shown in \figref{fig:Heating coefficient} in the appendix. 
It is found that $\kappa/\alpha^2$ for $\bar{T}<T^{\ast}$ only depends on $\bar{T}$, with the concrete value $0.1\sim 1$.

\begin{figure}[t]
	\centering
	\subfloat[CFT calculation]{\includegraphics[width=3in]{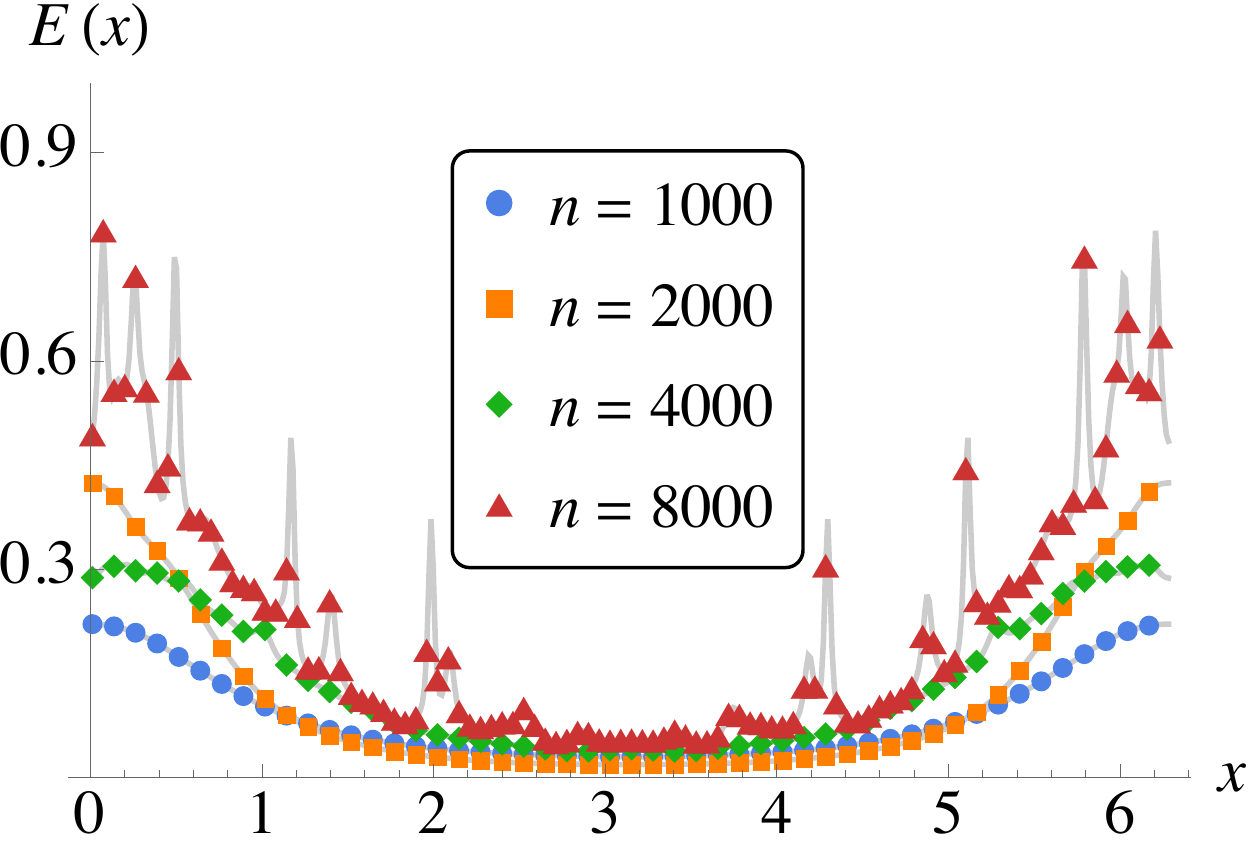}}
	\subfloat[Lattice simulation]{\includegraphics[width=3in]{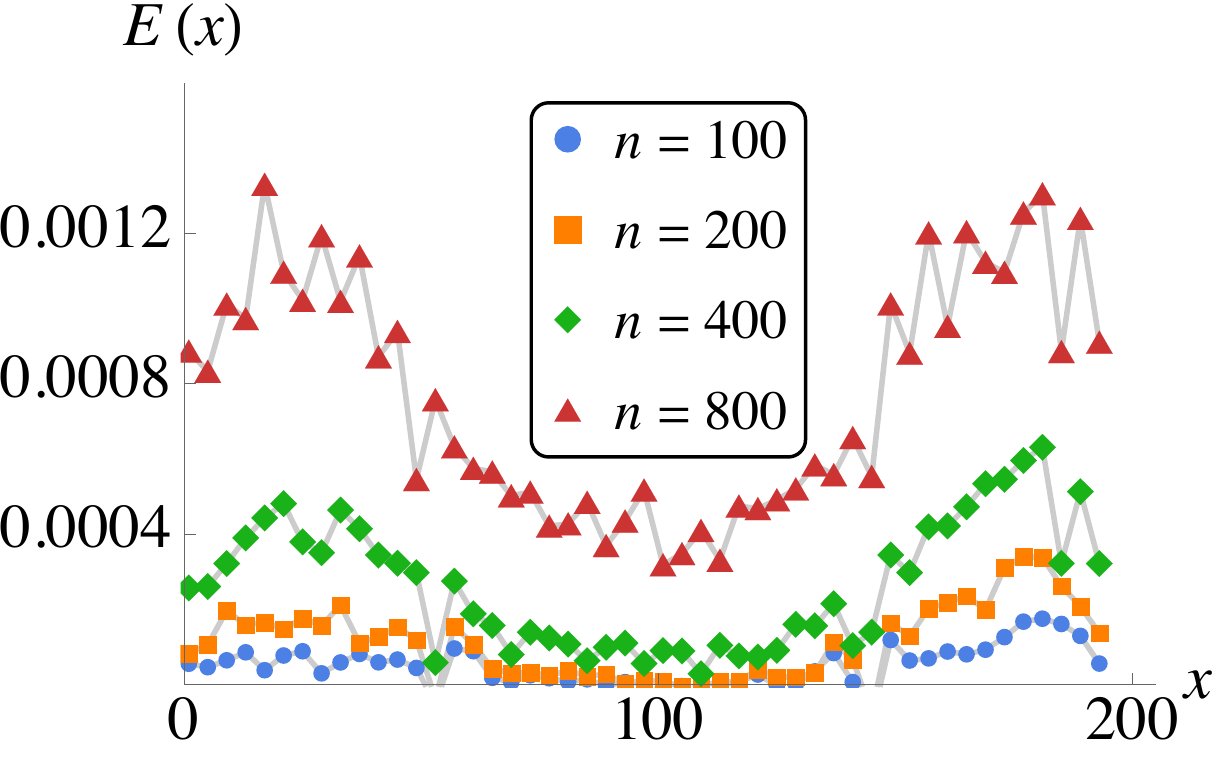}}
	\caption{Energy density for random driving. (a) CFT result on weak randomness. We choose $L=2\pi$ and $\bar{T_0}=\bar{T_1}=\bar{T}$, $\bar{T}/L=0.1, \alpha=0.01$. Each curve is calculated by doing random average 1000 times. (b) Lattice calculation on strong randomness. We simulate complex free fermion on the lattice with $L=200$, $\bar{T_0}=\bar{T_1}=\bar{T}$, $\bar{T}/L=0.3, \alpha=0.03$ and half filling. Each curve is calculated by doing random average $300$ times. In both calculations, we find the energy density peaks on the boundary.}
	\label{fig:random energy density}
\end{figure}

\subsection{Energy density distribution}

After determining how the randomness in the driving periods $T_0$ and $T_1$ affects the phase diagram, let us look at how it changes the energy density distribution.

If we start from the non-heating phase regime and add randomness, we find that the energy density peaks at the two ends of the system, as shown in \figref{fig:random energy density}. This phenomenon can be understood from the semi-classical quasi-particle picture. Without randomness, the quasi-particles are created and moved in the system coherently. After adding randomness, those motions become irregular. However, since the group velocity of the quasi-particles is smaller near the boundary, accordingly we have a higher probability to see more quasi-particles there.

\begin{figure}[t]
	\centering
	\subfloat[Growth of total energy]{\includegraphics[width=2.6in]{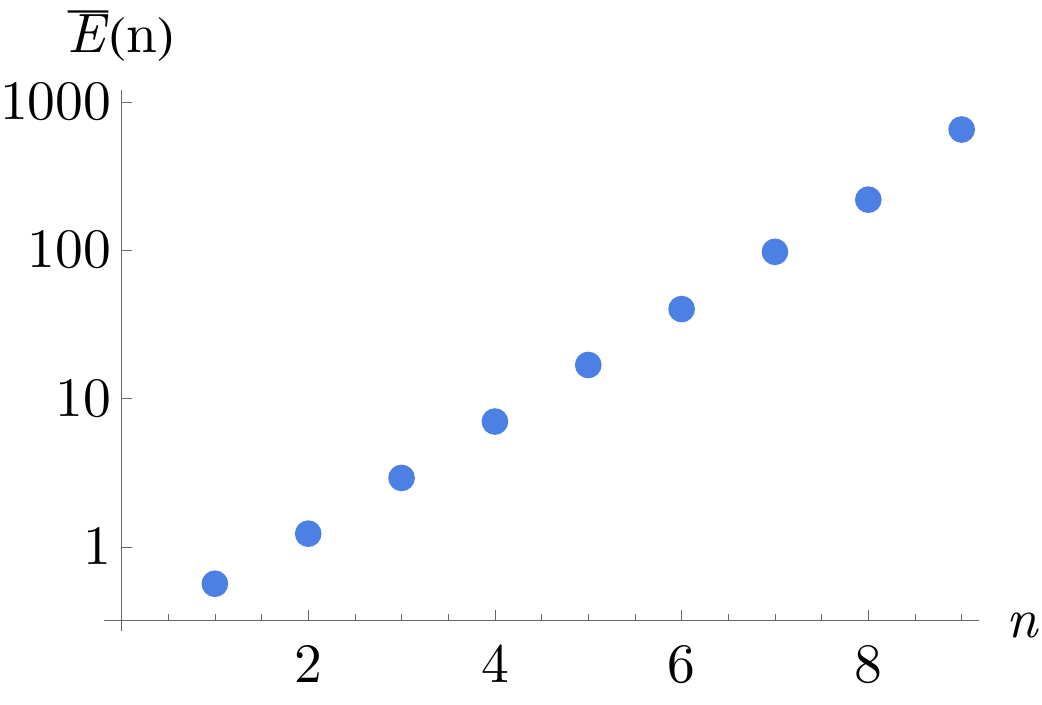}}
	\hspace{50pt}
	\subfloat[Energy density distribution]{\includegraphics[width=2.6in]{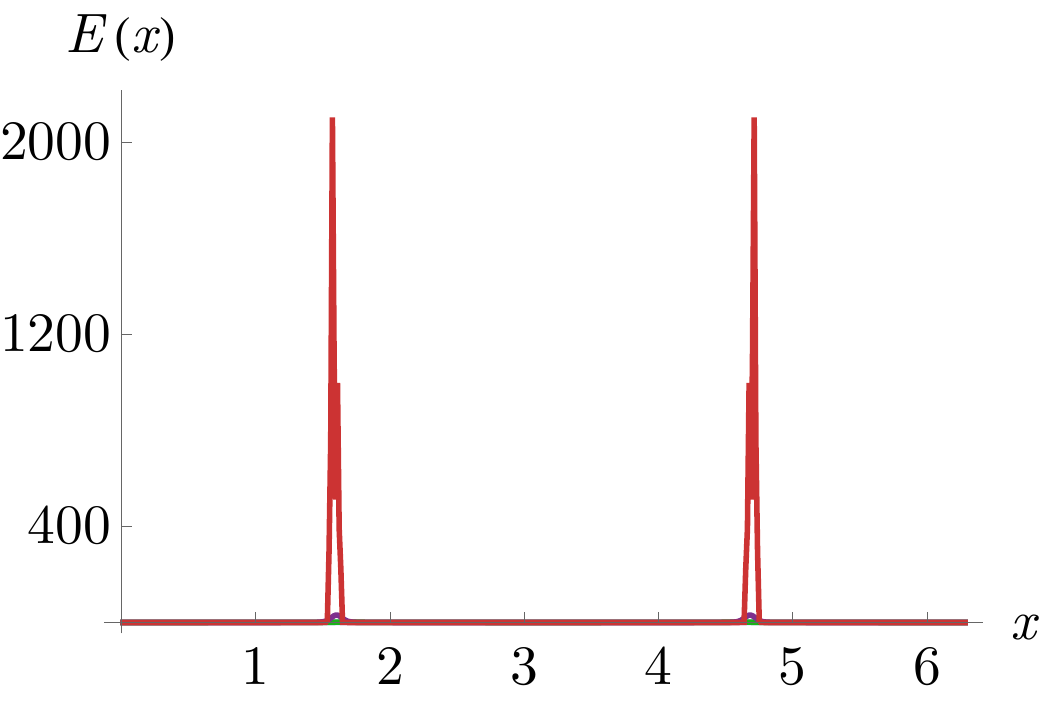}}
	\caption{Dynamics in the heating phase regime with small randomness. We choose $L=2\pi$, $T_0=0.9L$, $T_1=0.15L$, $\alpha=0.01$. Each data point is averaged over 100 times. (a) Averaged total energy still grows with time exponentially (b) The energy density distribution is the same as that without any randomness, except that the peaks are smeared out a little.
	}
	\label{fig:e heating random}
\end{figure}

However, if we start from the heating phase regime, the total energy keeps growing exponentially with time and the energy peaks will not disappear for moderate randomness, as shown in \figref{fig:e heating random}. We can first drive the system with a fixed period and let the energy peaks form. Then we turn on sufficiently weak randomness. Now the energy peaks will not be moved back perfectly but with a small discrepancy. As a result, the energy peaks will be smeared a little bit but still there. Therefore, just as the time crystal with MBL~\cite{normal2017timecrystal, lukin2017exp, zhang2017observation}, all the features for the heating phase we find here are also robust even though we slightly perturb it away from the fine-tuning (randomness-free) point.

Before we close this section, we want to emphasize that in our 
CFT calculation of the energy and entanglement evolution in the presence
of randomness (see \figref{fig:ee growth}), 
the average is performed numerically.
It is desirable to derive an analytic result, for example, for the heating
rate $\kappa$ in \figref{fig:Heating rate}. We leave this problem for  
future study.

\section{Generalization to other subalgebra}
\label{sec:generalization}

Appealing to the quasi-particle picture, our setup has a natural generalization, i.e. replacing the SSD with other arbitrary envelop functions $\int_0^L dx f(x) T_{00}(x)$. 
In this section, we consider a specific one, where $f(x)$ only involves a single Fourier component
\begin{equation}
    H_q = \int_0^L dx \left( 1 - \cos q\frac{2\pi x}{L} \right) T_{00}(x) = 2\int_0^L dx \sin^2 \left(q \frac{\pi x}{L} \right) T_{00}(x)\,,\quad q=2,3,4,\cdots\,.
\end{equation}
\emph{Periodic} boundary condition will be used in this section for the sake of simplicity but most phenomena shown below are qualitatively the same for the open boundary condition.\footnote{Just to remind that the reason we chose open boundary condition for the $q=1$ case is that in the periodic boundary condition, the ground state of $H_0$ will be annihilated by $H_1$ as well due to the $\SL(2)$ symmetry.}
Before detailing the results, let us first explain some intuition for this generalization from the algebraic viewpoint and the quasi-particle picture.

To understand it from the algebraic viewpoint, we rewrite $H_q$ in terms of the Virasoro generators
\begin{equation}
    H_q = \frac{2\pi}{L}\left( L_0 - \frac{L_{-q} + L_{q}}{2} + \bar{L}_0 - \frac{\bar{L}_{-q}+\bar{L}_{q}}{2} \right)\,,
\end{equation}
in which only $L_{0,\pm q}$ (and $\bar{L}_{0,\pm q}$) appear. They again form a $\fraksl(2,\RR)$ subalgebra
\footnote{Although the algebra is isomorphic, we may emphasis the different group action by denoting the subgroup as $\SL^{(m)}(2,\RR)$, which represents an $m$-fold 
cover of $\SL(2,\RR)$.}
and therefore follow the same classification scheme we have discussed before, i.e. non-heating, heating phase and the critical line. 
%to other subalgebras. Noticing that $[L_q,L_0]=qL_q$, $[L_q,L_{-q}]=2qL_0 + \frac{c}{12}(q^3-q)$, this subalgebra is actually isomorphic to $\fraksl(2,\RR)$ by a redefinition and thus should have the same classification. 
%It follows immediately that the dynamics also exhibits the non-heating, heating phase and the transition between them. 
%\XW{[Or we may follow Witten's notation here. That is, the Virasoro generators $L_0$, $L_n$ and $L_{-n}$ form the $SL^{(n)}(2,R)$ group, which is isomorphic to an $n$-fold cover of $SL(2,R)$.]}

From the quasi-particle picture, the SSD Hamiltonian $H_1$ (with periodic boundary condition) introduces one zero point at the identified edge for the spatial profile of the velocity $v(x)$, while $H_q$ puts $q$ zeros and arranges them with a equal spacing $L/q$. Let us denote the region $\frac{(m-1)L}{q} < x < \frac{mL}{q}$ as $R_m$, $m=1,2,\cdots, q$. For each $R_m$, the system can be treated as if being governed by $H_0$ and $H_1$. It implies that there will be $2$ energy peaks in each $R_m$, and the 
only difference is that the quasiparticles can move to nearby interval $R_{m\pm 1}$ after one cycle.

Next, we elaborate the details of the above intuitions with focus on the energy and entanglement patterns in the heating phase.
%The above intuition will be confirmed by more detailed analysis in the following sections with focus on the energy and entanglement patterns in the heating phase.

\subsection{Operator evolution}

In this section, we will derive the formula for the operator evolution by working in the Euclidean time and performing the analytical continuation at the end.
The whole procedure is similar to what have been shown in Sec.\ref{sec: mobius}.

To calculation the operator evolution after a single-cycle driving, we need a conformal mapping from the cylinder to a more convenient geometry. The above observation about each region $R_m$ leads to the following conformal transformation
\begin{equation}
    z = e^{q \frac{2\pi w}{L} } =e^{\frac{2\pi w}{l} } \,,\quad w=\tau+ix\,,
\end{equation}
where $l=L/q$ is the length of each region $R_n$. For a fixed $\tau$, $z$ will wind the origin $q$ times as $x$ increases from $0$ to $L$, which implies that $z$ describes a $q$-sheet Riemann surface with the $q$-fold branch cut being $[0,+\infty)$. Let us introduce $H_{0(q)}[R_m]$ as the part of the total Hamiltonian $H_{0(q)}$ supported on the region $R_m$, then its expression in the $z$ coordinate is
\begin{equation}
\label{eqn:generlized H on z plane}
\begin{aligned}
    H_0[R_m] =& \frac{2\pi}{l} \int_{C,n} \frac{dz}{2\pi i} z T(z) - (z\rightarrow \bar z) - \frac{c\pi}{6l}\\
    H_q[R_m] =& \frac{2\pi}{l} \int_{C,n} \frac{dz}{2\pi i} \left(-\frac{1}{2} + z - \frac{z^2}{2} \right) T(z) - (z\rightarrow \bar z) - \frac{c\pi}{6l}
\end{aligned}    
\end{equation}
which is locally the same as \eqnref{eqn:H on z plane} except that the total system size $L$ is replaced with the subregion size $l$ and $m$ is introduced as the Riemann sheet label. As a result, the operator evolution on this $q$-sheet Riemann surface is also described by an $\SL(2,\RR)$ transformation
\begin{equation}
    z_1 = \frac{a z+b}{c z+d}\,,\quad
    \begin{pmatrix} a & b \\ c & d \end{pmatrix} \in \SL(2,\RR)
\end{equation}
with the dimsionless coefficents being
\begin{equation}
\begin{aligned}
    & a = \left(1+\frac{\pi\tau_1}{l} \right) e^{\frac{\pi\tau_0}{l}}\,,\quad 
    b = -\frac{\pi\tau_1}{l} e^{-\frac{\pi\tau_0}{l}}\,, \\
    & c = \frac{\pi\tau_1}{l} e^{\frac{\pi\tau_0}{l}}\,,\quad\quad \qquad\,
    d = \left(1-\frac{\pi\tau_1}{l} \right) e^{-\frac{\pi\tau_0}{l}}\,.
\end{aligned}
\end{equation}
It has the same form as that for the simplest case derived in Sec.\ref{sec: mobius} with $L$ replaced by $l$.
The formula for multiple repeated cycles is the composition of the above transformation and will be denoted by the same equation \eqnref{eqn: zn formula} with $l$ used in the definition of parameters. 

Therefore, the operator evolution in this generalized protocol has the same classification as the previously discussed case ($q=1$), and the phase diagram of the dynamics is identical to \figref{fig: cft phase diagram} as long as the total system size $L$ is replaced with the subregion size $l$. 
On the other hand, the introduction of $q$ Riemann sheets will enrich the spatial structure of the operator evolution, e.g. the fixed points on one sheet will be duplicated to all the sheets and therefore the entanglement pattern will be enriched as we will see shortly. 

%it duplicates one copy of the fixed points on each of the Riemann sheets. Therefore, we have to keep track of which sheet the coordinate actually stays on, which will play an important role when we compute the expectation value of operators.

\subsection{Energy density}

%In this section, we will explicitly calculate the energy density and use the quasi-particle picture to explain their motion, which also helps us to have a qualitative understanding of their entanglement pattern.

\begin{figure}
    \centering
    \subfloat[The evolution of energy density profile for this generalized protocol.]{\includegraphics[width=7cm]{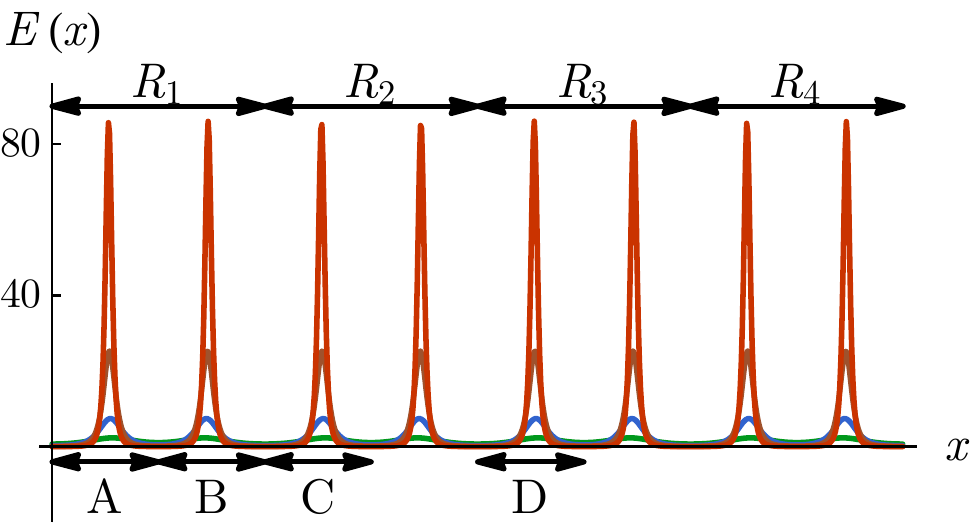}}
    \hspace{30pt}
    \subfloat[The evolution of mutual information between different peaks.]{\includegraphics[width=6cm]{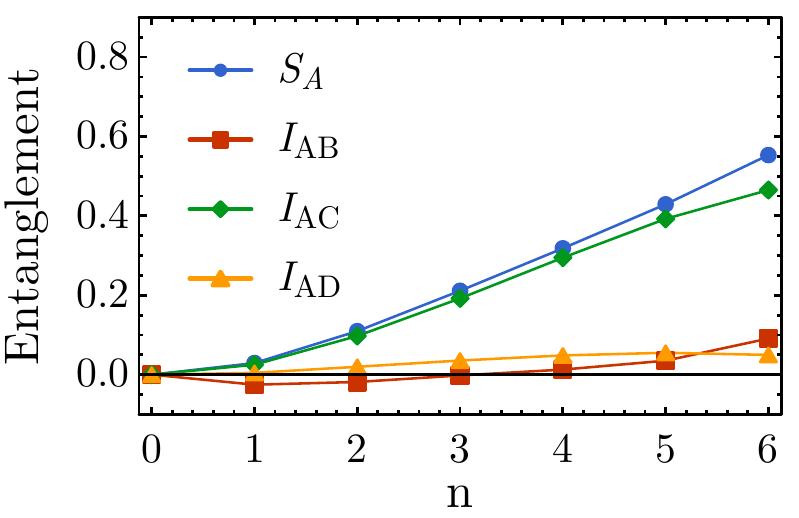}}
    \caption{(a) CFT calculation for the energy density. The system size is $L=2\pi$ and the central charge $c=1$. We choose $q=4$, and $T_0=0.92L/q$, $T_1=0.1L/q$ so that the system is in the heating phase. (b) Lattice calculation for the mutual information. We simulate complex free fermion on a $502$ site chain, $q=4$, $T_0=0.92L/q$, $T_1=0.1L/q$. The choice of subsystem $A,C,D$ is consistent with (a). All the quantities have been subtracted by their initial values respectively. The data for $n>6$ is spoiled by the lattice effect.}
    \label{fig:energy density H4}
\end{figure}

According to the above discussion, the time-evolved stress tensor is
\begin{equation}
    F^{-n} T(w) F^n = \left( \frac{\partial z}{\partial w} \right)^2 \left( \frac{\partial z_n}{\partial z} \right)^2 T(z_n) + \frac{c}{12}\text{Sch}(z,w)\,.
\end{equation}
Evaluated on the ground state of $H_0$, we obatin
%The conformal transformation $\xi=z^{1/q}$,\YG{[I didn't understand this because $\xi$ didn't appear in the formula here, do you want to say $z=e^{2\pi w/l}$ with $l=L/q$?]} which unfolds the Riemann surface to a normal complex plane yields the following expectation value, 
\begin{equation}\label{T_n_hol}
\langle T(x,t=nT)\rangle=
\frac{\pi^2 c}{6\,L^2}\cdot
(q^2-1)\cdot\frac{(AD-BC)^2z^2}
{(A z+B)^2(C z+D)^2}-\frac{q^2 \pi^2 c}{6\, L^2},\quad L=q\,l,
\end{equation}
%\begin{equation}
%    \braket{T}(x,t=nT) = \left(\frac{2\pi}{l} \right)^2 \frac{c}{24} \frac{q^2-1}{q^2} \frac{(AD-BC)^2 z^2}{(Az+B)^2(Cz+D)^2}\,,
%\end{equation}
Here $A,B,C,D$ also follows the prescription in Sec.\ref{sec: mobius} with $L$ replaced by $l$.
For $\langle \overline{T}(x,t=nT)\rangle$, one simply replaces $z$ with $\bar{z}$
in Eq.\eqref{T_n_hol}. The total energy $E(t=nT)=\int_0^L\frac{dx}{2\pi}(T+\bar{T})$ grows as
\begin{equation}
    E(t=nT)=-\frac{q^2 \pi c}{6L}
    +\frac{\pi c}{6L}\cdot(q^2-1)\cdot\frac{AD+BC}{AD-BC}.
\end{equation}
Several remarks are followed:
\begin{enumerate}
\item
For $q=1$, one can find that 
$\langle T(x,t=nT)\rangle=\langle \overline{T}(x,t=nT)\rangle=
-\frac{\pi^2 c}{6\, l^2}=-\frac{q^2\pi^2 c}{6\, L^2}$, which only contains the Casimir energy of the ground state. 
%In particular, 
%the energy density is independent of time. 
This is because for 
$q=1$, the ground states of $H_0$ and $H_{q=1}$ are the same, and therefore there is no nontrivial time evolution as mentioned in footnote 9. 
%This is also the reason why we consider a CFT with open boundary condition for the case $q=1$ in the 
%previous sections.
\item
For $q>1$, the feature of energy growth in each region $R_m$ is the same as those as discussed in Sec.\ref{sec:features}.\footnote{The prefactor differs by $\frac{4}{3}(q^2-1)$, where $\frac{4}{3}$ is due to the shift of the boundary condition from open to periodic in this section.} 
When the system is in the heating phase, we will observe two energy peaks in each of the $q$ regions, one from $T$ and the other from $\bar{T}$. An example with $q=4$ is shown in \figref{fig:energy density H4}, one can find 8 peaks in total.
\item
For $q>1$ in the heating phase, one can check that the energy density
away from the peaks will approach $-\frac{q^2\pi c}{6L^2}$ 
exponentially in time. It becomes `cooler' than the initial 
Casimir energy density $-\frac{\pi c}{6L^2}$.
This can be viewed as a dynamical Casimir effect\cite{Law1994,Dodonov1996,martin2019floquet}.
\end{enumerate}
%As an example, we consider $q=2$, and plot the energy density evolution in the heating phase in Fig.xx
%\textcolor{red}{[add a plot]}. The parameters we use
%are the same as that in Fig.\ref{fig:energydensity}, with 
%$T_0=0.9\,l$ and $T_1=0.1\,l$.
%]

\subsection{Entanglement pattern}

A more interesting question is how different energy peaks are entangled in the heating phase. Since we choose the boundary condition to be periodic and the initial state the ground state of $H_0$, the state remains a tensor product of the chiral and anti-chiral components. It immediately follows that the entanglement entropy between energy peaks with different chirality does not grow. The entanglement among peaks with the same chirality requires more detailed analysis.

We first study the entanglement entropy of a single interval $[x_1,x_2]$, which is related to the correlation function of two twist operators $\langle \psi(t)| \mathcal{T}_m(x_1) \mathcal{T}_m(x_2) |\psi(t) \rangle$. Given the recipe above, it can be mapped to the following two point function on the complex plane
\begin{equation}
    \left(\frac{\partial z_{1,n}}{\partial z_1} \right)^{h_m}
    \left(\frac{\partial \bar{z}_{1,n}}{\partial \bar{z}_1} \right)^{h_m}
    \left(\frac{\partial z_{2,n}}{\partial z_2} \right)^{h_m}
    \left(\frac{\partial \bar{z}_{2,n}}{\partial \bar{z}_2} \right)^{h_m} \frac{1}{(z_{1,n}^{1/q}-z_{2,n}^{1/q})^{2h_m}(\bar{z}_{1,n}^{1/q}-\bar{z}_{2,n}^{1/q})^{2h_m}}
\end{equation}
where we only keep the time dependent parts and $z_{j,n},\bar{z}_{j,n}$ denote the coordinates on the $q$-sheet Riemann surface after $n$-cycle driving. The result will depend on whether there are chiral/anti-chiral energy peaks between the $x_1$ and $x_2$. If there are no energy peaks between $x_1$ and $x_2$, then $z_{1,n}$ and $z_{2,n}$ will flow to the stable fixed point on the same sheet such that $z_{1,n}^{1/q}-z_{2,n}^{1/q}$ becomes exponentially small with time, which exactly cancels the time dependence from the $\partial z_{j,n}/\partial z_j$ prefactor. On the other hand, if there is a chiral energy peak between $x_1$ and $x_2$, $z_{1,n}$ and $z_{2,n}$ will go to different Riemann sheet such that $z_{1,n}^{1/q}-z_{2,n}^{1/q}$ becomes an $\calO(1)$ number at late time and the whole quantity has non-trivial time dependence. Similar argument works for $\bar{z}_{1,n}, \bar{z}_{2,n}$. Consequently, the entanglement entropy for a single region $[x_1,x_2]$ has a similar behavior as what has been shown in \eqnref{eqn:SA heating}
\begin{equation}
    S(t=nT) - S(0) = \left\{ \begin{array}{ll} \vspace{5pt}
        \calO(1) & [x_1,x_2]\, \text{does not include peaks} \\ \vspace{5pt}
        -\dfrac{c}{6} n \log \eta & [x_1,x_2]\,\text{includes one peak} \\
        -\dfrac{c}{3} n \log \eta & [x_1,x_2]\,\text{includes both chiral and anti-chiral peaks}
    \end{array}
    \right..
\end{equation}

To determine the structure of the entanglement, such as whether it has bipartite entanglement or multi-partite entanglement, we need to examine the mutual information between different peaks. 

For example, let us consider the mutual information between $A$ and $C$, as depicted in \figref{fig:energy density H4}(a), which covers two nearest neighbor chiral peaks (ignoring the anti-chiral ones). The entanglement entropy $S_{AC}$ is related to the correlation function of four twist operators
\begin{equation}
    \prod_{j=1}^{4} \left( \frac{\partial z_{j,n}}{\partial z_{j}} \right)^{h_m} 
    %\left( \frac{\partial \bar{z}_{j,n}}{\partial \bar{z}_{j}} \right)^{h_m} 
    \frac{1}{(z_{1,n}^{1/q}-z_{4,n}^{1/q})^{2h_m} (z_{2,n}^{1/q}-z_{3,n}^{1/q})^{2h_m}} F(\rho)\,,\quad \rho = \frac{(z_{1,n}^{1/q}-z_{4,n}^{1/q})(z_{2,n}^{1/q}-z_{3,n}^{1/q})}{(z_{1,n}^{1/q}-z_{2,n}^{1/q})(z_{4,n}^{1/q}-z_{3,n}^{1/q})}
\end{equation}
where the anti-holomorphic component is irrelevant to our discussion and thus ignored in the expression, $\rho$ is the cross ratio and $F(\rho)$ is the conformal block. In the long time limit, $z_{2,n}$ and $z_{3,n}$ flow to the same fixed point so that $z_{2,n}^{1/q}-z_{3,n}^{1/q}$ as well as $\rho$ becomes exponentially small while $z_{1,n}^{1/q}-z_{4,n}^{1/q}$ remains finite. This implies that $S_{AC}$ linearly grows with time as $-\frac{c}{6}n \log \eta$, so does the mutual information
\begin{equation}
    I_{AC}(t=nT)-I_{AC}(0) = -\frac{c}{6} n \log \eta\,.
\end{equation}
On the contrary, if we consider the mutual information between $A$ and $D$, as depicted in \figref{fig:energy density H4}(a), same analysis yields $S_{AD}(t=nT)-S_{AD}(0)=-\frac{c}{3}n\log \eta$ such that the mutual information $I_{AD}$ does not grow at all.

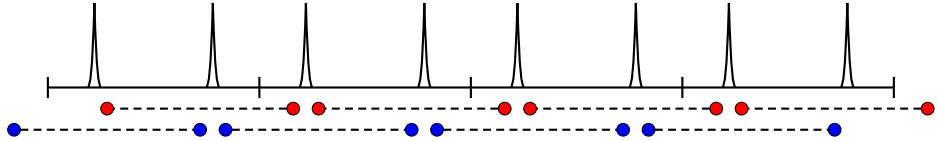
\begin{figure}[t]
\center
    \begin{tikzpicture}[scale=0.8, baseline={([yshift=-6pt]current bounding box.center)}]
    \draw [thick](-31pt,0pt)..controls (-30pt,2pt) and (-29pt,5pt)..(-28pt,40pt)..controls (-27pt,5pt) and (-26pt,2pt)..(-25pt,0pt);
    \draw [thick](31pt,0pt)..controls (30pt,2pt) and (29pt,5pt)..(28pt,40pt)..controls (27pt,5pt) and (26pt,2pt)..(25pt,0pt);
    \draw [thick] (-50pt,0pt)--(50pt,0pt);
    \draw [thick] (-50pt,-5pt) -- (-50pt,5pt);
    \draw[densely dashed, thick] (-22pt,-10pt) -- (66pt,-10pt);
    \filldraw[fill=red] (-22pt,-10pt) circle (3pt);
    \filldraw[fill=red] (66pt,-10pt) circle (3pt);
    \draw[densely dashed, thick] (22pt,-20pt) -- (-66pt,-20pt);
    \filldraw[fill=blue] (22pt,-20pt) circle (3pt);
    \filldraw[fill=blue] (-66pt,-20pt) circle (3pt);
    
     \draw [xshift=100pt,thick](-31pt,0pt)..controls (-30pt,2pt) and (-29pt,5pt)..(-28pt,40pt)..controls (-27pt,5pt) and (-26pt,2pt)..(-25pt,0pt);
    \draw [xshift=100pt,thick](31pt,0pt)..controls (30pt,2pt) and (29pt,5pt)..(28pt,40pt)..controls (27pt,5pt) and (26pt,2pt)..(25pt,0pt);
    \draw [xshift=100pt,thick] (-50pt,0pt)--(50pt,0pt);
    \draw [xshift=100pt,thick] (-50pt,-5pt) -- (-50pt,5pt);
    \draw[xshift=100pt,densely dashed, thick] (-22pt,-10pt) -- (66pt,-10pt);
    \filldraw[xshift=100pt,fill=red] (-22pt,-10pt) circle (3pt);
    \filldraw[xshift=100pt,fill=red] (66pt,-10pt) circle (3pt);
    \draw[xshift=100pt,densely dashed, thick] (22pt,-20pt) -- (-66pt,-20pt);
    \filldraw[xshift=100pt,fill=blue] (22pt,-20pt) circle (3pt);
    \filldraw[xshift=100pt,fill=blue] (-66pt,-20pt) circle (3pt);
    
    \draw [xshift=200pt,thick](-31pt,0pt)..controls (-30pt,2pt) and (-29pt,5pt)..(-28pt,40pt)..controls (-27pt,5pt) and (-26pt,2pt)..(-25pt,0pt);
    \draw [xshift=200pt,thick](31pt,0pt)..controls (30pt,2pt) and (29pt,5pt)..(28pt,40pt)..controls (27pt,5pt) and (26pt,2pt)..(25pt,0pt);
    \draw [xshift=200pt,thick] (-50pt,0pt)--(50pt,0pt);
    \draw [xshift=200pt,thick] (-50pt,-5pt) -- (-50pt,5pt);
    \draw[xshift=200pt,densely dashed, thick] (-22pt,-10pt) -- (66pt,-10pt);
    \filldraw[xshift=200pt,fill=red] (-22pt,-10pt) circle (3pt);
    \filldraw[xshift=200pt,fill=red] (66pt,-10pt) circle (3pt);
    \draw[xshift=200pt,densely dashed, thick] (22pt,-20pt) -- (-66pt,-20pt);
    \filldraw[xshift=200pt,fill=blue] (22pt,-20pt) circle (3pt);
    \filldraw[xshift=200pt,fill=blue] (-66pt,-20pt) circle (3pt);
    
    \draw [xshift=300pt,thick](-31pt,0pt)..controls (-30pt,2pt) and (-29pt,5pt)..(-28pt,40pt)..controls (-27pt,5pt) and (-26pt,2pt)..(-25pt,0pt);
    \draw [xshift=300pt,thick](31pt,0pt)..controls (30pt,2pt) and (29pt,5pt)..(28pt,40pt)..controls (27pt,5pt) and (26pt,2pt)..(25pt,0pt);
    \draw [xshift=300pt,thick] (-50pt,0pt)--(50pt,0pt);
    \draw [xshift=300pt,thick] (-50pt,-5pt) -- (-50pt,5pt);
    \draw[xshift=300pt,densely dashed, thick] (-22pt,-10pt) -- (66pt,-10pt);
    \filldraw[xshift=300pt,fill=red] (-22pt,-10pt) circle (3pt);
    \filldraw[xshift=300pt,fill=red] (66pt,-10pt) circle (3pt);
    \draw[xshift=300pt,densely dashed, thick] (22pt,-20pt) -- (-66pt,-20pt);
    \filldraw[xshift=300pt,fill=blue] (22pt,-20pt) circle (3pt);
    \filldraw[xshift=300pt,fill=blue] (-66pt,-20pt) circle (3pt);
    
     \draw [thick] (350pt,-5pt) -- (350pt,5pt);
       \end{tikzpicture}
    \caption{A cartoon of the entanglement pattern for $q=4$ with periodic boundary condition. Red and blue color stand for two different chiralities. Each peak is entangled with its nearest neighbour with the same chirality/color.}
    \label{fig:entanglement pattern generalized}
    
\end{figure}

Therefore, the system only develops bipartite entanglement, see Fig.~\ref{fig:entanglement pattern generalized} for an illustration of the pattern. Every two nearest neighbor and only nearest neighbor peaks of the same chirality share Bell pairs with each other. We want to point out that \figref{fig:entanglement pattern generalized} is a stroboscopic picture, all the energy peaks as well as Bell pairs keep moving towards the left/right in each cycle. If we choose $(k-1)l<T_0<kl$, each energy peak can move from one subregion to the $k$-th subregion on its left/right and only comes back to its original position after every $q/gcd(q,k)$ cycles of driving. This is a generalization to the peak-switching phenomena first discussed in Sec.~\ref{Sec: Quasi-particle}. We also simulate free fermion on the lattice. The results are shown in \figref{fig:energy density H4}(b), which supports our CFT argument. The deviation comes from the lattice effect.

%\textcolor{blue}{To polish:}

%Now let us give some physical picture before the calculation:

%To calculate the entanglement entropy evolution, we will study the two point correlation function of twist operators
%$\langle \psi(t)| \mathcal{T}(x_1) \mathcal{T}(x_2) |\psi(t) \rangle=\langle G|F^{-n} \mathcal{T}(x_1) F^n\cdot F^{-n} \mathcal{T}(x_2) F^n|G\rangle$.
%We will go to the complex $z$-plane, and calculate $\langle G|F^{-n} \mathcal{T}(z_1, \bar{z}_1) F^n\cdot F^{-n} \mathcal{T}(z_2, \bar{z}_2) F^n|G\rangle$.
%The result will depend on whether there are chiral/anti-chiral energy peaks between the $x_1$ and $x_2$.
%If there are no energy peaks between $x_1$ and $x_2$, then $z_1$ and $z_2$ will flow to the same fixed point, $\bar{z}_1$ and $\bar{z}_2$ will flow to another fixed points.
%On the other hand, if there are chiral/anti-chiral energy peaks between $x_1$ and $x_2$, at least one of the pairs $z_1$ and $z_2$ ($\bar{z}_1$ and $\bar{z}_2$) will flow
%to \textit{different} fixed points. This will lead to linear growth of entanglement entropy for the interval $[x_1,\, x_2]$.

\section{Summary}
\label{sec:conclusions}

In this paper, we presented a detailed study as well as a generalization of the Floquet CFT introduced in \cite{wen2018floquet}. The phase diagram obtained in that paper can be understood by mapping the problem to a Floquet harmonic oscillator. The reason for such a mapping arises from the fact that these two problems share a $\fraksl(2,\RR)$ algebra and the classification of dynamics becomes the classification of the linear combination of $\SL(2,\RR)$ generators. Because of that, although we use the entanglement entropy and total energy to explicitly determine the phase diagram, the calculation of which depends on the choice of the initial state, the result is actually a property of the driving Hamiltonian and doesn't depend on the initial state choice.

In the non-heating phase, the energy profile and total energy keep oscillating. In the heating phase, although the energy increases exponentially fast, the system is heated in an extremely non-uniform way, i.e. only two points absorb the heat. What is more, all the entanglement entropy is also shared by these two peaks. These peaks are determined by the fixed points of the relevant M\"obius transformation that is defined by the dynamics. On the phase boundary between the heating and non-heating phases, we still observe two peaks but the total energy only increases quadratically with time.

Although the question of whether the system absorbs energy relies on the detailed calculation, the energy density and entanglement structure can be understood by a quasi-particle picture. The questions of how energy distributes can be mapped to solving a pure classical motion. Inspired by this picture, we find a relation between the total energy and the entanglement between the two peaks, $E(t) \propto c \exp\left(\frac{6}{c}S(t)\right)$, which says the quasi-particle carries much more energy than entanglement. Such a relation, as contrary to the classic Cardy formula, is a clear manifestation of a non-equilibrium state. It will be interesting to understand whether this relation is special to this set-up that only involves $\SL(2,\RR)$ or is true for more general cases.

To make some connection to the real experiment, we examine the robustness of all these features against random driving. Even if we add tiny randomness to the driving period, the non-heating phase completely disappears and we only have the heating phase, where the total energy grows exponentially with time. After we know whether the total energy grows, the energy density can be analyzed perturbatively. If we start from a $(T_0,T_1)$ that is deep inside the non-heating phase and turn on the randomness, the energy density will peak near the boundary. This is because  the quasi-particles move incoherently and have smaller velocity near the boundary. On the other hand if we start from a $(T_0,T_1)$ that is deep inside the heating phase and turn on moderate randomness, we expect the energy peaks will remain although they are smeared out a little.

Most of the phenomena above, in particular including the existence of heating and non-heating phase and the features about the energy profile, not only hold for this special set-up but should also occur for any generic Floquet driving that only uses Virasoro generators as the Hamiltonian. The reason is that this type of Floquet driving can always be thought of as a conformal mapping on the complex plane. The energy density calculation to a large extent can be reduced to the problem of finding fixed points of the conformal mapping. If none of the fixed points is on the unit circle, the system must be at the non-heating phase and energy density just oscillates. Once it has a repulsive fixed point on the unit circle, we will see two energy peaks, one of which is purely chiral and the other is anti-chiral. The system is generically heated up. If there is also an attractive fixed point on the unit circle, the energy density will decrease to make the energy peaks sharper and sharper. For this case, one has to do a more detailed calculation to determine whether the system is heating or not. Consequently, the problem of classifying dynamics is equivalent to the problem of classifying conformal mappings. This Floquet CFT using sine-square deformed Hamiltonian is the first and simplest example that explicitly realizes this. It will be interesting to generalize this special set-up to more general protocols and give a more thorough discussion on the connection between dynamics and geometry. Furthermore, since generic many-body Floquet drives do not have this geometric interpretation, it is also important to consider driving protocols that go beyond the conformal transformation paradigm, which can help develop a more general understanding of Floquet dynamics.

Another interesting problem is to consider the Floquet CFT from a thermal initial state at finite temperature $\beta^{-1}$, which is closely related with experiments.
Since there are now three length scales, \textit{i.e.}, the
total length $L$ of the system, the driving periods $T$, and the 
finite temperature $\beta$, then the time evolution of entanglement and energy density may exhibit more rich features in particular in the early time of driving. In the long time limit, we expect 
there are still two phases, \textit{i.e.}, the heating and non-heating phases. One intuition is based on the quasi-particle 
picture as presented in Sec.\ref{Sec: Quasi-particle}. 
One can find that the existence of fixed point or not in the solution of equation of motion, which determines
the system is in heating or non-heating phases, is independent of the introduction of finite temperature $\beta^{-1}$. 
We expect these two phases will persist even if the system is prepared at a thermal initial state. We leave the detailed study in a future work. 

The heating phase discussed here realizes a highly nonequlibrium state where entangled EPR pairs are continuously produced and localized at specific locations. Given the utility of entanglement as a resource for quantum information processing, experimental realization of the protocols discussed here may be desirable. Indeed given the high tunability of ultracold atomic systems in optical lattices~\cite{bloch2005ultracold}, and the ability to measure both energy and entanglement entropy~\cite{islam2015measuring}, an important future direction will be to find routes to implement these protocols in the lab.

\section*{Acknowledgments}

We thank Liujun Zou, Shang Liu, Andrew Potter, Xie Chen, Adam Nahum, Meng Cheng, Jie-Qiang Wu, Shinsei Ryu, Tsukasa Tada, and Ivar Martin for helpful discussions. Y.G.\ is supported by the Gordon and Betty Moore Foundation EPiQS Initiative through Grant (GBMF-4306) and DOE grant, DE-SC0019030. X.W. is supported by the Gordon and Betty Moore Foundation’s EPiQS initiative through Grant (GBMF-4303) at MIT. AV and RF are supported by the DARPA DRINQS program  (award D18AC00033) and by a Simons Investigator Award. 

\appendix

\section{Phase diagram of the Floquet CFT}
\label{app:phase diagram cft}
\begin{figure}
    \centering
    \includegraphics[width=0.3\textwidth]{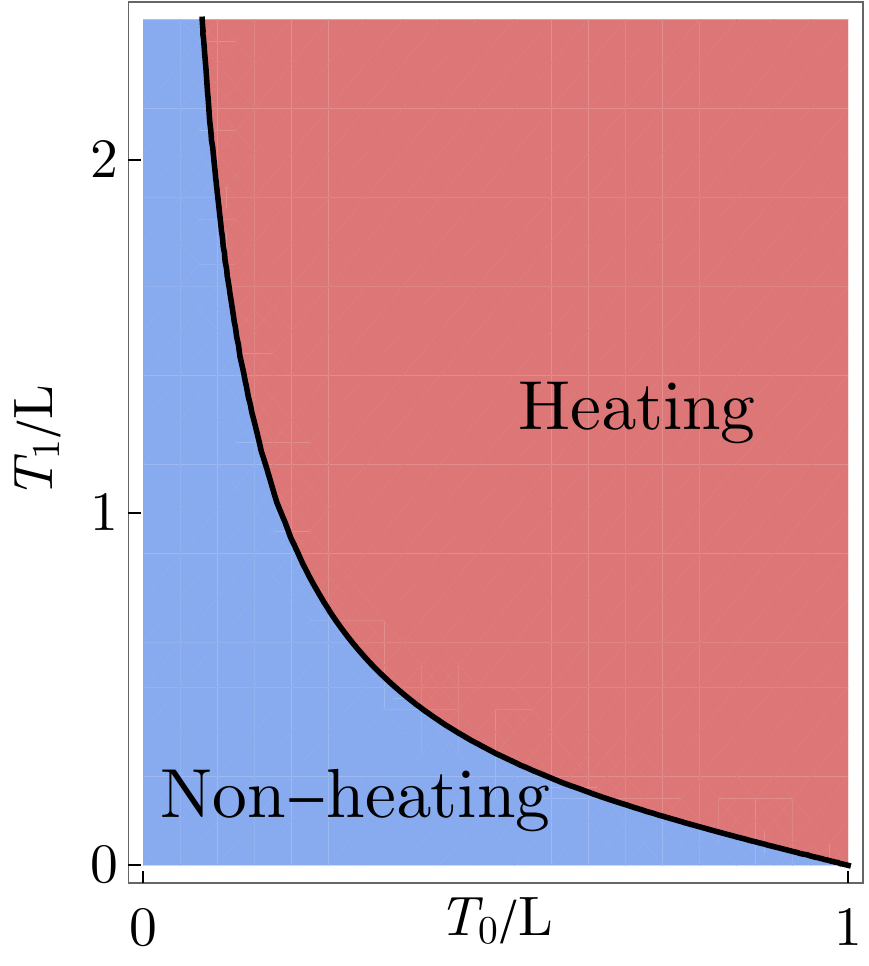}
    \caption{Phase diagram of the Floquet CFT defined in Sec.~\ref{sec:review}. The red regime is the heating phase and the blue regime is the non-heating phase.}
    \label{fig: cft phase diagram}
\end{figure}

The Floquet CFT defined in Sec.~\ref{sec:review} was known to have two different phases, which was first shown in \cite{wen2018floquet}. Here, we reproduce the phase diagram in \figref{fig: cft phase diagram} for the sake of being self-content. The high frequency regime is a non-heating phase, where the entanglement entropy and energy oscillate. The low frequency regime is a heating phase, where the entanglement entropy linearly grows and the energy exponentially grows. On the phase boundary, the entanglement entropy grows logarithmically and the energy grows quadratically. \figref{fig: cft phase diagram} only shows one domain, and the phase diagram repeats itself when we increase $T_0/L$ with a period $1$.

\section{Entanglement growth for a subsystem and branch cut crossing}
\label{app:entanglement}
In this section, we will present the details of the entanglement entropy growth calculation for a subsystem in the heating phase. As the early-time regime contains non-universal information, our analytical analysis will focus on the late-time regime (i.e. $n\gg1$). %As we will show below, the branch cut plays an important role in the discussions.

\subsection{Single entanglement cut}

For a given state on the interval $[0,L]$, we denote the (left) subsystem by $A=[0,x]$ and the corresponding reduced density matrix by $\rho_A$. Following Calabrese and Cardy's prescription \cite{Calabrese:2004eu}, the $m$-th R\'enyi entropy
\begin{equation}
	S_A^{(m)} = \frac{1}{1-m} \log \Tr \rho_A^m \,
\end{equation}
can be computed using the twist operator $\calT_m$, and the von Neumann entropy is the $m \rightarrow 1$ limit. More explicitly, the twist operator $ \calT_m$ is a primary with conformal dimension $h_m=\bar h_m=\frac{c}{24}(m-\frac{1}{m})$, whose one point function reproduces $\Tr \rho_A^m$
\begin{equation}
	\label{eqn: Trace rhoN}
	\Tr \rho_A^m = 
	\left(\frac{\partial z}{\partial w} \right)^{h_m} 
	\left(\frac{\partial \bar z}{\partial \bar w} \right)^{h_m} 
	\left(\frac{\partial z_n}{\partial z} \right)^{h_m} 
	\left(\frac{\partial \bar{z}_n}{\partial \bar z} \right)^{h_m}
	\braket{ \calT_m(z_n,\bar{z}_n)}.
\end{equation}
The one point correlation function $\braket{\calT_m(z_n,\bar{z}_n)}$ in a boundary CFT can be mapped to a two point function through the mirror trick on the whole plane, i.e.
\begin{equation}
	\label{eqn: Twist Single}
	\braket{ \calT_m(z_n,\bar{z}_n)} \propto
	\left( \frac{1}{4\sqrt{z_n} \sqrt{\bar{z}_n}}
	\right)^{h_m}
	\left( \frac{1}{\sqrt{z}_n - \sqrt{\bar{z}_n}}
	\right)^{2h_m} \,.
\end{equation}
%\begin{equation}
%	\label{eqn: Twist Single}
%	\braket{ \calT_m(z_n,\bar{z}_n)} = \mathcal{A}^b
%	\left( \frac{1}{4\sqrt{z_n} \sqrt{\bar{z}_n}}
%	\right)^{h_m}
%	\left( \frac{2\epsilon i}{\sqrt{z}_n - \sqrt{\bar{z}_n}}
%	\right)^{2h_m},
%\end{equation}
%where $\mathcal{A}^b$ is a constant which depends on the conformal boundary condition and $\epsilon\rightarrow 0$ is a short-distance cut-off. Since we are mainly interested in the time evolution of entanglement entropy, these constant pieces can be safely neglected for our purpose. 
Note the derivative term in \eqnref{eqn: Trace rhoN} decreases exponentially as a function of $n$ in the long time limit
\footnote{The intuitive reason is that in the heating phase, both $z_n$ and $\bar{z}_n$ will flow to the attractive fixed point $\gamma_1$ as an exponential function of $n$. Here we assume that neither $z_n$ nor $\bar{z}_n$ collides with the repulsive fixed point.}
\begin{equation}
	\left( \frac{\partial z_{n}}{\partial z} \right)^{h_m}
	\left( \frac{\partial \bar z_{n}}{\partial \bar z}  \right)^{h_m} \approx
	\left( \frac{\eta^{n}(\gamma_1-\gamma_2)^2}
	{(z-\gamma_2)(\bar z - \gamma_2)} \right)^{2h_m}\quad \text{at} \quad n\gg 1 \,,
\end{equation} 
which is related to the linear growth of entanglement entropy. While the behavior of $\braket{\calT_m(z_n, \bar z_n)}$ depends on an intersting branch cut structure that will lead to the spatial feature (the kink) of the entanglement entropy plotted in \figref{fig:entanglement pattern}. 

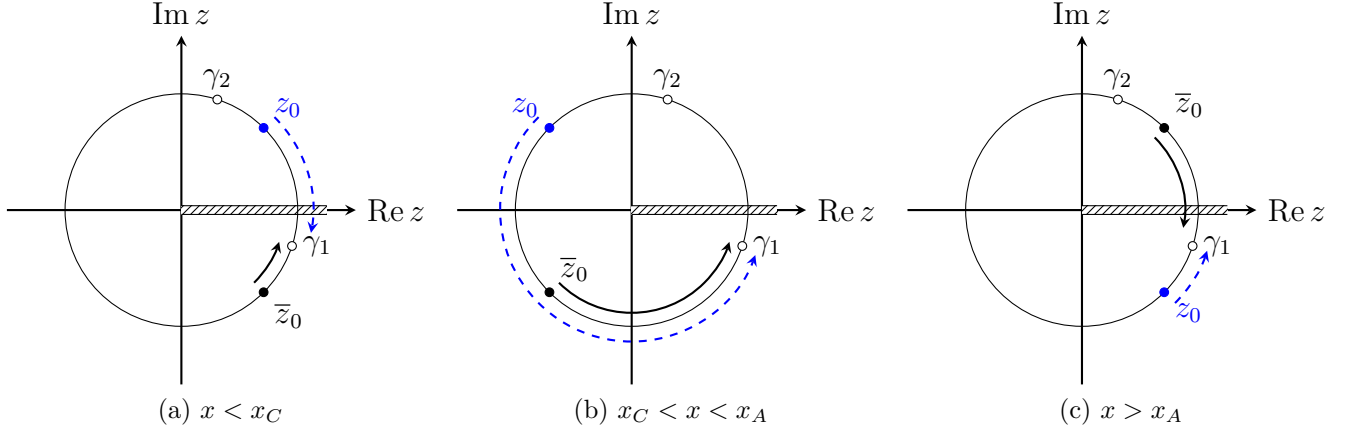
\begin{figure}[t]
    \centering
    \subfloat[$x<x_C$]{
    \begin{tikzpicture}[scale=1.1]
    \draw[thick, ->,>=stealth] (-60pt,0pt) -- (60pt,0pt) node[right]{ $\Re z$};
    \draw[thick, ->,>=stealth] (0pt,-60pt) -- (0pt,60pt) node[above]{ $\Im z$};
    \fill[fill=white] (0pt,-1.5pt) rectangle (50pt,1.5pt);
    \draw (0pt,1.5pt) -- (50pt,1.5pt);
    \draw (0pt,-1.5pt) -- (50pt,-1.5pt);
     \fill[pattern= south west lines] (0pt,-1.5pt) rectangle (50pt,1.5pt);
    \draw (0pt,0pt) circle (40pt);
    \filldraw[fill=white] (12.36pt , 38.04pt) circle (1.5pt) node[above]{ $\gamma_2$};
    \filldraw[fill=white] (38.04pt, -12.36pt) circle (1.5pt) node[right]{ $\gamma_1$};
    \filldraw[fill=blue,blue] (28.28pt, 28.28pt) circle (1.5pt) node[above right]{ $z_0$};
    \filldraw (28.28pt, -28.28pt) circle (1.5pt) node[below right]{ $\bar{z}_0$};
    \draw[dashed, blue, thick,->,>=stealth] (32pt,32pt) arc (45: -10: 45.25pt);
    \draw[thick,  ->,>=stealth] (25pt,-25pt) arc (-45: -18: 35.35pt);
    \end{tikzpicture}
    }
    \subfloat[$x_C<x<x_A$]{
    \begin{tikzpicture}[scale=1.1]
    \draw[thick, ->,>=stealth] (-60pt,0pt) -- (60pt,0pt) node[right]{ $\Re z$};
    \draw[thick, ->,>=stealth] (0pt,-60pt) -- (0pt,60pt) node[above]{ $\Im z$};
    \fill[fill=white] (0pt,-1.5pt) rectangle (50pt,1.5pt);
    \draw (0pt,1.5pt) -- (50pt,1.5pt);
    \draw (0pt,-1.5pt) -- (50pt,-1.5pt);
     \fill[pattern= south west lines] (0pt,-1.5pt) rectangle (50pt,1.5pt);
    \draw (0pt,0pt) circle (40pt);
    \filldraw[fill=white] (12.36pt , 38.04pt) circle (1.5pt) node[above]{ $\gamma_2$};
    \filldraw[fill=white] (38.04pt, -12.36pt) circle (1.5pt) node[right]{ $\gamma_1$};
    \filldraw[blue, fill=blue] (-28.28pt, 28.28pt) circle (1.5pt) node[above left]{ $z_0$};
    \filldraw (-28.28pt, -28.28pt) circle (1.5pt) node[above right]{ $\bar{z}_0$};
    \draw[dashed, blue, thick,->,>=stealth] (-32pt,32pt) arc (135: 340: 45.25pt);
    \draw[thick,  ->,>=stealth] (-25pt,-25pt) arc (-135: -18: 35.35pt);
    \end{tikzpicture}
    }
    \subfloat[$x>x_A$]{
    \begin{tikzpicture}[scale=1.1]
    \draw[thick, ->,>=stealth] (-60pt,0pt) -- (60pt,0pt) node[right]{ $\Re z$};
    \draw[thick, ->,>=stealth] (0pt,-60pt) -- (0pt,60pt) node[above]{ $\Im z$};
    \fill[fill=white] (0pt,-1.5pt) rectangle (50pt,1.5pt);
    \draw (0pt,1.5pt) -- (50pt,1.5pt);
    \draw (0pt,-1.5pt) -- (50pt,-1.5pt);
     \fill[pattern= south west lines] (0pt,-1.5pt) rectangle (50pt,1.5pt);
    \draw (0pt,0pt) circle (40pt);
    \filldraw[fill=white] (12.36pt , 38.04pt) circle (1.5pt) node[above]{ $\gamma_2$};
    \filldraw[fill=white] (38.04pt, -12.36pt) circle (1.5pt) node[right]{ $\gamma_1$};
    \filldraw (28.28pt, 28.28pt) circle (1.5pt) node[above right]{ $\bar{z}_0$};
    \filldraw[blue,fill=blue] (28.28pt, -28.28pt) circle (1.5pt) node[below right]{ $z_0$};
    \draw[thick,->,>=stealth] (25pt,25pt) arc (45: -10: 35.35pt);
    \draw[thick, dashed, blue, ->,>=stealth] (32pt,-32pt) arc (-45: -18: 45.25pt);
    \end{tikzpicture}
    }
    \caption{Schematic plots for the stroboscopic trajectories of $z_n$ and $\bar{z}_n$ on unit circle. %The gray circle is the unit circle on the $z$-plane. 
    Hollow dots are the fixed points, from which we know the chiral peak is on the left of the anti-chiral peak, i.e. $x_C<x_A$. Blue and black dots are the initial positions of $z_n$ and $\bar{z}_n$ respectively. In each cycle, both $z_n$ and $\bar{z}_n$ in each cycle will wind around the unit circle counterclockwise and stop at the next position, with their stroboscopic positions represented by blue dashed line and black line respectively. The dashed bar denotes the branch cut. (a) $x<x_C$. The blue line crosses the branch cut meaning that $z_n$ has a relative branch cut crossing to $\bar{z}_n$ therefore they stay on the same Riemann sheet in the long time limit. (b) $x_C<x<x_A$. neither line crosses the branch cut thus $z_n$ and $\bar{z}_n$ remain on different sheets. (c) $x>x_A$. This time it is the black line crosses the branch cut thus $z_n$ and $\bar{z}_n$ still end up getting on the same Riemann sheet.}
    \label{fig:trajectory}
\end{figure}

More explicitly, the branch cut arises from 
%For the evaluation of  $\braket{\calT_m(z_n,\bar{z}_n)}$, we have to pay special attention to 
the factor $\left(\sqrt{z}_n - \sqrt{\bar{z}_n}\right)$. The subtlety is that although both $z_{n}$ and $\bar z_{n}$ flow to the same attractive fixed point $\gamma_1$ at long time limit, their square roots can be different due to the branch cut, i.e. the sign structure arises from the square root. %This is because the $z$-plane has a branch cut at $[0,+\infty)$. 
%Therefore $z$ and $\bar z$ are actually defined on a two-sheet Riemann surface. 
To analyze the branch cut, let us use the two-layer Riemann sheet for $z$ and $\bar{z}$.
At $t=0$, $z$ sits on the first sheet while $\bar z$ sits on the second sheet.\footnote{
	This is consistent with the convention that at the imaginary time, when going to the UHP geometry, $\xi$ has to sit on the upper half plane while $\bar \xi$ has to sit on the lower half plane.
} 
Under the time evolution, we need to trace the trajectories of $z_n$ and $\bar{z}_n$, see \figref{fig:trajectory} for an illustration of the trajectories for different scenarios. The upshot is that only when the entanglement cut is between the two energy peaks, %do $z_n$ and $\bar{z}_n$ still remain on different Riemann sheets in the long time limit.
the factor $\left(\sqrt{z}_n - \sqrt{\bar{z}_n}\right)$ remains finite and therefore leads to the ``bump'' in the middle of \figref{fig:entanglement pattern}.

%As we turn on the time evolution, we have to trace their trajectories to determine which Riemann sheet they finally stay at, depending on which, $\sqrt{z}_n - \sqrt{\bar{z}_n}$ can be either exponentially small or exponentially close to $\pm 2\sqrt{\gamma_1}$. Therefore to determine the behavior of this quantity, we have to know whether the $z$ coordinates cross the branch cut or not.

To discuss this in more details, without loss of generality, we assume the chiral peak is on the left of the anti-chiral peak. Depending on the position of the entanglement cut, there are three different scenarios:
\begin{enumerate}
	\item $x<x_C$. In this case, $z_n$ will effectively cross the branch cut. Therefore, when we take the square root of $z$ and $\bar z$, we have
	\begin{equation}
		\sqrt{z_{n}} =  -\gamma_1^{1/2}\left(1 + \eta^n \frac{\gamma_1 - \gamma_2}{2\gamma_1}\frac{z - \gamma_1}{z - \gamma_2} \right)\,, \quad
		\sqrt{\bar{z}_{n}} = -\gamma_1^{1/2}\left(1 + \eta^n \frac{\gamma_1 - \gamma_2}{2\gamma_1}\frac{\bar z - \gamma_1}{\bar z - \gamma_2} \right) \,,
	\end{equation}
	so that their difference is exponentially small
	\begin{equation}
		\sqrt{z_{n}} - \sqrt{\bar{z}_{n}} = -\eta^n
		\frac{\gamma_1 - \gamma_2}{2\gamma_1^{1/2}}
		\frac{(z-\bar{z})(\gamma_1-\gamma_2)}{(z-\gamma_2)(\bar{z}-\gamma_2)}.
	\end{equation}
	This $\eta^n$ dependence will exactly cancel the $\eta^n$ dependence in the derivative term. Therefore the whole quantity and the entanglement entropy, to the leading order, does not grow with time.
	\item $x_C<x<x_A$. In this case, $z_n$ and $\bar z_n$ do not  cross the branch cut. When we calculate the square root, we have
	\begin{equation}
		\sqrt{z_{n}} =  \gamma_1^{1/2}\left(1 + \eta^n \frac{\gamma_1 - \gamma_2}{2\gamma_1}\frac{z - \gamma_1}{z - \gamma_2} \right), \quad
		\sqrt{\bar{z}_{n}} = -\gamma_1^{1/2}\left(1 + \eta^n \frac{\gamma_1 - \gamma_2}{2\gamma_1}\frac{\bar z - \gamma_1}{\bar z - \gamma_2} \right)
	\end{equation}
	and their difference converges to $2\gamma_1^{1/2}$ at the late time,
	\begin{equation}
		\sqrt{z_{n}} - \sqrt{\bar{z}_{n}} = 2\gamma_1^{1/2} + \calO(\eta^n)\,.
	\end{equation}
	Therefore, the whole quantity will depend on time through the $\eta^n$ in the derivative term. After taking the logarithm and $m\rightarrow 1$ limit, we can show that the entanglement entropy grows linearly with time 
	\begin{equation}
		S_A(t) = -\frac{c}{6} n \log \eta,
	\end{equation}
	the slope of which is independent of $x$. 
	\item $x>x_A$. In this case, $\bar z_n$ will cross the branch cut and all the calculation becomes the same as the first case. Therefore the entanglement doesn't grow with time, either.
\end{enumerate}
This explains the kinks that we observe in \figref{fig:entanglement pattern}.

\subsection{Entanglement entropy between two halves}

%There is one special case, where we can easily write down the expression for the entanglement entropy for the whole time regime. This is when the entanglement cut is made at the middle of the system and we calculate the entanglement entropy for the left/right half system.

For the special case $x=L/2$, we have $z=\bar z = e^{2\pi x/L}=-1$ and $z_n=\bar{z}_n$ for any integer $n$. 
%It is instructive to look at one special case, where the entanglement cut is made at the middle of the system and it is easy to write down an expression of the entanglement entropy for the whole time regime. $z=\bar z = e^{2\pi x/L}=-1$ implies $z_n=\bar{z}_n$ for any integer $n$. 
As a result, $z_n$ and $\bar{z}_n$ are always on the opposite Riemann surfaces, i.e. $\sqrt{z}_n = - \sqrt{\bar{z}_n}$. Hence the expression for $\Tr \rho_A^m$ can be simplified as
\begin{equation}
	\Tr \rho_A^m \propto 
	\left(\frac{\pi}{L} \right)^{2h_m}
	\left(\frac{\partial z_n}{\partial z} \right)^{2h_m}
	\left(\frac{-1}{z_n}\right)^{h_m}
	\left( \frac{1}{{z}_n}
	\right)^{h_m}
	\propto
	\left( \frac{\pi}{L}
	\frac{\partial z_n}{\partial z}
	\frac{1}{z_n}
	\right)^{2h_m},
\end{equation}
%\begin{equation}
%	\Tr \rho_A^m \propto 
%	\left(\frac{\pi}{L} \right)^{2h_m}
%	\left(\frac{\partial z_n}{\partial z} \right)^{2h_m}
%	\left(\frac{-1}{z_n}\right)^{h_m}
%	\left( \frac{\epsilon^2}{{z}_n}
%	\right)^{h_m}
%	\propto
%	\left( \frac{\pi}{L}
%	\frac{\partial z_n}{\partial z}
%	\frac{1}{z_n}
%	\right)^{2h_m},
%\end{equation}
where all the non-universal constants have been dropped. Recalling our expression \eqnref{eqn: zn formula} for $z_n$ and plugging in the initial condition that $z=-1$, we can write the universal part of the entanglement entropy as,
\begin{equation}
	S_A(t) = \frac{c}{6}
	\log\left[ \frac{L}{\pi}
	\frac{(A-B)(C-D)}{AD-BC}
	\right] + \text{(non-universal term)} \,,
\end{equation}
where the non-universal term refers to the $n$-indepedent contributions. 
In the non-heating phase, the universal part oscillates in a similar fashion as the total energy. In the heating phase, 
the leading growing behavior of the entanglement entropy is given as follows
%we can do a Taylor expansion with respect to $\eta^n$, which yields a linear growth behavior,
\begin{equation}
	S_{A,\text{heating}}(t) \sim  \frac{c}{6}
	\log \left[ \frac{L}{\pi}
	\frac{(1+\gamma_2)^2 \gamma_1}{(\gamma_1-\gamma_2)^2}\,
	\eta^{-n}
	\right] \quad \text{at $n\gg 1$.}
\end{equation}
%At the phase boundary, we can do an expansion with respect to $1/n$, which leads to a logarithmic growth behavior,
For the critical phase, we have the following logarithmic growing, 
\begin{equation}
	S_{A,\text{critical}}(t) \sim \frac{c}{6}
	\log \left[ \frac{L}{\pi}
	(1+\gamma)^2 \beta^2\gamma\, n^2
	\right]\quad \text{at $n\gg 1$.}
\end{equation}

\subsection{Two entanglement cuts}
In this section, we present the details of computing the entanglement entropy for a subsystem that does not end at the boundary, i.e. with ending points $x_1,x_2 \in (0,L)$. In other words, we need to insert two twist operators% to get the reduced density matrix
\begin{equation}
	\Tr \rho_A^m(t)= C_m(x_1,x_2,t)=
	\langle \psi(t) | \calT_m(x_1)  \calT_m(x_2)| \psi(t) \rangle = 
	\braket{G| \calT_m(x_1,t)  \calT_m(x_2,t)|G}.
\end{equation}
We follow the strategy in Sec.~\ref{sec:review} to do the calculation first in the imaginary time and analytic continue to real time in the end. The $n$-dependent part of the above formula is given as follows,
\begin{equation}
	\prod_{j=1,2}
	\left( \frac{\partial z_{j,n}}{\partial z_j} \right)^{h_m}
	\left( \frac{\partial \bar z_{j,n}}{\partial \bar z_j}  \right)^{h_m}
	\braket{ \calT_m(z_{1,n},\bar z_{1,n}) \calT_m(z_{2,n}, \bar z_{2,n})}\,.
\end{equation}
The mirror trick maps the two-point function $\braket{ \calT_m(z_{1,n},\bar z_{1,n}) \calT_m(z_{2,n}, \bar z_{2,n})}$ in a boundary CFT to a four-point function on the whole plane without boundary. The important $n$-dependent part is given by the following formula, 
\begin{equation}
	\label{eqn:twist 2pt uhp}
	\braket{ \calT_m(z_{1,n},\bar z_{1,n}) \calT_m(z_{2,n}, \bar z_{2,n})} \propto
	\frac{1}{\left( \sqrt{z_{1,n}} - \sqrt{z_{2,n}} \right)^{2h_m} \left( \sqrt{\bar{z}_{1,n}} - \sqrt{\bar{z}_{2,n}} \right)^{2h_m}}
	F(\rho)\,,
\end{equation}
where 
$\rho$ is the cross ratio of the four $\sqrt{z_n}$'s defined as follows, 
\begin{equation}
	\rho = \frac{(\sqrt{z_{1,n}} - \sqrt{z_{2,n}}) (\sqrt{\bar{z}_{1,n}} - \sqrt{\bar{z}_{2,n}})}{(\sqrt{z_{1,n}} - \sqrt{\bar{z}_{1,n}}) (\sqrt{z_{2,n}} - \sqrt{\bar{z}_{2,n}})}\,.
\end{equation}
$F(\rho)$ is a linear combination of the chiral conformal blocks with coefficients determined by the boundary condition. 

While analytically continuing to the real time, the derivative term shows an exponential decease as a function of $n$ (similar to the single entanglement cut case), 
%Now we perform the analytic continuation to the real time. As discussed in the single entanglement cut case, the derivative term in the long time limit exponentially deceases, 
\begin{equation}
	\label{eqn:twist derivatives}
	\prod_{j=1,2}
	\left( \frac{\partial z_{jn}}{\partial z_j} \right)^{h_m}
	\left( \frac{\partial \bar z_{jn}}{\partial \bar z_j}  \right)^{h_m} \approx
	\left( \frac{\eta^{2n}(\gamma_1-\gamma_2)^4}
	{(z_1-\gamma_2)(\bar z_1 - \gamma_2)(z_2 - \gamma_2)(\bar z_2 - \gamma_2)} \right)^{2h_m}\,.
\end{equation}
Note the exponential decrease of the correlation function is related to the linearly growth of entanglement entropy. 

For the analysis of the behavior of $\braket{ \calT_m(z_{1,n},\bar z_{1,n}) \calT_m(z_{2,n}, \bar z_{2,n})}$, there are two complications: first is the branch cut issue due to the $\sqrt{z}$ factor as we have discussed before; the second is the potential divergence caused by $F(\rho)$. In the following, we will show that $\rho$ will flow to a final value $\rho_{\text{final}}\neq 1$ which, for different choice of $x_1$ and $x_2$, is either $0$ or a constant finite value so that $F(\rho)$ converges to a constant non-zero value at late time and can be neglected.
 
Without loss of generality, let us assume the chiral peak is on the left of the anti-chiral peak. We fix $x_2$ to be between the two peaks so that $z_{2,n}$ and $\bar{z}_{2,n}$ stay on different Riemann sheets.
\begin{enumerate}
	\item $x_1<x_C$, the subsystem $A$ includes the chiral peak. As discussed in \figref{fig:trajectory}, only $z_{1,n}$ crosses the branch cut during the time evolution. Therefore, at the late time, the leading term of four $\sqrt{z}$ read,
	\begin{equation}
		\begin{aligned}
			\sqrt{z_{1,n}} =  -\gamma_1^{1/2}\left(1 + \eta^n \frac{\gamma_1 - \gamma_2}{2\gamma_1}\frac{z_1 - \gamma_1}{z_1 - \gamma_2} \right)\,, &\quad
			\sqrt{\bar{z}_{1,n}} = -\gamma_1^{1/2}\left(1 + \eta^n \frac{\gamma_1 - \gamma_2}{2\gamma_1}\frac{\bar z_1 - \gamma_1}{\bar z_1 - \gamma_2} \right)\,, \\
			\sqrt{z_{2,n}} =  \gamma_1^{1/2}\left(1 + \eta^n \frac{\gamma_1 - \gamma_2}{2\gamma_1}\frac{z_2 - \gamma_1}{z_2 - \gamma_2} \right)\,, &\quad
			\sqrt{\bar{z}_{2,n}} = -\gamma_1^{1/2}\left(1 + \eta^n \frac{\gamma_1 - \gamma_2}{2\gamma_1}\frac{\bar z_2 - \gamma_1}{\bar z_2 - \gamma_2} \right)\,.
		\end{aligned}
	\end{equation}
	These show that $\sqrt{z_{1,n}}-\sqrt{z_{2,n}}$ and $\sqrt{z_{2,n}}-\sqrt{\bar{z}_{2,n}}$ converge to $\pm 2\sqrt{\gamma_1}$ while $\sqrt{\bar{z}_{1,n}} - \sqrt{\bar{z}_{2,n}}$ and $\sqrt{z_{1,n}}-\sqrt{\bar{z}_{1,n}}$ become exponentially small. The cross ratio $\rho$ converges to a $\calO(1)$ value $\rho_{\text{final}}$
	\begin{align}
		\rho_{\text{final}} = \frac{(\bar{z}_1 - \bar{z}_2)(z_1 - \gamma_2)}{(z_1 - \bar{z}_1)(\bar{z}_2 - \gamma_2)}\,.
	\end{align} 
	The condition that neither $x_1$ or $x_2$ is at the energy peaks implies $\rho_{\text{final}} \neq 1$. Therefore, the conformal block term $F(\rho)$ only converges to an $\calO(1)$ value and does not contribute to the $n$ dependence. As a result, the late time behavior is controlled by the derivative terms which leads to the linear growth behavior of the entanglement entropy,
	\begin{equation}
		S_A(x_1,x_2,t) = \lim_{m\rightarrow 1} \frac{1}{1-m} \log C_m(x_1,x_2,t) \sim 
		-\frac{c}{6}n\log \eta.
	\end{equation}
	The slope only depends on the central charge and the characteristic constant $\eta$ but not on the positions of entanglement cuts, as long as $x_1<x_C$. 
	
	\item $x_C<x_1,x_2<x_A$, the subsystem is between the chiral and anti-chiral peak. None of the four $z$ coordinates have any relative branch cut crossing and they can be assumed to remain on their original Riemann sheets during the whole time evolution. Hence, the late time values of their square roots are,
	\begin{equation}
	\begin{aligned}
		\sqrt{z_{1,n}} = & \gamma_1^{1/2}\left(1 + \eta^n \frac{\gamma_1 - \gamma_2}{2\gamma_1}\frac{z_1 - \gamma_1}{z_1 - \gamma_2} \right) \,, \quad
		\sqrt{\bar{z}_{1,n}} = -\gamma_1^{1/2}\left(1 + \eta^n \frac{\gamma_1 - \gamma_2}{2\gamma_1}\frac{\bar z_1 - \gamma_1}{\bar z_1 - \gamma_2} \right), \\
		\sqrt{z_{2,n}} = & \gamma_1^{1/2}\left(1 + \eta^M \frac{\gamma_1 - \gamma_2}{2\gamma_1}\frac{z_2 - \gamma_1}{z_2 - \gamma_2} \right) \,, \quad
		\sqrt{\bar{z}_{2,n}} = -\gamma_1^{1/2}\left(1 + \eta^M \frac{\gamma_1 - \gamma_2}{2\gamma_1}\frac{\bar z_2 - \gamma_1}{\bar z_2 - \gamma_2} \right).
	\end{aligned}
	\end{equation}
	As a result, $\sqrt{z_{1,n}}-\sqrt{z_{2,n}}$, $\sqrt{\bar{z}_{1,n}}-\sqrt{\bar{z}_{2,n}}$ become exponentially small so that the prefactor in \eqnref{eqn:twist 2pt uhp} will cancel the time dependence in the derivative term. $\sqrt{z_{1,n}}-\sqrt{\bar{z}_{1,n}}$, $\sqrt{z_{2,n}}-\sqrt{\bar{z}_{2,n}}$ converge to $2\sqrt{\gamma_1}$ so that the cross ratio $\rho$ now will converge to $\rho=0$. However, in the way that we write \eqnref{eqn:twist 2pt uhp}, the conformal block term is already regularized at $\rho=0$ and takes an $\calO(1)$ value depending on the fusion from two twist operators to the identity channel. In this limit, the boundary two-point function should be reduced to a bulk two-point function, which is nonzero in our case. This implies $F(\rho=0)$ is a nonzero number and thus does not carry important time dependence.
	As a result, the late-time behavior of the entanglement entropy, to the leading order, is independent of time.
	\item $x_1>x_A$, the subsystem includes the anti-chiral peak. Now the $\bar z_1$ will cross the branch cut and the entanglement entropy linear grows again, which is the same as the first case.
\end{enumerate}

The analysis above confirms that the entanglement indeed only comes from the two energy peaks, which verifies our quasi-particle picture from a technical side. 

\section{Random driving Mathieu oscillator}
\label{app:random mathieu}

\begin{figure}[t]
	\centering
	\subfloat[Regular Mathieu oscillator]{\includegraphics[width=5cm]{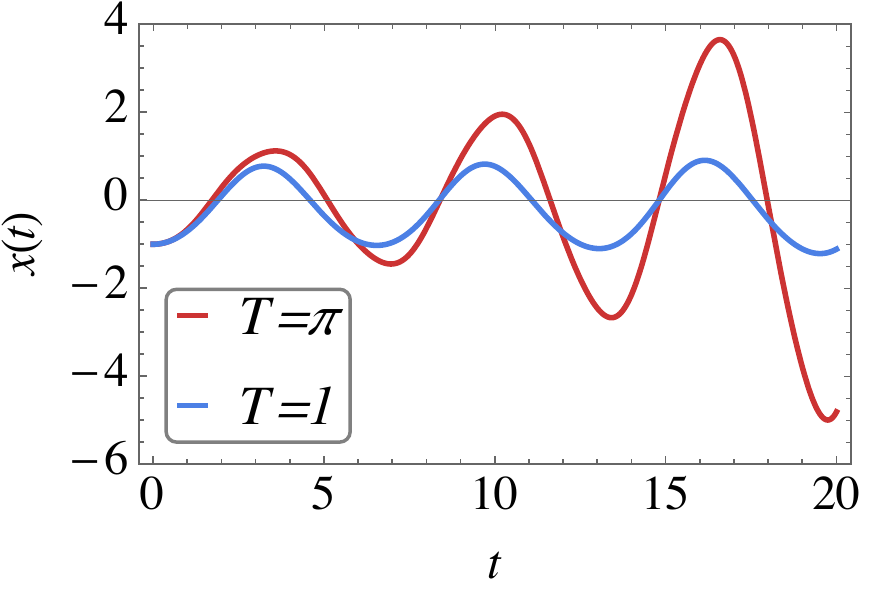}}
	\hspace{20pt}
	\subfloat[Random driving Mathieu oscillator]{\includegraphics[width=5cm]{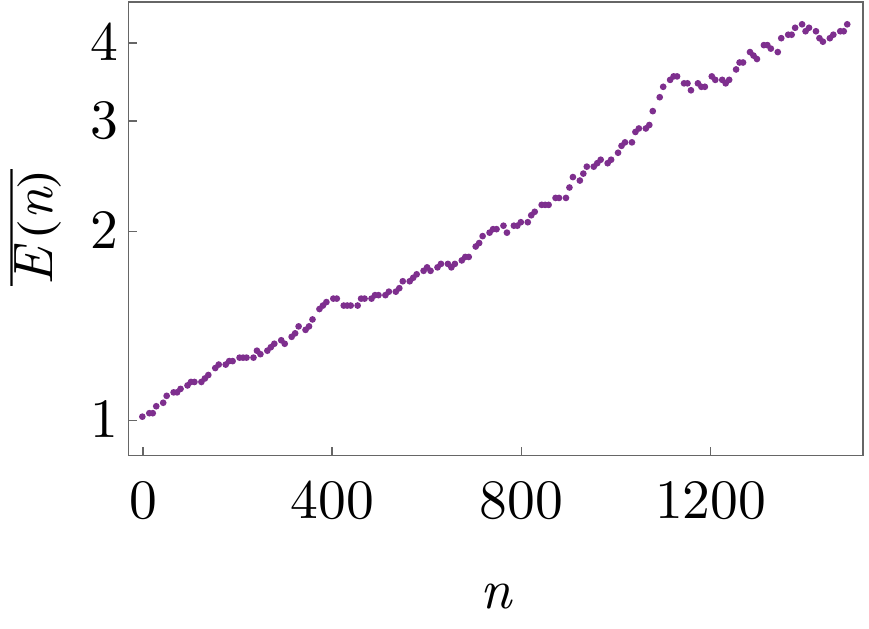}}
	\caption{Dynamics of a Mathieu oscillator without and with random driving. (a) Mathieu oscillator under weak driving force is stable for $T<1$ and unstable for $T=\pi$. We choose $h=0.2$. (b) For random driving, the energy will grow exponentially even for small randomness. We choose We choose $h=0.2$, $\bar{T}=1$, $\alpha=0.1$, Each data point is averaged over 100 times.
	}
	\label{fig:random mathieu}
\end{figure}

In this section, we discuss the random driving Mathieu oscillator. The classical Newton's equation for a Mathieu oscillator is
\begin{equation}
    \ddot{x}(t) + \left(1+h \cos \frac{2\pi t}{T}\right) x(t) = 0.
\end{equation}
$h$ controls the amplitude of the driving force and $T$ is driving period. The intrinsic period of the harmonic oscillator is $2\pi$. For a weak driving force $h\ll 1$, the first smallest unstable driving period is $T_{\text{unstable}}=\pi$. The system is stable(non-heating) for any $T<T_{\text{unstable}}$. These are shown in \figref{fig:random mathieu}~(a).

For a random driving Mathieu oscillator, we let the driving period $T$ uniformly distribute in an interval
\begin{equation}
    T = \bar{T} + \delta T,\quad
    \delta T = [-\alpha, \alpha],
\end{equation}
where $\alpha$ controls the strength of the randomness. In each cycle, we randomly choose a $T$ from the distribution and evolve the system accordingly. The final results will be averaged over ``disorder". If we choose $\bar{T}<T_{\text{unstable}}$ and $\alpha\ll 1$, we find that the averaged energy grows exponentially with time, as is demonstrated in \figref{fig:random mathieu}~(b).

\section{Spatial structures in lattice calculation}
\label{app:spatial structure lattice}

\begin{figure}
    \centering
    \subfloat[Energy density at early time]{
    \includegraphics[width=5cm]{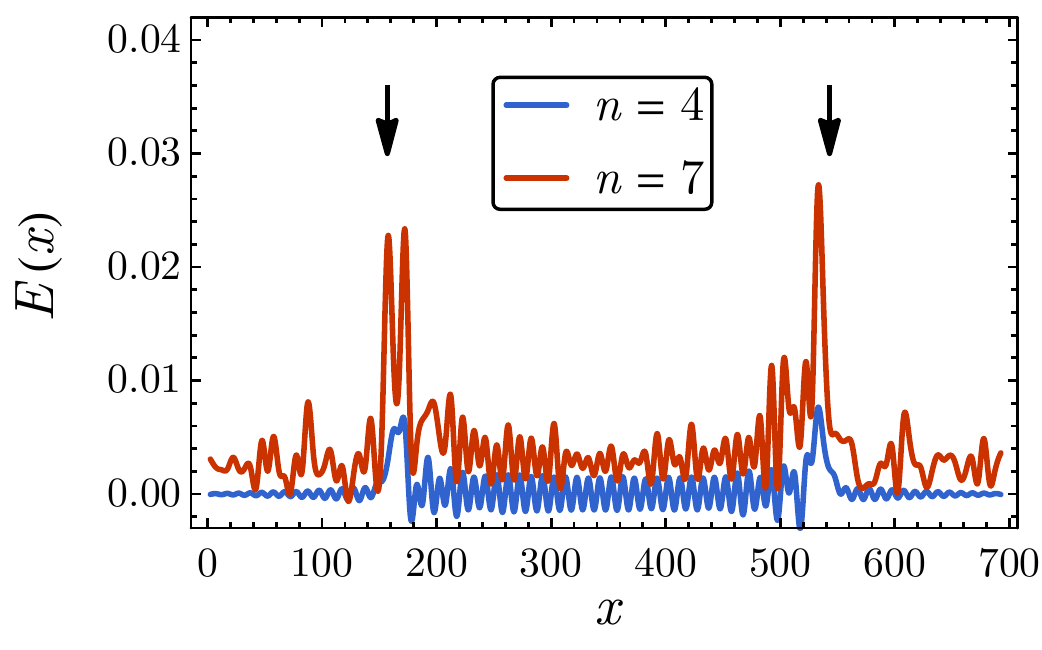}
    }
    \subfloat[Entropy density at early time]{
    \includegraphics[width=5cm]{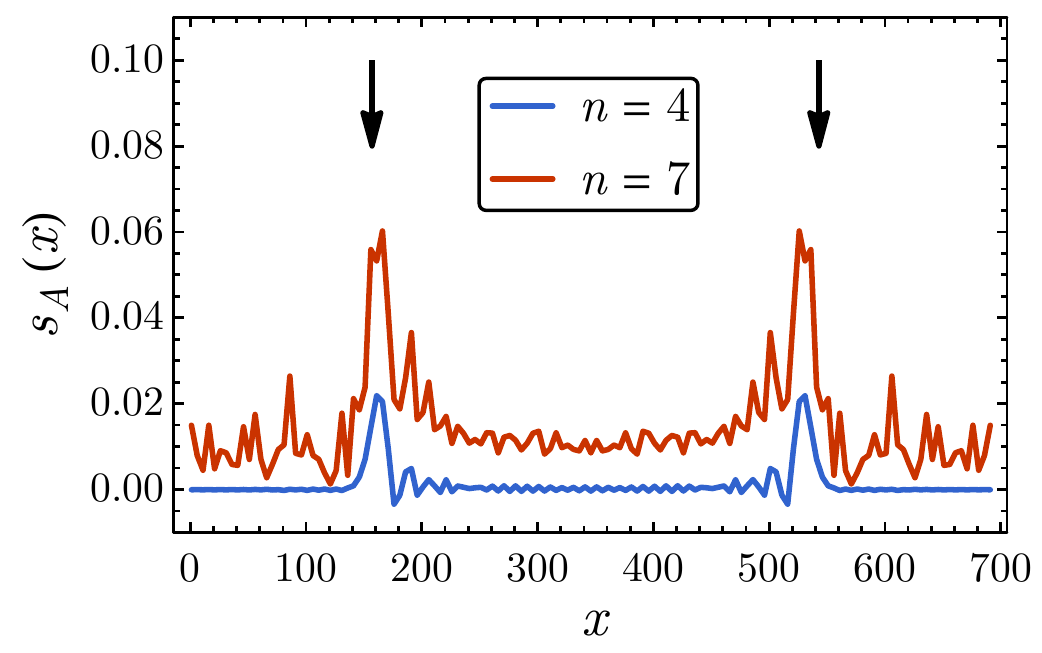}
    } \\
    \subfloat[Energy density at late time]{
    \includegraphics[width=5cm]{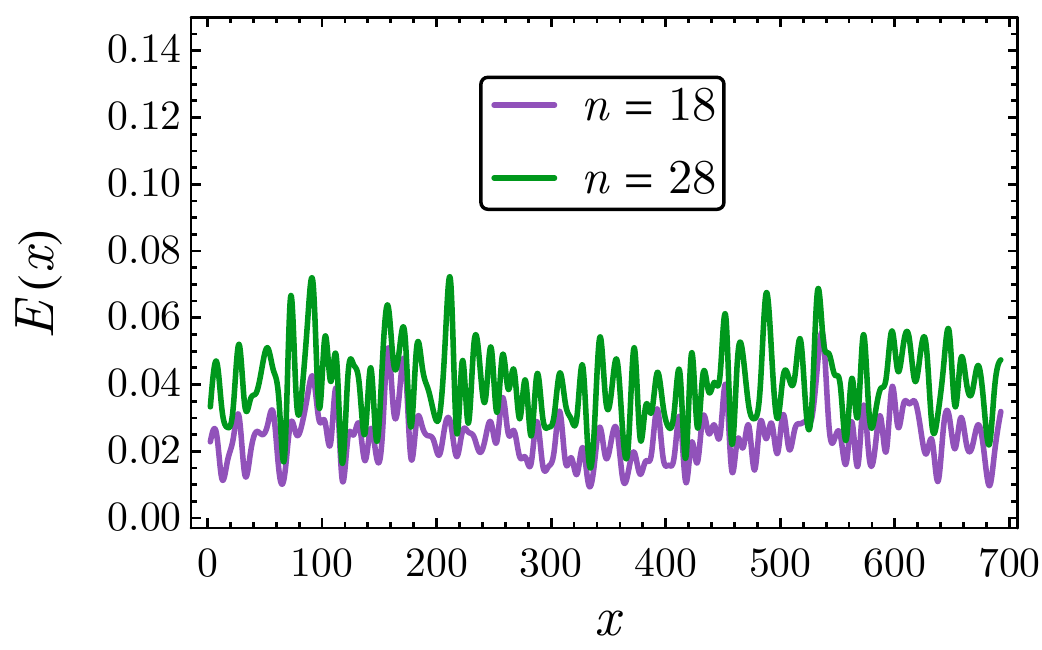}
    }
    \subfloat[Entropy density at late time]{
    \includegraphics[width=5cm]{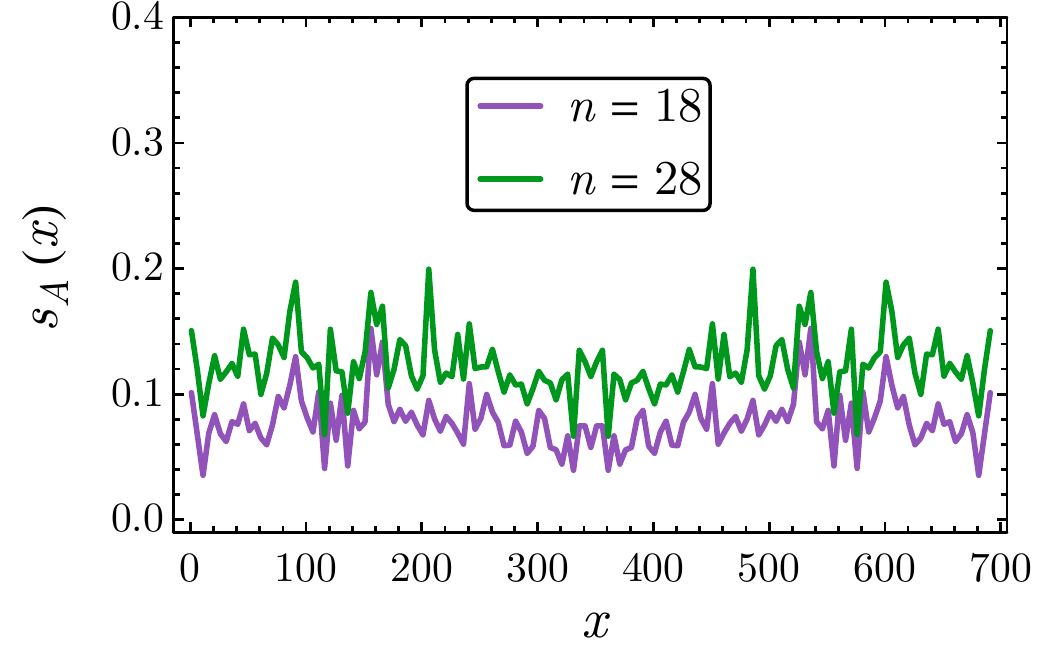}
    }
    \caption{Spatial structure in the lattice simulation. We choose the system size $L=700$, driving parameters $T_0/L=0.9$, $T_1/L=0.2$. We choose $\delta=5$ when performing the average. $n$ denotes the number of driving cycles. The black arrows in (a) and (b) indicates the peak positions predicted by the CFT calculation.}
    \label{fig:spatial structures lattice}
\end{figure}

In this section, we discuss the time evolution of energy density profile and the entanglement entropy density in the heating phase observed in a lattice calculation. 
The protocol is the one introduced by \cite{wen2018floquet} also reviewed in Sec.\ref{sec:review}. Simulation with the generalized setup in Sec.\ref{sec:generalization} yields the same results and thus is not included.

We simulate complex free fermion on an open chain with only nearest neighbor hopping at the half-filling. The results are shown in \figref{fig:spatial structures lattice}, with (a) (b) being the early time regime and (c) (d) being the late time regime. In the early time regime, both quantities show growing sharp peaks, whose positions are consistent with the CFT prediction as indicated by the arrows in the plots \figref{fig:spatial structures lattice}(a) and (b).
However, in the late time, as more and more excitations are created, the dynamics of the lattice system cannot be approximated by a CFT. One will see the spatial structure showing strong oscillation with time. The peaks also stop growing and finally give way to a smeared profile, as depicted in \figref{fig:spatial structures lattice}(c) and (d). Determining the timescale at which the prediction of conformal field theory begins to diverge from lattice calculations is a subtle question. Here we simply note that on comparing the energy or entropy density in this model using the parameters as in \figref{fig:spatial structures lattice}, the breakdown occurs around $n \sim 10$. It is roughly the time scale for $E_{\text{total}}/t \sim \calO(1)$ with $t$ being the hopping strength. On the other hand, the half-system entropy can agree with the CFT calculation for longer times, which in this model breaks down at $n\sim30$ (using the same parameters as \figref{fig:spatial structures lattice}).

%\RF{It is hard to identify the time scale for the breakdown of conformality precisely. Besides the calculation here, one can also study when the total energy and half-system entanglement for the lattice system start to deviate from the CFT prediction. That time scale for the total energy is quite close to what has been observed for the energy/entropy density (Using the same parameters as \figref{fig:spatial structures lattice}, it happens around $n\sim 10$, which is around the time when $E_{\text{total}}/t \sim \calO(1)$). However, the half-system entropy shows decent agreement up to a longer time. (Using the same parameters as \figref{fig:spatial structures lattice}, the breakdown time is around $n \lesssim 30$.)}

We close this section with some technical details of how the the data are extracted from the numerics.
The energy density $E(x)$ is obtained by computing the expectation value of the hopping term $-t\braket{c_i^\dag c_{i+1}+h.c.}$. We also perform an average over the nearest few sites to obtain a relatively smooth curve, i.e. $E(x) = \sum_{k=-\delta}^{\delta} -t\braket{c_{x+k}^\dag c_{x+k+1}+h.c.}$. The slight asymmetry of the plots with respect to the middle of the system is due to this average. Choosing different $\delta$ leads to results with the same qualitative features. 
The entanglement entropy density $s_A(x)$ is obtained by computing the entanglement entropy for the subsystem $[x-\delta,x+\delta]$. A similar average is also performed to obtain a smooth curve.

\bibliography{ref.bib}

\end{document}